\documentclass[a4paper, 11pt]{article}
\usepackage{jheppub}\makeatletter\gdef\@fpheader{}\makeatother
\usepackage{mathtools, amsfonts, microtype, dcolumn, booktabs}
\usepackage[bb=dsfontserif]{mathalpha} 
\usepackage[noautoscale]{youngtab}\Yboxdim 6.5pt\Ylinethick 0.4pt  
\usepackage[section]{placeins}\makeatletter\AtBeginDocument{\expandafter\renewcommand\expandafter\subsection\expandafter{\expandafter\@fb@secFB\subsection}}\makeatother
\usepackage[dvipsnames]{xcolor}\definecolor{darkgreen}{rgb}{0,0.4,0.2} 
\usepackage{orcidlink}\makeatletter\gdef\@OrigHeightRecip{0.005}\makeatother
\usepackage{tikz} 
\usetikzlibrary{arrows.meta}  
\hypersetup{linkcolor=blue, urlcolor=teal, citecolor=teal}

\graphicspath{{Figures/}} 

\newcommand\mc[1]{\multicolumn{1}{c|}{#1}}
\newcommand{\T}{\mathbb{T}} 
\newcommand{\Z}{\mathbb{Z}}  
\newcommand{\R}{\mathcal{R}}
\newcommand{\U}{\textrm{U}}
\newcommand{\SU}{\textrm{SU}}
\newcommand{\USp}{\textrm{USp}} 
  
\def\K3{\ensuremath{\mathbb{K}3}}

\title{Three-Family Supersymmetric Pati-Salam Flux Models from Rigid D-Branes}

\author[a]{Adeel Mansha,\,\orcidlink{0000-0002-1183-0355}}
\author[b]{Mudassar Sabir,\,\orcidlink{0000-0002-8551-2608}}
\author[c]{Tianjun Li,\,\orcidlink{0000-0003-1583-5935}} 
\author[a]{Luyang Wang\,\orcidlink{0000-0001-9400-7331}}

\affiliation[a]{College of Physics and Optoelectronic Engineering, Shenzhen University, 3688 Nanhai Avenue, Shenzhen, P. R. China}
\affiliation[b]{School of Physics, University of Electronic Science and Technology of China, 2006 Xiyuan Avenue, Chengdu, P. R. China} 
\affiliation[c]{School of Physics, Henan Normal University, 
Xinxiang 453007, P. R. China}

\emailAdd{adeelmansha@alumni.itp.ac.cn}
\emailAdd{mudassar.sabir@uestc.edu.cn}
\emailAdd{tli@itp.ac.cn}  
\emailAdd{wangly@szu.edu.cn}
 
\arxivnumber{2512.21141}

\abstract{Intersecting D-brane model building often suffer from the unstabilized open-string moduli, leading to the unwanted massless adjoint scalars. In our previous work~\cite{Mansha:2025yxm}, 
this issue was resolved by employing the rigid D6-branes on the $\mathbb{T}^6/(\mathbb{Z}_2 \times \mathbb{Z}_2^\prime)$ orientifold with discrete torsion, where fractional cycles eliminate all adjoint scalars. In this paper, we construct new three-family flux models in the Type IIB setup on $\mathbb{T}^6/(\mathbb{Z}_2 \times \mathbb{Z}_2)$, T-dual to the Type IIA rigid D6-brane construction with discrete torsion, by introducing the quantized background $G_3$ flux that stabilizes the closed-string complex structure moduli and axio-dilaton. The resulting Pati-Salam gauge symmetry can be spontaneously broken down to the Standard Model via a supersymmetry-preserving Higgs mechanism. All the consistency conditions, including $\mathcal{N}=1$ supersymmetry, RR tadpole cancellation, and K-theory constraints, are satisfied. We present the complete particle spectra for these models and discuss how exotic states dynamically decouple through strong dynamics in the hidden sector.}

\begin{document} 
\maketitle

\section{Introduction}\label{sec:Intro}

A central goal of string phenomenology is to construct the realistic string vacua that reproduce the observed features of the Standard Model (SM) at low energies, while ensuring the consistent stabilization of all geometric and scalar moduli. In particular, intersecting D6-brane\footnote{A D$p$-brane in a compactification on $\mathbb{R}^{1,D-1} \times X^{10-D}$ can wrap an internal cycle as long as its worldvolume is equal to spacetime, i.e. $WV(Dp) = \mathbb{R}^{1,D-1} \times \Pi_{p-(D-1)}$, where $\Pi_{p-(D-1)}$ is a $(p-(D-1))$-cycle in $X^{10-D}$.} models in Type IIA string theory on the $\mathbb{T}^6/(\mathbb{Z}_2 \times \mathbb{Z}_2)$ orientifold have proven to be a fertile ground for realizing the $\mathcal{N}=1$ supersymmetric chiral spectra with SM–like features, cf.~\cite{Cvetic:2004ui, Chen:2007zu, Li:2019nvi, Li:2021pxo, Sabir:2022hko, Mansha:2022pnd, Mansha:2023kwq, Mansha:2024yqz}. More recently, the complete landscape of such vacua has been mapped in \cite{He:2021kbj} and their phenomenology has been worked out in \cite{Sabir:2024cgt, Sabir:2024jsx}, leading to precise predictions of Dirac-type neutrino masses~\cite{Sabir:2024mfv}. However, such models typically suffer from the unstabilized open-string moduli, arising from fluctuations of the D-branes' positions in the internal space and Wilson lines, which generically introduce non-chiral matter in adjoint, symmetric, or antisymmetric representations. These states can spoil realistic spectra unless the moduli are completely frozen.

Within Type IIA string theory, open-string moduli corresponding to D-brane positions and Wilson lines can be completely {\it frozen} on the mirror $\mathbb{T}^6/(\mathbb{Z}_2 \times \mathbb{Z}_2')$ orientifold with discrete torsion\footnote{The mirror $\mathbb{T}^6/(\mathbb{Z}_2 \times \mathbb{Z}_2')$ orientifold \emph{with} discrete torsion has Hodge numbers $(h^{1,1}, h^{2,1}) = (3,51)$, whereas the corresponding orientifold \emph{without} discrete torsion, $\mathbb{T}^6/(\mathbb{Z}_2 \times \mathbb{Z}_2)$, has the exchanged cohomology $(h^{1,1}, h^{2,1}) = (51,3)$.}, as explicitly demonstrated in \cite{Blumenhagen:2005tn} through the construction of four-family Pati-Salam models. In our recent works \cite{Mansha:2025gvr, Mansha:2025yxm}, we have constructed several consistent three-family Pati-Salam models in the same setup, incorporating appropriate Higgs fields that enable the spontaneous breaking of the Pati-Salam gauge symmetry down to the SM gauge group. In this setup with torsion, the fractional D6-branes wrap rigid 3-cycles fixed under all twisted sectors, preventing any non-chiral adjoint matter from appearing, while maintaining the non-trivial intersection numbers that generate the desired chiral spectrum.

Moduli stabilization can also be addressed in the T-dual Type IIB framework, where the same models correspond to D-branes on a $\mathbb{T}^6/(\mathbb{Z}_2 \times \mathbb{Z}_2)$ orientifold without discrete torsion. In Type IIB string theory, background bulk three-form fluxes $G_3 = F_3 - \tau H_3$ stabilize the axio-dilaton and complex structure moduli~\cite{Blumenhagen:2003vr}. As usual $F_3=d C_2$ and $H_3=d B_2$ denote the Ramond-Ramond and Neveu-Schwarz three-form fluxes, respectively, and $\tau = C_0 + i e^{-\phi}$ is the axio-dilaton, with $C_0$ the RR scalar and $\phi$ the dilaton. The Kähler moduli, however, are not stabilized by $G_3$ flux at tree level and require additional mechanisms 
for their stabilization. Since no fully consistent three-family Pati-Salam models with discrete torsion and rigid branes existed prior to \cite{Mansha:2025gvr, Mansha:2025yxm}, the existing flux models in the literature, cf.~\cite{Blumenhagen:2003vr, Cascales:2003zp, Marchesano:2004xz, Cvetic:2005bn, Chen:2005mj, Chen:2006gd, Chen:2007af, Chen:2011jba}, are T-dual to Type IIA models without discrete torsion, where twisted sectors are absent, D-branes are generally non-rigid, and exotic O3$^{++}$-planes do not appear. 

In this paper, we fill this gap by presenting the explicit supersymmetric three-family Pati-Salam flux models that include exotic O3$^{++}$-planes\footnote{The sign $++$ correspond to the charge and tension, respectively \cite{Blumenhagen:2006ci}.}, T-dual to the Type IIA $\mathbb{T}^6/(\mathbb{Z}_2 \times \mathbb{Z}_2')$ orientifold with discrete torsion, {\it i.e.}, including the twisted sectors. These models satisfy all the consistency conditions, including the RR tadpole cancellation (taking into account the flux-induced contributions), $\mathcal{N}=1$ supersymmetry, and K-theory constraints. For general $\mathbb{Z}_N$ orbifolds with $N>2$, fractional D6-branes wrapping twisted 3-cycles may naively seem to remove adjoint scalars; however, additional adjoint fields can arise from non-trivial intersections of orbifold images, $\Theta_g \Pi_a$ and $\Theta_h \Pi_a$ with $g \neq h$ \cite{Blumenhagen:2001te}, each producing a chiral multiplet in the adjoint. This effect has been explicitly demonstrated in, e.g., the $\T^6/\mathbb{Z}_4$ orbifold \cite{Blumenhagen:2002gw}. In contrast, the $\T^6/(\mathbb{Z}_2 \times \mathbb{Z}_2)$ orientifold with discrete torsion provides a special setup where fractional D6-branes wrap rigid 3-cycles fixed under all twisted sectors, and different orbifold images of a given cycle do not intersect, preventing unwanted adjoint matter from appearing. This makes it the optimal choice for constructing $\mathcal{N}=1$ chiral models without adjoint scalars.

From a model-building perspective, these constructions are similar in spirit to those in \cite{Mansha:2025yxm}, but include quantized background $G_3$-flux. The flux modifies the RR tadpole cancellation conditions and invariably makes the construction of consistent three-family models technically challenging. Our construction employs rigid branes for the visible sector, combined with semi-rigid or non-rigid hidden-sector branes, to ensure the cancellation of both twisted and untwisted RR tadpoles, including flux contributions, while simultaneously realizing three chiral families. A key feature is the presence of  Higgs fields, which spontaneously break the Pati-Salam gauge symmetry to the SM gauge group via a supersymmetry-preserving Higgs mechanism. Following \cite{Mansha:2025yxm}, only rectangular two-tori are used, and the symmetry breaking is achieved through the controlled recombination of hidden-sector branes with the $\SU(2)_R$ stack, effectively giving Vacuum Expectation Values (VEVs) to the Higgs fields while preserving $\mathcal{N}=1$ supersymmetry. The complete perturbative massless spectra of all models are presented, including chiral multiplets and vector-like states. All the exotic states not part of the SM acquire masses dynamically via strong hidden-sector dynamics (confinement, gaugino condensation, or stringy instantons), decoupling from the low-energy theory. Consequently, the low-energy effective theory retains exactly three chiral families with the desired SM quantum numbers, free from non-chiral adjoint or exotic matter.

The paper is organized as follows. In section~\ref{sec:model-building}, we review the rules for constructing three-family Pati-Salam models in Type IIA $\T^6/(\mathbb{Z}_2 \times \mathbb{Z}_2^\prime)$ orientifold with discrete torsion including the computation of intersection numbers, associated consistency conditions from $\mathcal{N}=1$ supersymmetry, K-theory constraints, and RR tadpole cancellation. In section~\ref{sec:IIBflux}, we consider the T-dual Type~IIB description on the $\T^6/(\mathbb{Z}_2 \times \mathbb{Z}_2)$ orientifold, where quantized $G_3$ flux is introduced. We discuss its impact on tadpole cancellation, the Chern-Simons coupling to $C_4$, and the stabilization of closed-string moduli. In section~\ref{sec:models}, we present the explicit examples of consistent three-family models, detailing the complete particle spectra and gauge groups. We also discuss phenomenological aspects, including the decoupling of exotic states via hidden-sector dynamics. Section~\ref{sec:conclusion} summarizes our findings and outlines directions for future work.
 
\section{Flux model building on rigid cycles}\label{sec:model-building}
Let us consider Type IIA string theory compactified on the factorized six-torus $\T^6 = \T^2_1 \times \T^2_2 \times \T^2_3$, modded by the orbifold group $\mathbb{Z}_2 \times \mathbb{Z}_2'$ with generators $\theta$ and $\omega$, associated with twist vectors $v_\theta = (1/2,-1/2,0)$ and $v_\omega = (0,1/2,-1/2)$, specifying a rotation $z_i \mapsto e^{2\pi i v_i} z_i$ on the corresponding torus $T^2_i$ as \cite{Blumenhagen:2005mu},
\begin{align}
	\theta & : (z_1, z_2, z_3) \mapsto (-z_1, -z_2, z_3), \nonumber \\
	\omega & : (z_1, z_2, z_3) \mapsto (z_1, -z_2, -z_3).           
\end{align}
Orientifold projection corresponds to gauging the $\Omega \R$ symmetry, where $\Omega$ is worldsheet parity that interchanges the left- and right-moving sectors of a closed string and swaps the two ends of an open string whereas $\R$ acts as complex conjugation on $z_i$:
\begin{equation}
	\begin{array}{ccccc}
		\Omega :      & (\sigma_1, \sigma_2) & \mapsto & (2\pi - \sigma_1, \sigma_2) & \text{(Closed)} \\
		              & (\tau, \sigma)       & \mapsto & (\tau, \pi - \sigma)        & \text{(Open)}   \\
		\mathcal{R} : & z_i                  & \mapsto & \overline{z}_i.             &                 
	\end{array}
\end{equation}
The orientifold action $\R$ acts on the real coordinates $(x_i, y_i)$ of each two-torus $\T^2_i$ as
\begin{equation}
	\R: (x_i, y_i) \mapsto (x_i, -y_i) \mod 1.
\end{equation} 
Thus, the fixed locus of $\R$ on each $\T^2_i$ consists of two lines, $y_i = 0$ and $y_i = \tfrac{1}{2}$, corresponding respectively to the one-cycles $[a^i]$ on a rectangular torus and $[a'^i] = [a^i] + \tfrac{1}{2}[b^i]$ on a tilted torus. Denoting the wrapping numbers on the rectangular and tilted two-tori as $n_a^i [a^i] + m_a^i [b^i]$ and $n_a^i [a'^i] + m_a^i [b^i]$, respectively, a generic one-cycle $(n_a^i, l_a^i)$ satisfies $l_a^i = m_a^i$ on a rectangular torus and $l_a^i = 2\tilde m_a^i = 2m_a^i + n_a^i$ on a tilted torus, implying that $l_a^i - n_a^i$ is even for the tilted case. The two bases $(n^i, m^i)$ and $(n^i, l^i)$ are related by
\begin{align}\label{basis-l-m}
	l^i = 2^{\beta_i} \Bigl( m^i + \frac{\beta_i}{2} n^i \Bigr), \quad
	\beta_i = \begin{cases} 
	0, & \text{rectangular } \T^2, \\ 
	1, & \text{tilted } \T^2.      
	\end{cases}                                           
\end{align}
In the $\mathbb{Z}_2 \times \mathbb{Z}_2'$ orbifold, there are four orientifold projections, $\Omega \R$, $\Omega \R \theta$, $\Omega \R \omega$, and $\Omega \R \theta\omega$, each producing its own set of fixed 3-cycles. Since each two-torus $\T^2_i$ has two distinct fixed one-cycles under $\R$, a bulk 3-cycle on $(\T^2)^3$ can be chosen independently along each torus, giving $2^3 = 8$ fixed 3-cycles per orientifold action before accounting for tilted tori. When a two-torus $\T^2_i$ is tilted, the basis cycles are related by $[a'^i] = [a^i] + \tfrac{1}{2}[b^i]$, so that the two fixed lines under $\R$ become identified, thus reducing the number of distinct orientifolds by a factor of 2 per tilted torus \cite{Cvetic:2001nr},
\begin{equation}
	N_{\rm O6}^{\rm bulk} = 2^{-k}\cdot 32\,, \qquad \because k=\sum_{i=1}^3\beta^i.
\end{equation} 
In orientifold constructions, the presence of O6-planes is described in the closed string channel by {\it crosscap states} $|\Omega \mathcal{R} \, g\rangle$, where the overlap of two crosscap states  
\begin{align}
	\langle \Omega \mathcal{R} \, g_1 | \, e^{-l \mathcal{H}_{\text{cl}}} \, | \Omega \mathcal{R} \, g_2 \rangle 
\end{align}  
computes the {\it tree-level propagation} of a closed string between the two O6-planes $g_1$ and $g_2$, with $\mathcal{H}_{\text{cl}}$ the closed string Hamiltonian and $l$ the proper length of the cylinder.

By worldsheet duality (modular transformation), the same amplitude can be expressed in the open string channel as a {\it one-loop trace over open strings} in the sector twisted by $g_2 g_1^{-1}$:  
\begin{align}\label{crosscap}
	\langle \Omega \mathcal{R} | e^{-l \mathcal{H}_{\text{cl}}} | \Omega \mathcal{R} \omega \rangle 
	  & = \text{Tr}_{\omega} \Big( \Omega \mathcal{R} \, e^{-2\pi t H} \Big), \nonumber \\
	\langle \Omega \mathcal{R} \theta | e^{-l \mathcal{H}_{\text{cl}}} | \Omega \mathcal{R} \theta \omega \rangle 
	  & = \text{Tr}_{\omega} \Big( \Omega \mathcal{R} \theta \, e^{-2\pi t H} \Big),    
\end{align}  
where $t = 1/(2l)$ is the modular parameter in the open string channel, $\text{Tr}_{\omega}(\dots)$ denotes the trace over open string states in the $\omega$-twisted sector, and $\Omega \mathcal{R}$ acts as an involution on the open string Hilbert space.

Physically, eqs.~\eqref{crosscap} encode the \emph{crosscap consistency condition}, ensuring that the orientifold projection is compatible with both the orbifold group and the chosen discrete torsion $\eta = \pm 1$. For the $\mathbb{Z}_2 \times \mathbb{Z}_2$ orbifold, this requires
\begin{align}
	\eta_{\Omega \mathcal{R}} \, \eta_{\Omega \mathcal{R}\theta} \, \eta_{\Omega \mathcal{R}\omega} \, \eta_{\Omega \mathcal{R}\theta\omega} = \eta \,, \label{opsigns} 
\end{align}
where $\eta_{\Omega \mathcal{R} g} = \pm 1$ denotes the type of O6-plane (ordinary O6$^{(+,+)}$ or exotic O6$^{(-,-)}$). In particular, consistency with discrete torsion $\eta = -1$ necessitates that an odd number of exotic O6$^{(+,+)}$-planes~\cite{Angelantonj:1999ms, Blumenhagen:2005tn}.

\subsection{Untwisted sector and 4D $\mathcal{N}=2$ multiplets}

In ten dimensions, the graviton $g_{MN}$ has $\frac{D(D+1)}{2}=55$ components for $D=10$. Imposing the traceless condition $h^M_M = 0$ removes one degree of freedom, the transverse condition $\partial^M h_{MN}=0$ removes 10 more, and residual gauge transformations $\xi_M$ satisfying $\Box \xi_M=0$ and $\partial^M \xi_M=0$ remove an additional 9 components, leaving 35 physical degrees of freedom.\footnote{This yields $N_{\rm graviton} = \frac{D(D+1)}{2} - 1 - D - (D-1) = \frac{D(D-3)}{2}$, giving 35 in ten dimensions.} Under $\mathrm{SO}(3,1) \times \mathrm{SO}(6)$, the metric decomposes into a 4D graviton $g_{\mu\nu}$ (10 components), six 4D vectors $g_{\mu m}$ (24 components), and 21 internal scalars $g_{mn}$.

The $\mathbb{Z}_2 \times \mathbb{Z}_2'$ orbifold acts as discrete reflections on the internal coordinates. Using real coordinates, the three tori correspond to $(x^1,x^2)$, $(x^3,x^4)$, and $(x^5,x^6)$. Each metric component $g_{mn}$ transforms with parity $P_m P_n$ under each $\mathbb{Z}_2$, where $P_m = \pm 1$ is the intrinsic parity of $x^m$. Only components even under all generators survive. This eliminates all off-diagonal elements mixing different tori, e.g., $g_{13}$ or $g_{24}$, leaving only the intra-torus components $g_{11}, g_{22}, g_{33}, \dots, g_{66}$. Consequently, there are three independent metric deformations per two-torus, giving nine real internal scalars from $g_{mn}$.

These scalars organize into geometric moduli. Six of them combine into three complex structure moduli $U^i$ $(i=1,2,3)$, controlling the shapes of the tori $\mathbb{T}^2_i$. The remaining three pair with scalars from the NS--NS two-form $B_2$ to form complexified Kähler moduli
\begin{equation}
	T^i = b^i + i\, t^i \,,
\end{equation}
where $b^i=\int_{\Sigma_2^i} B_2$ arises from the NS--NS two-form integrated over the corresponding two-cycle $\Sigma_2^i$, and $t^i=\text{Vol}(\Sigma_2^i)$ comes from the geometric volume of that two-cycle. The R-R 4-form $C_4$ also contributes scalars through components with all indices internal, of which there are $\binom{6}{4} = 15$ possibilities. Only three specific components survive the orbifold projection: $C_{1234}$, $C_{1256}$, and $C_{3456}$. Including the axio-dilaton $\tau = C_0 + i\, e^{-\phi}$, these fields complete the untwisted closed-string sector.
 
The untwisted spectrum gives a four-dimensional $\mathcal{N}=2$ supergravity multiplet $(g_{\mu\nu}, \psi_{\mu\alpha}, \psi_{\mu\dot{\alpha}}, A_\mu^\text{grav})$ containing the graviton $g_{\mu\nu}$, two opposite chirality gravitini $\psi_{\mu\alpha}, \psi_{\mu\dot{\alpha}}$, and the graviphoton $A_\mu^\text{grav}$; one universal hypermultiplet $(\phi, \zeta_\alpha, \tilde{\zeta}_{\dot{\alpha}}, S)$ containing the dilaton $\phi$, two fermions $\zeta_\alpha, \tilde{\zeta}_{\dot{\alpha}}$, and one complex scalar $S$; three vector multiplets $(A_\mu^i, \phi^i, \lambda_\alpha^i, \tilde{\lambda}_{\dot{\alpha}}^i)$ corresponding to $T^i$, each containing a vector $A_\mu^i$, a complex scalar $\phi^i$, and two gaugini $\lambda_\alpha^i, \tilde{\lambda}_{\dot{\alpha}}^i$; and three hypermultiplets $(Q^i, \tilde{Q}^i, \psi_\alpha^i, \tilde{\psi}_{\dot{\alpha}}^i)$ associated with $U^i$, each containing two complex scalars $Q^i, \tilde{Q}^i$ and two fermions $\psi_\alpha^i, \tilde{\psi}_{\dot{\alpha}}^i$. These multiplets collectively parameterize the moduli space before fluxes or orientifold projections \cite{Grana:2005jc, Blumenhagen:2006ci}.

Upon performing the orientifold projection with $\Omega \mathcal{R}$ in the $\mathbb{Z}_2 \times \mathbb{Z}_2'$ orientifold with discrete torsion ($\eta = -1$), the $\mathcal{N}=2$ supersymmetry is broken to $\mathcal{N}=1$. The supergravity multiplet reduces to $(g_{\mu\nu}, \psi_{\mu\alpha})$, while the second gravitino $\psi_{\mu\dot{\alpha}}$ and the graviphoton $A_\mu^\text{grav}$ are projected out.

Each of the three $\mathcal{N}=2$ vector multiplets $(A_\mu^i, \phi^i, \lambda_\alpha^i, \tilde{\lambda}_{\dot{\alpha}}^i)$ decomposes into an $\mathcal{N}=1$ vector multiplet $(A_\mu^i, \lambda_\alpha^i)$ and a chiral multiplet $(\phi^i, \tilde{\lambda}_{\dot{\alpha}}^i)$. The orientifold projection acts such that the vectors $A_\mu^i$ are projected out, leaving only three $\mathcal{N}=1$ chiral multiplets $(T^i, \chi^i)$.

Similarly, the three hypermultiplets $(Q^i, \tilde{Q}^i, \psi_\alpha^i, \tilde{\psi}_{\dot{\alpha}}^i)$ corresponding to the complex structure moduli $U^i$ split into two chiral multiplets each, with one surviving the projection, yielding three $\mathcal{N}=1$ chiral multiplets $(U^i, \psi_{U^i})$. The universal hypermultiplet $(\phi, \zeta_\alpha, \tilde{\zeta}_{\dot{\alpha}}, S)$ reduces to a chiral multiplet $(S, \chi_S)$, where $S = e^{-\phi} + i C_0$.

After the orientifold, the untwisted closed string sector contains the $\mathcal{N}=1$ gravity multiplet and seven chiral multiplets $(S, T^i, U^i)$. Nonperturbative effects generate a superpotential for the Kähler moduli:
\begin{equation}
	W_\text{np} \sim A \, e^{-a T^i} = A \, e^{- a t^i} \, e^{- i a b^i},
\end{equation}
with $T^i = b^i + i t^i$, $t^i$ the geometric volume of the corresponding two-cycle, and $b^i$ the NS--NS axion. Here $a = 2\pi$ for a single E3-instanton or $a = 2\pi/N$ for gaugino condensation on an SU($N$) stack. This stabilizes the Kähler moduli alongside the flux-induced superpotential for $(U^i, \tau)$ \cite{Kachru:2003aw, Grana:2005jc}.

\subsection{Twisted sectors and fractional branes}

Twisted sectors generated by $\theta$, $\omega$, and $\theta\omega$ arise from strings localized at orbifold fixed points. Each $\mathbb{Z}_2$ acts nontrivially on two tori, leaving the third invariant. Each twisted torus has four fixed points, giving $4\times4=16$ fixed tori per twist. Locally, the geometry is $\mathbb{C}^2/\mathbb{Z}_2\times\T^2$, supporting localized modes.

Without discrete torsion, blowing up these singularities contributes 48 complex structure moduli, giving Hodge numbers $(h^{1,1},h^{2,1})=(51,3)$. With discrete torsion, the twisted sectors contribute 48 Kähler moduli instead. Upon T-duality between IIA and IIB frames,
\begin{align}
	(h^{1,1},h^{2,1})_\text{IIA} = (3,51),\quad (h^{1,1},h^{2,1})_\text{IIB} = (51,3), 
\end{align}
reflecting a mirror-like exchange between Kähler and complex structure moduli.

Under the $\mathbb{Z}_2\times\mathbb{Z}_2'$ action, a factorizable 3-cycle on $\T^6$ has three orbifold images with the same wrapping numbers. Hence a bulk 3-cycle in the orbifold can be identified as $[\Pi_a^B] = 4[\Pi_a^{\T^6}]$. The bulk intersection number is then
\begin{align} \label{prod}
	[\Pi_a^B]\circ[\Pi_b^B] & =4[\Pi_a^{\T^6}]\circ[\Pi_b^{\T^6}] =4\prod_{i=1}^3(n_a^i\widetilde m_b^i-\widetilde m_a^in_b^i),\nonumber \\
	                        & =4\cdot 2^{-k}\prod_{I=1}^3(n_a^il_b^i-l_a^in_b^i),                                                        
\end{align}
where we have identified intersection points related by the $\mathbb{Z}_2\times\mathbb{Z}_2'$ action.

In addition to the untwisted cycles, there are 32 independent collapsed three-cycles for each twisted sector $g=\theta,\omega,\theta\omega$. For example, in the $\theta$-twisted sector we denote the 16 fixed points on $(\T^2_i\times\T^2_j)/\mathbb{Z}_2$ by $[e_{ij}^\theta]$, with $i,j\in\{1,2,3,4\}$. After blowing up the orbifold singularities, these become two-cycles with $\mathbb{S}^2$ topology. Each such $\T^4/\mathbb{Z}_2$ is locally a $\K3$ surface before taking the orientifold action into account. With discrete torsion, these two-cycles combine with a one-cycle $(n^3,\widetilde m^3)$ of $\T^2_3$ to form a three-cycle in the $\theta$-twisted sector:
\begin{align}
	[\alpha^\theta_{ij,\,n}] = 2[e^\theta_{ij}]\otimes[a^3],\qquad 
	[\alpha^\theta_{ij,\,m}] = 2[e^\theta_{ij}]\otimes[b^3],       
\end{align}
where the factor of two arises from the second $\mathbb{Z}_2$ action. Analogously, in the $\omega$ and $\theta\omega$ sectors we define
\begin{align}
	[\alpha^{\omega}_{ij,\,n}]       & =2[e^{\omega}_{ij}]\otimes[a^1],       &   
	[\alpha^{\omega}_{ij,\,m}]&=2[e^{\omega}_{ij}]\otimes[b^1],\nonumber\\
	[\alpha^{\theta\omega}_{ij,\,n}] & =2[e^{\theta\omega}_{ij}]\otimes[a^2], &   
	[\alpha^{\theta\omega}_{ij,\,m}]&=2[e^{\theta\omega}_{ij}]\otimes[b^2].
\end{align}

The collapsed two-cycles of the \K3 orbifold satisfy $[e_{ij}]\circ[e_{kl}]=-2\,\delta_{ik}\delta_{jl}$, and two-cycles of different twisted sectors do not intersect. For three-cycles 
\begin{align}
	[\Pi^g_{ij,\,a}] & = n_a^{I_g}[\alpha_{ij,\,n}] + \widetilde m_a^{I_g}[\alpha_{ij,\,m}],\qquad \nonumber \\
	[\Pi^h_{kl,\,b}] & = n_b^{I_h}[\alpha_{kl,\,n}] + \widetilde m_b^{I_h}[\alpha_{kl,\,m}],                 
\end{align}
with $g,h=\theta,\omega,\theta\omega$, we obtain
\begin{align}\label{inttwist}
	[\Pi^g_{ij,\,a}]\circ[\Pi^h_{kl,\,b}]
	  & =4\,\delta_{ik}\delta_{jl}\delta^{gh}(n_a^{I_g}\widetilde m_b^{I_g}-\widetilde m_a^{I_g}n_b^{I_g})\nonumber \\
	  & =4\,\delta_{ik}\delta_{jl}\delta^{gh}2^{-\beta^g}(n_a^{I_g}l_b^{I_g}-l_a^{I_g}n_b^{I_g}),                   
\end{align}
where $\widetilde m_a^{I_g}\equiv2^{-\beta^g}l_a^{I_g}$, and $I_g=3,1,2$ for $g=\theta,\omega,\theta\omega$, respectively.

\begin{table}[t]
	\renewcommand{\arraystretch}{1.35}\centering
	\begin{tabular}{|c|c|}
		\hline
		$(n_a, m_a)$ & Fixed points $S^a_g$   \\
		\hline \hline
		(odd, odd)   & $\{1,4\}$ or $\{2,3\}$ \\
		(odd, even)  & $\{1,3\}$ or $\{2,4\}$ \\
		(even, odd)  & $\{1,2\}$ or $\{3,4\}$ \\
		\hline
	\end{tabular}
	\caption{Fixed points of a one-cycle on $\T^2/\Z_2$ in terms of its wrapping numbers.}
	\label{fixed}
\end{table}

\subsection{Spectrum from rigid branes (\texorpdfstring{$\eta =-1$}{η = -1})}

To construct rigid D6-branes, one considers \emph{fractional} D6-branes that carry charges under all three $\Z_2$ twisted sectors of the orbifold. We begin with a factorizable three-cycle, characterized by three pairs of wrapping numbers $(n_a^i, \widetilde m_a^i)$. A fractional D6-brane must be invariant under the orbifold action, and hence it passes through four fixed points on each of the three twisted sectors. 

We denote by $S_g^a$ the set of four fixed points associated with a given orbifold element $g$, where each fixed point is labeled by a pair $(i,j)$. The pattern of fixed points $S_g^a$ can be directly determined from the parity of the wrapping numbers $(n_a^i, m_a^i)$ on the corresponding two-torus, as summarized in Table~\ref{fixed}.

The full three-cycle wrapped by such a fractional D6-brane is then expressed as
\begin{align} \label{rigid}
	\Pi_a 
	  & = \frac{1}{4}\, \Pi^B_a 
	+ \frac{1}{4} \sum_{(i,j) \in S_\theta^a} 
	\epsilon^\theta_{a,ij}\, \Pi^\theta_{ij,\,a} 
	+ \frac{1}{4} \sum_{(j,k)\in S_{\omega}^a} 
	\epsilon^{\omega}_{a,jk}\, \Pi^{\omega}_{jk,\,a}  + \frac{1}{4} \sum_{(i,k)\in S_{\theta\omega}^a} 
	\epsilon^{\theta\omega}_{a,ik}\, \Pi^{\theta\omega}_{ik,\,a}\ ,
\end{align}
where $\Pi^B_a$ denotes the bulk three-cycle and the coefficients 
$\epsilon^\theta_{a,ij}$, $\epsilon^{\omega}_{a,jk}$, and $\epsilon^{\theta\omega}_{a,ik} = \pm 1$ 
encode the brane's twisted charges with respect to the localized massless fields at the corresponding fixed points. Geometrically, these signs specify the orientation with which the brane wraps each exceptional $\mathbb{S}^2$ at the resolved fixed points.

Only those fixed points that the D6-brane passes through contribute to (\ref{rigid}). Because the brane is localized at the orbifold singularities in all three $\T^2_i$ factors, it cannot move away from them, and thus no adjoint scalars appear in the massless spectrum. Note that it is \emph{necessary} to use $m^i$ (and not $l^i$) to compute the fixed points in \eqref{rigid}, while $\widetilde{m}^i$ (or equivalently $2^{-\beta}l^i$) is useful in the calculation of intersection numbers, tadpole cancelation and the supersymmetry conditions \cite{Forste:2010gw}. 

\begin{table}[t]
	\centering\renewcommand{\arraystretch}{1.35}
	$\begin{array}{|c|c|}
		\hline
		\text{Representation}            & \text{Multiplicity}                                                       \\
		\hline\hline
		(\yng(1)_a,\overline{\yng(1)}_b) & \Pi_a\circ \Pi_{b}                                                        \\
		(\yng(1)_a,\yng(1)_b)            & \Pi_a\circ \Pi'_{b}                                                       \\
		\yng(1,1)_a                      & \frac{1}{2}\left(\Pi_a\circ \Pi'_a + \Pi_{a}\circ \Pi_{\text{O}6}\right)  \\
		\yng(2)_a                        & \frac{1}{2}\left(\Pi_a\circ \Pi'_a - \Pi_{a} \circ \Pi_{\text{O}6}\right) \\
		\hline
	\end{array}$
	\caption{Chiral spectrum for intersecting D6-branes}
	\label{tcs}
\end{table} 

For D6-branes on 3-cycles not invariant under $\R$, the gauge group is $\prod_a \U(N_a)$. The massless left-handed chiral spectrum is then determined by 3-cycle intersection numbers, including fermions in symmetric and anti-symmetric representations of $\U(N)$, as summarized in table~\ref{tcs}.

Given \eqref{prod} and \eqref{inttwist}, it is now easy to compute the intersection number between two rigid D6-branes of the form (\ref{rigid}).
\begin{align}\label{Iab}
	\Pi_a^F \circ \Pi_b^F & = \frac{1}{4}\Bigg(2^{-k}\prod_I (n^i_a\,l_b^i - l_a^i\,n^i_b)  + \sum _{g\in \{3,1,2\}} \delta_{ab}^g\,2^{-\beta^{g}}(n^g_a\,l_b^g - l_a^g\,n^g_b)\Bigg), 
\end{align} 
where $k$ is the number of tilted tori and $\delta_{ab}^g$ is the number of common fixed points where the branes $a$ and $b$ intersect for each twisted sector $g \in \theta, \omega,\theta\omega$ and can be read from \eqref{inttwist} as,
\begin{align}\label{eq:deltaij}
	\delta_{a b}^g\equiv \delta^g(\alpha^a_{ij}, \alpha^b_{kl}) & = \delta^g_{S^a_{i},S^b_{k}} \delta^g_{S^a_{j},S^b_{l}}. 
\end{align}
Assuming that every fractional brane intersects the origin the above relation simplifies as,
\begin{align}
	\delta_{ab}^g & = \sum _{i=1}^2 \sum _{j=1}^2 \sum _{k=1}^2 \sum _{l=1}^2 \delta^g_{S^a_{1,i},S^b_{1,k}} \delta^g_{S^a_{2,j},S^b_{2,l}}~. \label{eq:delta} 
\end{align}

\subsubsection{Orientifold action}  

\begin{figure}[t]
	\centering
	\begin{tikzpicture}[x=3cm, y=3cm, scale=1] 
		\begin{scope}
			\draw[-Latex] (0,0) -- (1.5,0) node[right] {$x^i$};
			\draw[-Latex] (0,0) -- (0,1.5) node[above] {$y^i$};
			\draw[dotted] (0,.5) -- (1,.5) ;
																																																  
			\draw (0,0) rectangle (1,1);
			\draw[thick, Red] (0,0) -- (1,0) node[midway, below] {$[a^i]$}; 
			\draw[thick, RoyalBlue] (0,0) -- (0,1) node[midway, left] {$[b^i]$}; 
																																																   
			\node at (.5,1.5) {$\beta^i = 0$}; 
			\node at (1,0) [below]{$R_1^i$};
			\node at (0,1) [left]{$R_2^i$};
			\fill[darkgreen] (0,0) circle (2pt) node[above right]{1};
			\fill[darkgreen] (0,.5) circle (2pt) node[above right]{2};
			\fill[darkgreen] (.5,0) circle (2pt) node[above]{3};
			\fill[darkgreen] (.5,.5) circle (2pt) node[above]{4};
		\end{scope}
		\begin{scope}[xshift=7cm]
			\draw[-Latex] (0,0) -- (1.5,0) node[right] {$x^i$};
			\draw[-Latex] (0,0) -- (0,1.5) node[above] {$y^i$};
			\draw[dotted] (0,.5) -- (1,.5) ;
			\draw[dotted] (0,1) -- (1,1) ;
																																																
			\draw (0,0) -- (1,.5) -- (1,1.5) -- (0,1);
			\draw[thick, Red] (0,0) -- (1,0) node[midway, below] {$[a^i]$}; 
			\draw[thick, RoyalBlue] (0,0) -- (0,1) node[midway, left] {$[b^i]$}; 
			\draw[thick, Green] (0,0) -- (1,.5) node[near end, below] {$[a'^i]$};
																																																  
			\node at (.5,1.5) {$\beta^i = 1$}; 
			\draw (1,-0.025) -- (1,0.025) ;
			\node at (1,0) [below]{$R_1^i$}; 
			\node at (0,1) [left]{$R_2^i$}; 
			\fill[darkgreen] (0,0) circle (2pt) node[above right, yshift=0.05cm]{1};
			\fill[darkgreen] (0,.5) circle (2pt) node[above right]{2};
			\fill[darkgreen] (.5,.25) circle (2pt) node[above]{3};
			\fill[darkgreen] (.5,.75) circle (2pt) node[above]{4};
		\end{scope}
	\end{tikzpicture}  
	\caption{The $\Z_2$ invariant \textbf{a}-type (left) and \textbf{b}-type (right) lattices.
		$\Z_2$ fixed points $\{1,2,3,4\}$ are shown as blobs.
		The $\R$ invariant $x^i$ axis is along the 1-cycle $[a^i] - \frac{\beta^i}{2} [b^i]$ with $\beta^i=0$ for the \textbf{a}-type lattice and $\beta^i=1$ for the \textbf{b}-type lattice.
		$\R$ acts as reflection along the $y^i$ axis, which is spanned by the 1-cycle $[b^i]$.
		For the \textbf{a}-type lattice, all $\Z_2$ fixed points are invariant under $\R$, whereas
		for the \textbf{b}-type lattice, only 1 and 2 are invariant while $3 \stackrel\R{\leftrightarrow} 4$.}
	\label{Fig:Z2-lattice}
\end{figure}
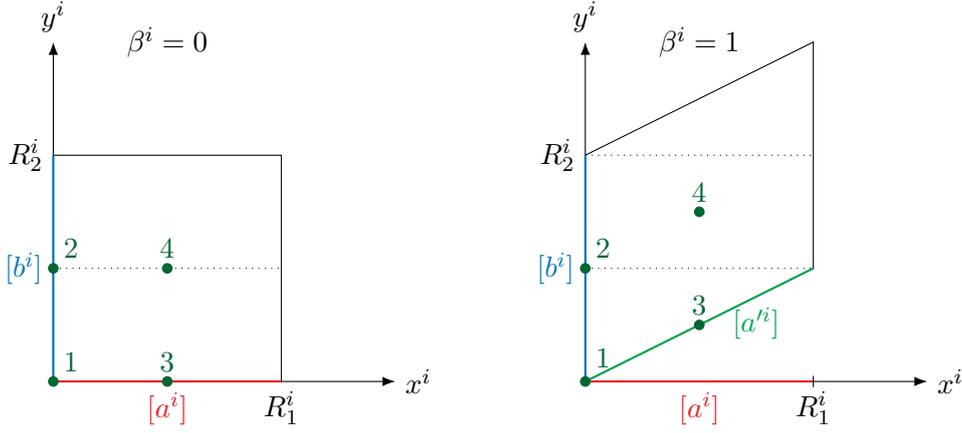

Let us determine how $\Omega \R$ acts on the various 3-cycles. For the untwisted cycles this is straightforward because the horizontally placed O-planes reflect the vertical axis only,
\begin{align}\label{untwR}
	\Omega \R: \begin{cases} 
	~[a^i]\to [a^i] ~,       \\
	~[b^i]\to -[b^i] ~.      
	\end{cases}              
\end{align}
Therefore, the wrapping numbers are mapped as $\Omega \R:(n^i_a,\widetilde m^i_a)\to (n^i_a,-\widetilde m^i_a)$, respectively $\Omega \R:(n^i_a,m^i_a)\to (n^i_a,-m^i_a-\beta^i\,n^i_a)$.

For the twisted sector 3-cycles, the canonical action  $\Omega \R$ corresponding to the models without vector structure including the signs $\eta_{\Omega \R}, \eta_{\Omega \R g}$ consistent with \eqref{opsigns} reads as,
\begin{align}\label{twR}
	\Omega \R:\begin{cases}                                                                   
	\alpha^g_{ij,\,n} \to -\eta_{\Omega \R}\, \eta_{\Omega \R g} \, \alpha^g_{\R(i)\R(j),\,n} \\
	\alpha^g_{ij,\,m} \to  \eta_{\Omega \R}\, \eta_{\Omega \R g}\, \alpha^g_{\R(i)\R(j),\,m}  
	\end{cases}                                                                               
\end{align}
where for $\beta^i=0$ the reflection $\R$ leaves all four fixed points $i\in\{1,2,3,4\}$ invariant, whereas for $\beta^i=1$ the action $\R$ interchanges the fixed points 3 and 4 while leaving 1 and 2 unchanged,
\begin{equation}\label{eq:R34}
	\R:  \begin{cases}
	1 \rightarrow 1, \\
	2 \rightarrow 2, \\
	3 \rightarrow 3+\beta^i ,\\
	4 \rightarrow 4-\beta^i ,
	\end{cases} \qquad \beta^i \in \{0,1\},
\end{equation}
as can be seen from the figure \ref{Fig:Z2-lattice}.

Implementing the $3\leftrightarrow 4$ interchange \eqref{eq:R34} coming from the $\Omega \R$ action on the number of common fixed points \eqref{eq:deltaij}, we get,
\begin{align}\label{eq:deltapij}
	\delta_{ab'}^g\equiv \delta^g(\alpha^{a}_{ij}, \alpha^{b'}_{kl}) & = \delta^g_{S^{a}_{i},S^{b'}_{k}} \delta^g_{S^{a}_{j},S^{b'}_{l}}. 
\end{align}
Again assuming that every fractional brane intersects the origin the above relation simplifies as,
\begin{align}\label{eq:deltap}
	\delta_{ab'}^g & = \sum _{i=1}^2 \sum _{j=1}^2 \sum _{k=1}^2 \sum _{l=1}^2 \delta^g_{S^{a}_{1,i},S^{b'}_{1,k}} \delta^g_{S^{a}_{2,j},S^{b'}_{2,l}}~. 
\end{align}

Using equations \eqref{untwR}, \eqref{twR} and \eqref{eq:deltapij} the intersection number of $\Pi_a^F$ with the $\Omega \R$ image of $\Pi_b^F$ can be computed as,
\begin{align} \label{Iabp}
	\Pi_a^F \circ \left(\Pi_b^F\right)' & = \eta_{\Omega \R}\,\frac{1}{4}\Bigg(-\eta_{\Omega \R}\,2^{-k}\prod_I (n^i_a l_b^i + l_a^i n^i_b)  +\!\!\! \sum _{g\in \{\theta, \omega,\theta\omega\}} \!\!\!\eta_{\Omega \R g}\, \delta_{ab'}^g \,2^{-\beta^{g}} (n^g_a\,l_b^g + l_a^g\,n^g_b) \Bigg). 
\end{align}
As a special case of \eqref{eq:deltap}, setting $a=b$ implies,
\begin{align}\label{eq:gij}
	\delta_{aa'}^g & = \sum_{h<i} \delta_{a a}^g \frac{\left|\epsilon^{ghi}\right|\,Q^h_{a}\, Q^i_{a}}{4}, 
\end{align}
where $Q^i_{a}$ denotes the number of invariant fixed points on the $i^\text{th}$ $\T^2$ under the $\Omega \R$ given as,
\begin{align}\label{Q}
	Q^i_a & \equiv \frac{3}{2} + \frac{(-1)^{\beta^i n^i_a}}{2}, \qquad \beta^i \in \{0,1\}, 
\end{align}
which is equal to two, except for the case $\beta^i=1$ and $n_a^i=$ odd, where it is equal to one.

Setting $a=b$ in \eqref{Iabp} and using \eqref{eq:gij} the intersection number of $\Pi_a^F$ with its own $\Omega \R$ image is,
\begin{align} \label{Iaap}
	\Pi_a^F\circ \left(\Pi_a^F\right)'  = \eta_{\Omega \R}\, & \Bigl(-\eta_{\Omega \R}\,2^{1-k}\prod_I n^i_a\, l_a^i   +\eta_{\Omega \R\theta}\, \frac{Q^1_{a}\, Q^2_{a}}{ 2}\,2^{-\beta^{3}} n_a^3\, l_a^3  \nonumber                                    \\ 
	                                                         & +\eta_{\Omega \R\omega}\, \frac{Q^2_{a}\, Q^3_{a}}{2}\,2^{-\beta^{1}} n_a^1\, l_a^1   +\eta_{\Omega \R\theta\omega}\, \frac{Q^1_{a}\, Q^3_{a}}{ 2}\,2^{-\beta^{2}} n_a^2\, l_a^2 \Bigr)\,. 
\end{align} 

For the intersection with the orientifold plane one obtains,
 
\begin{align} \label{IaO6}
	\Pi_{a}\circ \Pi_\textrm{O6}^F & = 2^{1-k}\,\Bigl(-\eta_{\Omega \R}\,\prod_I l_a^i + \eta_{\Omega \R\theta}\,  n_a^1\, n_a^2\, l_a^3  + \eta_{\Omega \R\omega}\, l_a^1 \, n_a^2\, n_a^3 + \eta_{\Omega \R\theta\omega}\, n_a^1\, l_a^2 \, n_a^3 \Bigr). 
\end{align}

Using the expressions in Eqs.~\eqref{Iab}, \eqref{Iabp}, \eqref{Iaap}, and \eqref{IaO6}, one can determine the chiral spectrum involving symmetric and antisymmetric representations of the gauge group $\prod_a \U(N_a)$. In particular, a stack of $N_a$ fractional D6-branes satisfying the condition $\Omega \mathcal{R} \Pi_a^F = \Pi_a^F$ gives rise to a $\USp(2N_a)$ gauge symmetry.

In our construction of Pati-Salam models, we employ four visible sector rigid branes denoted by $a, b, c, d$, in contrast to the usual case which involves only three branes and does not include twisted sectors. Consequently, the three-family condition is modified and takes the following form:
\begin{equation}\label{eq:3families}
	I_{ab} + I_{ab'} = - \left( I_{ac} + I_{ac'} + I_{ad} + I_{ad'} \right) = \pm 3 \,,
\end{equation}
where positive intersection numbers, in our convention, refer to left-chiral supermultiplets.

\subsection{Gauge couplings from complex structure moduli}

Dynamical supersymmetry breaking in D6-brane models derived from Type IIA orientifolds has been explored in~\cite{Cvetic:2003yd}. The Kähler potential is given by
\begin{equation}\label{eq: Kaehler_potential}
	K = - \ln(S + \overline{S}) - \sum_{i=1}^3 \ln(U^i + \overline{U}^i) \, .
\end{equation}
The complex structure moduli $U^i$ can be extracted from the supersymmetry conditions, as shown in~\cite{Sabir:2022hko},
\begin{align}\label{U-moduli}
	U^i & = \frac{4i \chi^i + 2\beta_i^2 \chi_i^2}{4 + \beta_i^2 \chi_i^2}, \qquad \chi^i \equiv \frac{R_2^i}{R_1^i} \, . 
\end{align}
These moduli, expressed in the string theory basis, can be mapped to the field theory basis using $\{s, u^i\}$ as follows~\cite{Lust:2004cx}:
\begin{equation}\label{eq:sugra-string-basis}
	\begin{split}
		\mathrm{Re}(s)   &= \frac{e^{-\phi_4}}{2\pi} \frac{\sqrt{\mathrm{Im}(U^1) \, \mathrm{Im}(U^2) \, \mathrm{Im}(U^3)}}{\left|U^1 U^2 U^3\right|} , \\
		\mathrm{Re}(u^1) &= \frac{e^{-\phi_4}}{2\pi} \sqrt{\frac{\mathrm{Im}(U^1)}{\mathrm{Im}(U^2) \, \mathrm{Im}(U^3)}} \left| \frac{U^2 U^3}{U^1} \right| , \\
		\mathrm{Re}(u^2) &= \frac{e^{-\phi_4}}{2\pi} \sqrt{\frac{\mathrm{Im}(U^2)}{\mathrm{Im}(U^1) \, \mathrm{Im}(U^3)}} \left| \frac{U^1 U^3}{U^2} \right| , \\
		\mathrm{Re}(u^3) &= \frac{e^{-\phi_4}}{2\pi} \sqrt{\frac{\mathrm{Im}(U^3)}{\mathrm{Im}(U^1) \, \mathrm{Im}(U^2)}} \left| \frac{U^1 U^2}{U^3} \right| .
	\end{split}
\end{equation}
The four-dimensional dilaton is related to the moduli via
\begin{equation}
	2\pi e^{\phi_4} = \left( \mathrm{Re}(s) \, \mathrm{Re}(u^1) \, \mathrm{Re}(u^2) \, \mathrm{Re}(u^3) \right)^{-1/4}.
\end{equation}

The holomorphic gauge kinetic function for a D6--brane stack $x$ wrapping a supersymmetric 3--cycle is~\cite{Blumenhagen:2006ci}:
\begin{align}\label{eq:gauge-kinetic-f}
	f_x & = \frac{1}{4 k_x} \left( n_x^1 n_x^2 n_x^3 \, s 
	- \frac{n_x^1 l_x^2 l_x^3 \, u^1}{2^{\beta_2 + \beta_3}} 
	- \frac{l_x^1 n_x^2 l_x^3 \, u^2}{2^{\beta_1 + \beta_3}} 
	- \frac{l_x^1 l_x^2 n_x^3 \, u^3}{2^{\beta_1 + \beta_2}} \right),
\end{align}
where $s$ and $u^i$ are the four--dimensional dilaton and complex structure moduli, respectively, and $k_x$ is the Kac--Moody level of $G_x$: $k_x = 1$ for $\text{U}(N_x)$ and $k_x = 2$ for $\text{USp}(2N_x)$ or $\text{SO}(2N_x)$ \cite{Ginsparg:1987ee, Hamada:2014eia}. With this convention, the gauge coupling is given by
\begin{equation}
	g_x^{-2} = \mathrm{Re}(f_x).
\end{equation}

When two gauge factors $G_c$ and $G_d$ are Higgsed to their diagonal subgroup $G_R$, canonical normalization of the gauge kinetic terms implies
\begin{equation}
	g_R^{-2} = g_c^{-2} + g_d^{-2}
	= \mathrm{Re}(f_c) + \mathrm{Re}(f_d).
\end{equation}
Therefore, the holomorphic gauge kinetic function for the diagonal is simply \cite{Dienes:1996yh}
\begin{equation}
	f_R = f_c + f_d \,.
\end{equation}   
No factor of $1/2$ appears here: the dependence on the Kac–Moody levels $k_x$ is already included in each $f_x$ through the $1/(4k_x)$ prefactor in its definition \eqref{eq:gauge-kinetic-f}. For $k_c = k_d = 1$ one finds $k_R = k_c + k_d = 2$ for the diagonal group, but this is automatically encoded in $f_R$ above.

At tree level, the gauge couplings satisfy
\begin{equation}
	\begin{split}
		g_a^2 &= \frac{\mathrm{Re}(f_b)}{\mathrm{Re}(f_a)}\, g_b^2
		= \frac{\mathrm{Re}(f_c) + \mathrm{Re}(f_d)}{\mathrm{Re}(f_a)} \, g_R^2 \\
		&= k_Y\!\left(\frac{5}{3}\right) g_Y^2
		= \gamma \,\pi e^{\phi_4} \,,
	\end{split}
\end{equation}
where $g_a$, $g_b$, and $g_Y$ are the strong, weak, and hypercharge couplings, $g_R$ is the coupling of the diagonal $\mathrm{SU}(2)_R$, $k_Y$ is the effective Kac–Moody level of the canonically normalized hypercharge, and $\gamma$ is a model-dependent constant fixed by the internal moduli.

\subsection{Supersymmetry conditions}\label{susyconstraints}

Let us define the following products of wrapping numbers,
\begin{alignat}{2}
	A_a & \equiv -\,n_a^1 n_a^2 n_a^3, & \qquad \tilde{A}_a & \equiv -\,l_a^1 l_a^2 l_a^3, \nonumber       \\
	B_a & \equiv n_a^1 l_a^2 l_a^3,    & \tilde{B}_a        & \equiv l_a^1 n_a^2 n_a^3,   \nonumber        \\
	C_a & \equiv l_a^1 n_a^2 l_a^3,    & \tilde{C}_a        & \equiv n_a^1 l_a^2 n_a^3,   \nonumber        \\
	D_a & \equiv l_a^1 l_a^2 n_a^3,    & \tilde{D}_a        & \equiv n_a^1 n_a^2 l_a^3.  \label{variables} 
\end{alignat}
Preserving $\mathcal{N}=1$ supersymmetry in four dimensions after compactification from ten-dimensions restricts the rotation angle of any D6-brane with respect to the orientifold plane to be an element of $\SU(3)$, i.e.
\begin{equation}
	\theta^a_1 + \theta^a_2 + \theta^a_3 = 0 \mod 2\pi ,
\end{equation}
where $\theta_i$ is the angle between the D6-brane and orientifold-plane in the $i^\text{th}$ 2-torus and $\chi_i=R^2_i/R^1_i$ are the complex structure moduli for the $i^\text{ th}$ 2-torus,
\begin{align}\label{theta}
	\tan \theta^a_j = \chi_j \frac{\widetilde{m}^a_j}{n^a_j} = \chi_j \frac{2^{- \beta_j}l^a_j}{n^a_j} \,. 
\end{align}
$\mathcal{N}=1$ supersymmetry conditions are given as,
\begin{eqnarray}
	x_A\tilde{A}_a+x_B\tilde{B}_a+x_C\tilde{C}_a+x_D\tilde{D}_a=0,\nonumber\\
	\frac{A_a}{x_A}+\frac{B_a}{x_B}+\frac{C_a}{x_C}+\frac{D_a}{x_D} < 0, \label{susyconditions}
\end{eqnarray}
where $x_A=\lambda,\; x_B=2^{\beta_2+\beta_3}\cdot\lambda /\chi_2\chi_3,\; x_C=2^{\beta_1+\beta_3}\cdot\lambda /\chi_1\chi_3,\; x_D=2^{\beta_1+\beta_2}\cdot\lambda /\chi_1\chi_2$.

\subsection{Tadpole cancellation}\label{tadpoleconstraints}

Since D6-branes and O6-orientifold planes are the sources of Ramond-Ramond charges they are constrained by the Gauss's law in compact space implying the sum of D-brane and cross-cap RR-charges must vanishes
\begin{eqnarray}\label{RRtadpole}
	\sum_a N_a [\Pi_a]+\sum_a N_a \left[\Pi_{a'}\right]-4[\Pi_\textrm{O6}]=0,
\end{eqnarray}
where the last terms arise from the O6-planes, which have $-4$ RR charges in D6-brane charge units. RR tadpole constraint is sufficient to cancel the $\SU(N_a)^3$ cubic non-Abelian anomaly while $\U(1)$ mixed gauge and gravitational anomaly or $[\SU(N_a)]^2 \U(1)$ gauge anomaly can be cancelled by the Green-Schwarz mechanism, mediated by untwisted RR fields \cite{Green:1984sg}.

The twisted and the untwisted tadpole cancellation conditions (where the first condition will later be modified after inclusion of $G_3$ flux) are given by
\begin{align}
	  & \sum_a N_a n_a^1 n_a^2 n_a^3 = 16 \eta_{\Omega \R}\,, \nonumber                                                                                 \\
	  & \sum_a N_a n_a^i \widetilde{m}_a^j \widetilde{m}_a^k = -2^{4-\beta^j-\beta^k} \eta_{\Omega \R i},  \quad (i,j,k) = (\overline{1,2,3}) \nonumber \\
	  & \sum_a N_a n_a^i (\epsilon_{a,kl}^i -\eta_{\Omega \R} \eta_{\Omega \R i} \epsilon_{a,\R(k)\R(l)}^i) = 0 \,, \nonumber                           \\
	  & \sum_a N_a \widetilde{m}_a^i (\epsilon_{a,kl}^i + \eta_{\Omega \R} \eta_{\Omega \R i} \epsilon_{a,\R(k)\R(l)}^i) = 0 \,,\label{twtadp}          
\end{align}
where $N_a$ denotes the number of D6-branes on stack $a$ and the sum is a sum over all stacks of D6-branes. $\R(k)=k$ in case of an untilted torus and $\R(\{1,2,3,4\})=\{1,2,4,3\}$ in the tilted case. The twisted charge $\epsilon_{ij, a}^{\omega}$ is non-zero if and only if $ij \in S_{\omega}^a$, i.e., if the brane $a$ passes through the fixed point $ij$ in the $\omega$-twisted sector, and so on. The orientifold projection acts on the wrapping numbers and twisted charges as follows,
\begin{eqnarray}
	\widetilde{m}^i &\rightarrow & - \widetilde{m}^i \,, \nonumber \\
	\epsilon_{kl}^i &\rightarrow & - \eta_{\Omega \R} \eta_{\Omega \R i} \epsilon^i_{\R(k)\R(l)}\, .
\end{eqnarray}

Cancellation of RR tadpoles requires introducing a number of orientifold planes also called ``filler branes'' that trivially satisfy the four-dimensional $\mathcal{N}=1$ supersymmetry conditions. The filler branes belong to the hidden sector USp group and carry the same wrapping numbers as one of the O6-planes as shown in table \ref{tab:Oplanes}. USp group is hence referred with respect to the non-zero $A$, $B$, $C$ or $D$-type.

Orientifolds also have discrete D-brane RR charges classified by the $\mathbb{Z}_2$ K-theory groups, which are subtle and invisible by the ordinary homology \cite{Cascales:2003zp, Marchesano:2004yq, Marchesano:2004xz}, which should also be taken into account \cite{Uranga:2000xp}. The K-theory conditions are,
\begin{eqnarray}
	\sum_a \tilde{A}_a  = \sum_a  N_a  \tilde{B}_a = \sum_a  N_a  \tilde{C}_a = \sum_a  N_a \tilde{D}_a = 0 \textrm{ mod }4. \label{K-charges}
\end{eqnarray}

\begin{table}[th]
	\centering
	\renewcommand{\arraystretch}{1.35} 
	\begin{tabular}{|c|c|c|c|}
		\hline
		\textbf{Orientifold} & \multicolumn{2}{c|}{\textbf{Op-Planes}} & \textbf{Wrapping Numbers} \\
		\cline{2-3}
		\textbf{Action}        & \textbf{~IIA~} & \textbf{IIB} & $(n^1,l^1) \times (n^2,l^2) \times (n^3,l^3)$                    \\
		\hline\hline
		$\Omega\R$             & O6$_1$         & O3           & $(2^{\beta_1},0) \times (2^{\beta_2},0) \times (2^{\beta_3},0)$  \\
		\hline
		$\Omega\R\omega$       & O6$_2$         & O7$_1$       & $(2^{\beta_1},0) \times (0,-2^{\beta_2}) \times (0,2^{\beta_3})$ \\
		\hline
		$\Omega\R\theta\omega$ & O6$_3$         & O7$_2$       & $(0,-2^{\beta_1}) \times (2^{\beta_2},0) \times (0,2^{\beta_3})$ \\
		\hline
		$\Omega\R\theta$       & O6$_4$         & O7$_3$       & $(0,-2^{\beta_1}) \times (0,2^{\beta_2}) \times (2^{\beta_3},0)$ \\
		\hline
	\end{tabular}
	\caption{Wrapping numbers of the four orientifold planes in type~IIA and their T-dual type~IIB counterparts. An Op-plane carries $2^{p-4}$ units of the charge of a Dp-brane.}
	\label{tab:Oplanes}
\end{table}

\section{T-dual Type IIB flux-induced stabilization}\label{sec:IIBflux}

Up to now, we have focused on Type~IIA $\mathcal{N}=1$ chiral models consisting of rigid intersecting D6-branes on orientifolds with discrete torsion. We now turn to their T-dual Type~IIB flux compactification counterpart. For our case with discrete torsion $\eta = -1$, containing an odd number of exotic O$6^{++}$ planes. By mirror symmetry transformation, this background translates into a $\Z_2 \times \Z_2$ type IIB orientifold without discrete torsion containing O$3^{++}$ and O$7^{--}_i$-planes, see table \ref{tab:Oplanes}. This yields a four-dimensional $\mathcal{N}=1$ effective theory with stabilized complex structure and dilaton, non-perturbatively stabilized Kähler moduli, and chiral matter arising from intersecting or magnetized D-branes.

\subsection{Three-form fluxes and the Gukov--Vafa--Witten superpotential}

In the Type~IIB frame, moduli stabilization is achieved by turning on background three-form fluxes
\begin{equation}
	G_3 = F_3 - \tau H_3 \,,
\end{equation}
where $F_3$ and $H_3$ denote the RR and NS--NS three-form field strengths, respectively, and $\tau = C_0 + i\, e^{-\phi}$ is the axio-dilaton combining the RR scalar $C_0$ and the dilaton $\phi$. Flux quantization requires that the periods of $F_3$ and $H_3$ over a basis of fractional three-cycles satisfy
\begin{equation}
	\int_{\Sigma_2 \times \Sigma_1} F_3 \,, \;\;
	\int_{\Sigma_2 \times \Sigma_1} H_3 \;\in\; 4\,\mathbb{Z} \,,
\end{equation}
where $\Sigma_2$ denotes a collapsed two-cycle at a $\mathbb{Z}_2$ fixed point, and
\begin{equation}
	\Sigma_2 \in \Big\{ [\alpha_1]\otimes[\alpha_2] \;\Big|\; [\alpha_i] = [a_i] \text{ or } [b_i], \; i=1,2 \Big\},
\end{equation}
with $[a_i], [b_i]$ being the canonical one-cycles of the two-tori. The factor of $4$ reflects the normalization of the twisted three-cycles inherited from the $\mathbb{Z}_2\times\mathbb{Z}_2$ orbifold, where each collapsed two-cycle contributes a multiplicity of four fixed points.

The presence of these fluxes generates a superpotential of the Gukov--Vafa--Witten (GVW) form~\cite{Gukov:1999ya}:
\begin{equation}
	W = \int_{X^6} G_3 \wedge \Omega_3 \,,
\end{equation}
where $\Omega_3$ is the holomorphic $(3,0)$ form of the internal Calabi--Yau threefold. Supersymmetric Minkowski or AdS vacua arise when $G_3$ is imaginary self-dual (ISD), satisfying $\star_6 G_3 = i G_3$. This condition fixes the axio-dilaton $\tau$ and the complex structure moduli through the F-term conditions $D_I W = 0$ \cite{Grana:2005jc, Douglas:2006es}, where the index $I$ here runs over the set of moduli fields in the compactified theory, which include the axio-dilaton $\tau$ and the complex structure moduli $\phi^I$. The ISD condition ensures that the fluxes preserve $\mathcal{N}=1$ supersymmetry in four dimensions.

\subsection{Flux contribution to the D3-brane tadpole}

Background RR and NS--NS three-form fluxes not only induce gravitational backreaction but also source the RR four-form potential $C_4$, thereby carrying both D3-brane charge and tension. In D3-brane units, the net RR charge contributed by the fluxes is a topological quantity. The contribution of quantized $G_3$ flux to the D3-brane tadpole is
\begin{align}
	N_\text{flux} & = \frac{1}{(2\pi)^4 \alpha'^2} \int_{X^6} H_3 \wedge F_3 = \frac{1}{(2\pi)^4 \alpha'^2} \int_{X^6} \frac{i}{2 \,\mathrm{Im}\,\tau} G_3 \wedge \overline{G}_3 \,. 
\end{align}
With fractional branes, the Dirac quantization conditions for $F_3$ and $H_3$ on the $\T^6/(\Z_2 \times \Z_2')$ orientifold require the flux contribution $N_\text{flux}$ to be an integer multiple of $16$. Imposing the imaginary self dual condition $\star_6 G_3 = i G_3$ further restricts $N_\text{flux}$ to be strictly positive.\footnote{By contrast, supersymmetric flux vacua constructed with non-rigid D-branes require $N_\text{flux}$ to be a multiple of $64$ \cite{Blumenhagen:2003vr,Cvetic:2005bn}. This difference originates from the structure of the underlying three cycles. Each of the three two tori contains four fixed points. In the absence of discrete torsion, bulk three cycles wrap all fixed points, leading to a quantization in units of $4^3$. In the presence of discrete torsion, rigid fractional branes instead involve collapsed two cycles localized at fixed points, effectively reducing the quantization unit to $4^2$ \cite{Blumenhagen:2005tn}.} Consequently, the flux-induced D3-brane charge contributes to the total tadpole, modifying the first RR tadpole cancellation condition in~\eqref{twtadp} to
\begin{align}
	\sum_a N_a n_a^1 n_a^2 n_a^3 + \frac{1}{2} N_\text{flux} = 16 \, \eta_{\Omega \R} \,. \label{modtadpole1} 
\end{align}
Here, the factor $1/2$ arises because the integral $H_3 \wedge F_3$ counts flux quanta on the covering space, while the orientifold projection identifies pairs of cycles, effectively halving the contribution to the physical D3-brane charge. The complex structure moduli are fixed at values consistent with these quantization conditions, whereas the Kähler moduli remain unfixed by the flux alone.

We emphasize that our analysis is carried out within the standard four-dimensional effective field theory framework of type IIB flux compactifications. We work in a regime where a parametric separation of scales is assumed, so that the low-energy dynamics is described by the effective four-dimensional $\mathcal{N}=1$ theory, with background flux effects incorporated at the level of the low-energy effective action. In particular, we do not attempt a full worldsheet quantization of strings in $G_3$ flux backgrounds and do not determine the corresponding massive string spectrum. Instead, we focus on the low-energy effective description relevant for moduli stabilization.

\subsection{Explicit ISD $(2,1)$ flux and primitivity}

A general supersymmetric flux consistent with the orbifold symmetries is of type $(2,1)$ in the complex coordinates:
\begin{equation}
	G_3 = g_1\, d\bar z_1 \wedge dz_2 \wedge dz_3 
	+ g_2\, dz_1 \wedge d\bar z_2 \wedge dz_3 
	+ g_3\, dz_1 \wedge dz_2 \wedge d\bar z_3 \,,
\end{equation}
where $g_i$ are complex constants specifying the flux quanta. Supersymmetry further requires the \emph{primitivity condition}:
\begin{equation}
	G_3 \wedge J = 0 \,,
\end{equation}
with the Kähler form of the factorized torus
\begin{equation}
	J = J_1 \, dz_1 \wedge d\bar z_1 + J_2 \, dz_2 \wedge d\bar z_2 + J_3 \, dz_3 \wedge d\bar z_3 \,.
\end{equation}
Because off-diagonal metric components are projected out by the orbifold, this flux is automatically primitive and preserves $\mathcal{N}=1$ supersymmetry.  

A supersymmetric, primitive ISD $(2,1)$ flux that contributes exactly one unit of D3-brane charge ($N_\text{flux}=16$) in the discrete torsion case, assuming torus normalization $\int dz_i\wedge d\bar z_i = 2i (2\pi)^2 \alpha'$, can be chosen as
\begin{equation}
	G_3 = \frac{1}{\sqrt{3}} \Big( d\bar z_1 \wedge dz_2 \wedge dz_3 
	+ dz_1 \wedge d\bar z_2 \wedge dz_3 
	+ dz_1 \wedge dz_2 \wedge d\bar z_3 \Big) \,,
\end{equation}
which satisfies $G_3 \wedge J = 0$ and stabilizes the complex structure moduli at
\begin{equation}
	\tau_1 = \tau_2 = \tau_3 = \tau = e^{2\pi i/3} \,.
\end{equation}

\subsection{Couplings of $G_3$ to $C_4$ in Type IIB supergravity}

The coupling between the three-form flux $G_3$ and the four-form potential $C_4$ arises both from the Chern-Simons term in the Type IIB supergravity action and from the Wess--Zumino part of the D-brane action. The Wess--Zumino (WZ) term for a D$p$-brane is
\begin{equation}
	I_{\text{WZ}} = \mu_p \int_{W_{p+1}} C \wedge e^{\mathcal{F}} \,,
\end{equation}
where $C = \sum_n C^{(n)}$ denotes the formal sum of RR potentials of all degrees, and the gauge-invariant worldvolume field strength is
\begin{equation}
	\mathcal{F} = F - B \,,
\end{equation}
with $F = dA$ the worldvolume $U(1)$ field strength and $B$ the pullback of the NS--NS two-form onto the brane\footnote{Gauge invariance under $B \to B + d\Lambda$ and $A \to A + \Lambda$ requires the combination $F-B$ to appear in the action.}.

The exponential $e^{\mathcal{F}}$ is defined in the algebra of differential forms:
\begin{equation}
	e^{\mathcal{F}} = 1 + \mathcal{F} + \frac{1}{2!} \mathcal{F}\wedge\mathcal{F} + \frac{1}{3!} \mathcal{F}\wedge\mathcal{F}\wedge\mathcal{F} + \cdots,
\end{equation}
so that $C\wedge e^{\mathcal{F}}$ produces forms of various degrees, allowing a D$p$-brane to couple to lower-dimensional RR potentials. For example,
\begin{align}
	I_{\text{WZ}}^{(D4)} & = \mu_4 \int_{W_5} 
	\left( C^{(5)} + C^{(3)} \wedge \mathcal{F} + \frac{1}{2} C^{(1)} \wedge \mathcal{F}\wedge\mathcal{F} \right), \\[2mm]
	I_{\text{WZ}}^{(D3)} & = \mu_3 \int_{W_4} 
	\left( C^{(4)} + C^{(2)} \wedge \mathcal{F} + \frac{1}{2} C^{(0)} \mathcal{F}\wedge\mathcal{F} \right).
\end{align}
Hence a D3-brane with worldvolume flux $\mathcal{F}$ carries induced D1- and D(${-}$1)-brane charges. This structure is essential for gauge invariance under $B$-field transformations and for capturing the hierarchy of induced lower-dimensional D-brane charges.

The Chern-Simons term in the supergravity action is
\begin{align}
	S_{\text{CS}} & = \frac{1}{4\kappa_{10}^2} \int C_4 \wedge H_3 \wedge F_3  = \frac{1}{4\kappa_{10}^2} \int C_4 \wedge G_3 \wedge \overline{G}_3 + \cdots, 
\end{align}
showing a direct coupling of $C_4$ to the three-form fluxes. The five-form field strength is modified as
\begin{equation}
	F_5 = dC_4 + \frac{1}{2} B_2 \wedge F_3 - \frac{1}{2} C_2 \wedge H_3,
\end{equation}
and satisfies the self-duality constraint
\begin{equation}
	F_5 = \star F_5 \,,
\end{equation}
which is imposed at the level of the equations of motion. When 3-form fluxes are turned on, $C_4$ is sourced via
\begin{equation}
	d \star F_5 = H_3 \wedge F_3 = \frac{1}{2i} G_3 \wedge \overline{G}_3,
\end{equation}
or in components,
\begin{equation}
	\nabla^M F_{MNOPQ} = \frac{1}{2i} (G \wedge \overline{G})_{NOPQ},
\end{equation}
which explicitly shows how $G_3$ acts as a source for $C_4$. Decomposing $C_4$ along harmonic 4-forms $\omega_I^{(4)}$ of the internal manifold,
\begin{equation}
	C_4 = \sum_I D^i(x) \omega_I^{(4)},
\end{equation}
leads to a flux-induced potential for the Kähler moduli associated with the $D^i(x)$ scalars:
\begin{equation}
	V_\text{flux} \sim \frac{1}{\mathcal{V}^2} \int_M G_3 \wedge \star \overline{G}_3,
\end{equation}
where $\mathcal{V}$ is the internal volume, depending on the Kähler moduli. This potential stabilizes the complex structure moduli and the dilaton, but Kähler moduli require additional non-perturbative effects for stabilization.

\begin{table}[t]
	\centering\renewcommand{\arraystretch}{1.35}
	\begin{tabular}{lcccc}
		\toprule
		Brane  & $(\T^2)_1$       & $(\T^2)_2$       & $(\T^2)_3$       \\
		\midrule
		D9     & $(n_a^1, m_a^1)$ & $(n_a^2, m_a^2)$ & $(n_a^3, m_a^3)$ \\
		D7$_1$ & $(1,0)$          & $(n_a^2, m_a^2)$ & $(n_a^3, m_a^3)$ \\
		D5$_1$ & $(n_a^1, m_a^1)$ & $(1,0)$          & $(1,0)$          \\
		D3     & $(1,0)$          & $(1,0)$          & $(1,0)$          \\
		\bottomrule
	\end{tabular}
	\caption{Magnetized D-branes wrapping numbers and fluxes on factorized tori.}
	\label{tab:magnet}
\end{table}

\subsection{Magnetized D-branes and T-duality}

Magnetized D9-branes in Type~IIB are characterized by seven integers: the number of D9-branes $N_a$ and six `magnetic numbers' $(n_a^i,m_a^i)$, $i=1,2,3$, where $m_a^i$ denotes the number of times the D9's wrap the $i$th $\T^2$, and $n_a^i$ the unit of magnetic flux on that torus:
\begin{equation}
	\frac{m_a^i}{2\pi} \int_{\T^2_i} F_a^i = n_a^i.
	\label{magnet}
\end{equation}
This notation also allows description of lower-dimensional D-branes in the factorized torus basis as shown in table \ref{tab:magnet}. These magnetized D-branes correspond via T-duality to intersecting D6-branes in Type~IIA, and their flux-induced charges contribute to tadpole cancellation exactly as in~\eqref{modtadpole1}. The WZ couplings in~\eqref{magnet} ensure that worldvolume fluxes induce lower-dimensional D-brane charges, reproducing the complete D-brane charge lattice.
  
The T-dual Type~IIB description of rigid D6-brane models with discrete torsion combines ISD $(2,1)$ three-form fluxes with magnetized D-branes to yield fully $\mathcal{N}=1$ vacua. Fluxes stabilize the axio-dilaton and complex structure moduli, induce a D3-brane tadpole contribution, and source the four-form potential $C_4$ through WZ and Chern--Simons couplings. The Kähler moduli remain flat at tree level and require non-perturbative effects. Magnetized D-branes carry induced lower-dimensional charges, matching the intersecting D6-brane picture and ensuring consistent tadpole cancellation. This framework allows the construction of chiral vacua with all moduli stabilized at the perturbative level except for the Kähler moduli \cite{Blumenhagen:2006ci, Giddings:2001yu}.
 
\section{Three family Pati Salam flux models from rigid branes}\label{sec:models}

We construct three family Pati Salam models using rigid, semi rigid, and non rigid D6 branes, following the general strategy developed in \cite{Forste:2008ex}. The construction is carried out on the $\mathbb{Z}_2 \times \mathbb{Z}_2^{\prime}$ orbifold, where fractional branes invariant under $\Omega \mathcal{R}$ must be located on top of an exotic O$6^{++}$ plane, identified in our conventions as O$_{\Omega\mathcal{R}}$ in \eqref{opsigns}. For rigid branes, adjoint chiral multiplets from the $aa$ sector are absent. Throughout, all three two tori are taken to be rectangular.

The central challenge in obtaining an odd number of chiral families is that some twisted sector fixed point contributions $\delta^{g}_{ab}$ must deviate from their maximal values, namely $\delta^{g}_{ab} \neq (4,4,4)$, where each entry equals the total number of fixed points on a given two torus. As a consequence, twisted RR tadpoles are not automatically canceled at all fixed points. This necessitates the introduction of additional D6 brane stacks beyond those appearing in simpler configurations where all branes share all fixed points, corresponding to $\delta^{g}_{ab} = (4,4,4)$ \cite{Forste:2008ex}. These extra branes must be chosen so as not to introduce exotic chiral matter in the visible sector. They are therefore assigned to a hidden sector. 

However, in order to generate massless GUT Higgs pairs in the open string spectrum, one of these additional stacks must later recombine with the stack initially realizing the $\SU(2)_R$ gauge symmetry of the Pati Salam model. We further restrict attention to configurations preserving four dimensional $\mathcal{N}=1$ supersymmetry, which imposes constraints on the brane wrapping numbers and leads to partial stabilization of the closed string moduli. The introduction of quantized units of $G_3$-flux achieves stabilization of the axio-dilaton and the closed-string complex structure moduli.  

The starting point is a visible sector consisting of four rigid D-brane stacks labeled $\{a,b,c,d\}$. These branes generate the initial gauge symmetry and intersect at fixed points that are only partially shared among the stacks. This generically results in uncanceled twisted tadpoles. To minimize the number of additional branes required for tadpole cancellation, the three stacks $\{b,c,d\}$ responsible for the $\SU(2)_L \times \SU(2)_R$ gauge symmetry are chosen to wrap identical fixed points in each twisted sector, so that $\delta^{g}_{bc} = (4,4,4)$. Their twisted tadpole contributions are therefore identical and can be arranged to cancel among themselves. After this choice, the remaining uncanceled twisted tadpoles arise solely from the stack $a$. To cancel these residual twisted tadpoles, we introduce two additional stacks ${e_1,e_2}$. These semi rigid branes are engineered to have identical wrapping numbers and twisted charges with respect to one of the $\mathbb{Z}_2$ orbifold factors, ensuring that their mutual twisted tadpole contributions cancel internally. 

With the inclusion of $\{e_1,e_2\}$, the total twisted tadpole contributions from the combined set $\{a,b,c,d,e_1,e_2\}$ vanish. The cancellation of untwisted tadpoles is then achieved by adding further hidden sector branes that do not carry twisted charges. These branes, denoted $\{f_1,f_2,f_3,f_4\}$, can recombine into bulk branes and thus form non-rigid stacks. The requirement of unbroken $\mathcal{N}=1$ supersymmetry restricts all hidden sector branes to be either semi rigid or non rigid.

The four global $\U(1)$ symmetries embedded in $\U(4)_C$, $\U(2)_L$, $\U(2)_{R_1}$, and $\U(2)_{R_2}$ are anomalous and are rendered massive via the Green Schwarz mechanism through linear $B \wedge F$ couplings. As a result, the effective gauge symmetry at this stage is $\SU(4)_C \times \SU(2)_L \times \SU(2)_{R_1} \times \SU(2)_{R_2}$. When the fields $\Delta_i$ acquire vacuum expectation values, the product $\SU(2)_{R_1} \times \SU(2)_{R_2}$ is broken to the diagonal subgroup $\SU(2)_R$, yielding the Pati Salam gauge group $\SU(4)_C \times \SU(2)_L \times \SU(2)_R$.

The chiral spectrum consists of three generations of left handed fermions $F_L^i$ and three generations of right handed fermions arising from appropriate linear combinations of $F_R^i$ and $F_R^{\prime i}$, as specified in \eqref{eq:3families}. By assigning suitable vacuum expectation values to $F_R^{ci}$ and to a linear combination of $F_R^i$, the Pati Salam symmetry can be further broken to the Standard Model gauge group. This symmetry breaking can preserve $\mathcal{N}=1$ supersymmetry provided the D-flatness and F-flatness conditions are satisfied.

Standard Model fermion masses and mixing angles, as well as vector like masses for $F_R^{\prime ci}$ and for the two remaining linear combinations of $F_R^i$, are generated by superpotential terms of the form $W \supset Y_{ijk} F_L^i F_R^j \Phi^k + Y_{ijk}^{\prime} F_R^{\prime ci} F_R^j \Delta^k$. In addition, the hidden sector $\SU(4)$ gauge groups exhibit negative beta functions, making supersymmetry breaking via gaugino condensation possible.

Explicit realizations following this construction are presented in appendix~\ref{appA}. In appendix~\ref{appB}, we tabulate the decoupling of chiral exotic states through strong coupling dynamics associated with the non abelian hidden sector gauge factors.

Finally, in presenting the full spectra of the models, we implement a specific deformation to eliminate additional massless states in the hidden sector. Concretely, the hidden sector D branes $e,f,g,\ldots$ are displaced away from the origin of the internal space $\mathbb{T}^2 \times \mathbb{T}^2 \times \mathbb{T}^2$ to alternative orbifold fixed points, as summarized in table~\ref{fixed}. This displacement causes certain intersection numbers between the hidden and visible sector branes ${a,b,c,d}$ to vanish, thereby removing unwanted massless matter. An equivalent effect can alternatively be achieved through the introduction of non trivial discrete Wilson lines \cite{Blumenhagen:2005tn, Forste:2010gw}. 

We now turn to a detailed discussion of each model individually. In particular, we analyze flux vacua with $N_\text{flux}=16,32,48$ and $64$, corresponding respectively to one, two, three, and four units of quantized flux. The explicit models presented below are obtained from a supervised-random scan subject to all consistency conditions, including tadpole cancellation, supersymmetry, and K-theory constraints. They therefore constitute representative examples rather than an exhaustive classification of all possible solutions.

\subsection{Model~\texorpdfstring{\hyperref[spec:r15f1]{r15f1}}{r15f1}}

\begin{table}[t]
	\centering
	$\begin{array}{|c|D{.}{\times}{20}|c|}
		\hline
		\text{Model~\hyperref[exotic:r15f1]{r15f1}} & \mc{\text{Quantum Numbers}}            & \text{Fields}     \\ 
		\hline
		ab                                          & 1. (\bar{4},2,1,1,1,1,1,1,1)           & F_L^i(Q_L,L_L)    \\
		ab'                                         & 4. (4,2,1,1,1,1,1,1,1)                 & F_L^i(Q_L,L_L)    \\
		ac                                          & 2. (\bar{4},1,2,1,1,1,1,1,1)           & F_R^i(Q_R,L_R)    \\
		ad'                                         & 1. (\bar{4},1,1,\bar{2},1,1,1,1,1)     & F_R^{'i}(Q_R,L_R) \\
		b_{\yng(1,1)_{}}                            & 76. (1,1_{\yng(1,1)_{}},1,1,1,1,1,1,1) & S_L^i             \\
		d_{\yng(1,1)_{}}                            & 2. (1,1,1,1_{\yng(1,1)_{}},1,1,1,1,1)  & S_R^i             \\
		b_{\yng(2)}                                 & 18. (1,3_{\yng(2)},1,1,1,1,1,1,1)      & T_L^i             \\
		cd                                          & 1. (1,1,\bar{2},2,1,1,1,1,1)           & \Delta ^i         \\
		bc                                          & 3. (1,\bar{2},2,1,1,1,1,1,1)           & \Phi ^i(H_u,H_d)  \\
		bc'                                         & 4. (1,\bar{2},\bar{2},1,1,1,1,1,1)     & \Phi ^i(H_u,H_d)  \\
		bd                                          & 6. (1,\bar{2},1,2,1,1,1,1,1)           & \Xi ^{i}          \\
		be                                          & 2. (1,2,1,1,\bar{2},1,1,1,1)           & X_L^{i}           \\
		be'                                         & 1. (1,2,1,1,2,1,1,1,1)                 & X_L^{i}           \\
		bf_1                                        & 1. (1,2,1,1,1,\bar{2},1,1,1)           & X_L^{i}           \\
		bf_1'                                       & 4. (1,2,1,1,1,2,1,1,1)                 & X_L^{i}           \\
		bf_2                                        & 2. (1,2,1,1,1,1,\bar{2},1,1)           & X_L^{i}           \\
		bf_2'                                       & 2. (1,2,1,1,1,1,2,1,1)                 & X_L^{i}           \\
		bg_1                                        & 3. (1,2,1,1,1,1,1,\bar{4},1)           & X_L^{i}           \\
		bg_1'                                       & 3. (1,2,1,1,1,1,1,4,1)                 & X_L^{i}           \\
		bg_2                                        & 3. (1,2,1,1,1,1,1,1,\bar{4})           & X_L^{i}           \\
		bg_2'                                       & 3. (1,2,1,1,1,1,1,1,4)                 & X_L^{i}           \\
		cf_1                                        & 1. (1,1,2,1,1,\bar{2},1,1,1)           & X_R^{i}           \\
		cf_2                                        & 1. (1,1,2,1,1,1,\bar{2},1,1)           & X_R^{i}           \\
		cg_1                                        & 1. (1,1,\bar{2},1,1,1,1,4,1)           & X_R^{i}           \\
		cg_2                                        & 1. (1,1,\bar{2},1,1,1,1,1,4)           & X_R^{i}           \\
		de'                                         & 1. (1,1,1,2,2,1,1,1,1)                 & X_R^{i}           \\
		df_1'                                       & 1. (1,1,1,\bar{2},1,\bar{2},1,1,1)     & X_R^{i}           \\
		\hline
	\end{array}$
	\caption{Particle spectrum of Model~\hyperref[model:r15f1]{r15f1} with gauge symmetry $\SU(4)_{C}\times\SU(2)_{L}\times\SU(2)_{R_1}\times\SU(2)_{R_2}\times\SU(2)^3\times\SU(4)^2$.}
	\label{spec:r15f1}
\end{table}  

The model~\hyperref[model:r15f1]{r15f1} is a gauge theory of rank 15 with 1 unit of flux. The construction involves four rigid D6-branes
$\{a,b,c,d\}$ sharing identical fixed points,
$\delta_{ab}^g \neq (4,4,4)$, and realizes the gauge symmetry
$\SU(4)_{C}\times\SU(2)_{L}\times\SU(2)_{R_1}\times\SU(2)_{R_2}\times\SU(2)^3\times\SU(4)^2.$

The matter content is summarized in table~\ref{spec:r15f1}, where fields are
organized by their quantum numbers under the gauge symmetry. The spectrum contains chiral fermions, scalar fields, and Higgs-like states arising from
brane intersections such as $ab$, $ac$, and $ad'$, yielding fundamental
representations of the corresponding gauge factors. The model~\hyperref[spec:r15f1]{r15f1} features the torus moduli
$\chi_1 = 2 \sqrt{6}$,
$\chi_2 = \sqrt{6}$,
$\chi_3 = 2 \sqrt{6}$,
and the tree-level gauge coupling relation
$g_a^2=\frac{35}{2}g_b^2=\frac{7}{9}g_{cd}^2=\frac{35}{41}\frac{5 g_Y^2}{3}=\frac{4\ 2^{3/4}}{\sqrt[4]{3}}\, \pi \,  e^{\phi _4}$.

The chiral sector consists of left-handed multiplets
$(Q_L,L_L)$ and right-handed multiplets
$(Q_R,L_R)$ arising from distinct D-brane intersections. The left-handed states accommodate the quark and lepton doublets and originate from the $ab$ (-1), $ab'$ (4) sectors. The right-handed states provide the corresponding singlet partners and receive contributions from the $ac$ (-2), $ad'$ (-1) sectors.

At the GUT scale, there is 1 scalar field $\Delta^i$ responsible for breaking the Pati-Salam symmetry
to the Standard Model gauge group. In addition, there are 7 Higgs-like fields arising from the $bc$, $bc'$ sectors. The spectrum further contains several chiral exotic states
$X_L^i$ and $X_R^i$ charged under hidden sector gauge groups.

Table~\ref{exotic:r15f1} presents the composite spectrum obtained after confinement in the hidden
sector. States charged under non-abelian hidden gauge factors experience strong coupling dynamics, which leads to the formation of bound states neutral under the hidden gauge group. Consequently, these states decouple from the low-energy effective spectrum.

 
\subsection{Model~\texorpdfstring{\hyperref[spec:r17f1]{r17f1}}{r17f1}}

\begin{table}[t]
	\centering
	$\begin{array}{|c|D{.}{\times}{21}|c|}
		\hline
		\text{Model~\hyperref[exotic:r17f1]{r17f1}} & \mc{\text{Quantum Numbers}}                               & \text{Fields}     \\ 
		\hline
		ab'                                         & 3. (4,2,1,1,1,1,1,1,1,1)                                  & F_L^i(Q_L,L_L)    \\
		ac'                                         & 2. (\bar{4},1,\bar{2},1,1,1,1,1,1,1)                      & F_R^i(Q_R,L_R)    \\
		ad                                          & 1. (\bar{4},1,1,2,1,1,1,1,1,1)                            & F_R^{'i}(Q_R,L_R) \\
		b_{\overline{\yng(1,1)_{}}}                 & 10. (1,\bar{1}_{\overline{\yng(1,1)_{}}},1,1,1,1,1,1,1,1) & S_L^i             \\
		c_{\overline{\yng(1,1)_{}}}                 & 40. (1,1,\bar{1}_{\overline{\yng(1,1)_{}}},1,1,1,1,1,1,1) & S_R^i             \\
		d_{\yng(1,1)_{}}                            & 40. (1,1,1,1_{\yng(1,1)_{}},1,1,1,1,1,1)                  & S_R^i             \\
		cd                                          & 24. (1,1,\bar{2},2,1,1,1,1,1,1)                           & \Delta ^i         \\
		cd'                                         & 4. (1,1,\bar{2},\bar{2},1,1,1,1,1,1)                      & \Delta ^i         \\
		bc                                          & 4. (1,2,\bar{2},1,1,1,1,1,1,1)                            & \Phi ^i(H_u,H_d)  \\
		bc'                                         & 4. (1,\bar{2},\bar{2},1,1,1,1,1,1,1)                      & \Phi ^i(H_u,H_d)  \\
		bd                                          & 2. (1,\bar{2},1,2,1,1,1,1,1,1)                            & \Xi ^{i}          \\
		be_1                                        & 2. (1,\bar{2},1,1,4,1,1,1,1,1)                            & X_L^{i}           \\
		be_1'                                       & 2. (1,\bar{2},1,1,\bar{4},1,1,1,1,1)                      & X_L^{i}           \\
		be_2                                        & 2. (1,\bar{2},1,1,1,2,1,1,1,1)                            & X_L^{i}           \\
		be_2'                                       & 2. (1,\bar{2},1,1,1,\bar{2},1,1,1,1)                      & X_L^{i}           \\
		bf_2                                        & 1. (1,\bar{2},1,1,1,1,1,4,1,1)                            & X_L^{i}           \\
		bf_2'                                       & 1. (1,\bar{2},1,1,1,1,1,\bar{4},1,1)                      & X_L^{i}           \\
		ce_1                                        & 2. (1,1,\bar{2},1,4,1,1,1,1,1)                            & X_R^{i}           \\
		ce_2'                                       & 2. (1,1,\bar{2},1,1,\bar{2},1,1,1,1)                      & X_R^{i}           \\
		cf_4                                        & 1. (1,1,2,1,1,1,1,1,1,\bar{4})                            & X_R^{i}           \\
		cf_4'                                       & 1. (1,1,2,1,1,1,1,1,1,4)                                  & X_R^{i}           \\
		de_1                                        & 4. (1,1,1,2,\bar{4},1,1,1,1,1)                            & X_R^{i}           \\
		de_1'                                       & 2. (1,1,1,2,4,1,1,1,1,1)                                  & X_R^{i}           \\
		de_2                                        & 2. (1,1,1,2,1,\bar{2},1,1,1,1)                            & X_R^{i}           \\
		de_2'                                       & 4. (1,1,1,2,1,2,1,1,1,1)                                  & X_R^{i}           \\
		df_3                                        & 1. (1,1,1,\bar{2},1,1,1,1,4,1)                            & X_R^{i}           \\
		df_3'                                       & 1. (1,1,1,\bar{2},1,1,1,1,\bar{4},1)                      & X_R^{i}           \\
		\hline
	\end{array}$
	\caption{Particle spectrum of Model~\hyperref[model:r17f1]{r17f1} with gauge symmetry $\SU(4)_{C}\times\SU(2)_{L}\times\SU(2)_{R_1}\times\SU(2)_{R_2}\times\SU(4)\times\SU(2)\times\USp(4)^4$.}
	\label{spec:r17f1}
\end{table}  

The model~\hyperref[model:r17f1]{r17f1} realizes a gauge sector of rank 18 with 1 unit of flux. The construction involves four rigid D6-branes
$\{a,b,c,d\}$ sharing identical fixed points,
$\delta_{ab}^g \neq (4,4,4)$, and realizes the gauge symmetry
$\SU(4)_{C}\times\SU(2)_{L}\times\SU(2)_{R_1}\times\SU(2)_{R_2}\times\SU(4)\times\SU(2)\times\USp(4)^4.$

The matter content is presented in table~\ref{spec:r17f1}, where fields are
organized by their quantum numbers under the gauge symmetry. The resulting spectrum exhibits chiral fermions, scalar fields, and Higgs-like states arising from
brane intersections such as $ab$, $ac$, and $ad'$, yielding fundamental
representations of the corresponding gauge factors. The model~\hyperref[spec:r17f1]{r17f1} features the torus moduli
$\chi_1 = 2 \sqrt{611}$,
$\chi_2 = \sqrt{\frac{47}{13}}$,
$\chi_3 = \sqrt{\frac{47}{13}}$,
and the tree-level gauge coupling relation
$g_a^2=\frac{6120}{47}g_b^2=\frac{2460}{47}g_{cd}^2=\frac{4100}{1687}\frac{5 g_Y^2}{3}=\frac{8 \sqrt{2} 13^{3/4}}{\sqrt[4]{47}}\, \pi \,  e^{\phi _4}$.

The chiral matter content comprises left-handed multiplets
$(Q_L,L_L)$ and right-handed multiplets
$(Q_R,L_R)$ arising from distinct D-brane intersections. The left-handed multiplets naturally realize the quark and lepton doublets and originate from the $ab'$ (3) sector. The right-handed multiplets supply the associated singlet states and receive contributions from the $ac'$ (-2), $ad$ (-1) sectors.

At high energies, there are 28 scalar fields $\Delta^i$ responsible for breaking the Pati-Salam symmetry
to the Standard Model gauge group. At lower energies, there are 8 Higgs-like fields arising from the $bc$, $bc'$ sectors. The spectrum further contains several chiral exotic states
$X_L^i$ and $X_R^i$ charged under hidden sector gauge groups.

Table~\ref{exotic:r17f1} presents the composite spectrum obtained after confinement in the hidden
sector. Fields transforming under non-abelian hidden gauge groups are subject to strong coupling effects, which results in the emergence of bound states neutral under the hidden gauge group. As a result, these degrees of freedom are absent from the low-energy effective theory.

 
\subsection{Model~\texorpdfstring{\hyperref[spec:r43f1]{r43f1}}{r43f1}}

\begin{table}[t]
	\centering
	$\begin{array}{|c|D{.}{\times}{28}|c|}
		\hline
		\text{Model~\hyperref[exotic:r43f1]{r43f1}} & \mc{\text{Quantum Numbers}}                                      & \text{Fields}     \\ 
		\hline
		ab'                                         & 3. (4,2,1,1,1,1,1,1,1,1,1,1,1)                                   & F_L^i(Q_L,L_L)    \\
		ac                                          & 1. (4,1,\bar{2},1,1,1,1,1,1,1,1,1,1)                             & F_R^i(Q_R,L_R)    \\
		ac'                                         & 6. (\bar{4},1,\bar{2},1,1,1,1,1,1,1,1,1,1)                       & F_R^i(Q_R,L_R)    \\
		ad                                          & 2. (4,1,1,\bar{2},1,1,1,1,1,1,1,1,1)                             & F_R^{'i}(Q_R,L_R) \\
		c_{\overline{\yng(1,1)_{}}}                 & 100. (1,1,\bar{1}_{\overline{\yng(1,1)_{}}},1,1,1,1,1,1,1,1,1,1) & S_R^i             \\
		b_{\yng(2)}                                 & 16. (1,3_{\yng(2)},1,1,1,1,1,1,1,1,1,1,1)                        & T_L^i             \\
		c_{\overline{\yng(2)}}                      & 18. (1,1,\bar{3}_{\overline{\yng(2)}},1,1,1,1,1,1,1,1,1,1)       & T_R^i             \\
		d_{\overline{\yng(2)}}                      & 8. (1,1,1,\bar{3}_{\overline{\yng(2)}},1,1,1,1,1,1,1,1,1)        & T_R^i             \\
		cd                                          & 12. (1,1,\bar{2},2,1,1,1,1,1,1,1,1,1)                            & \Delta ^i         \\
		bc                                          & 9. (1,\bar{2},2,1,1,1,1,1,1,1,1,1,1)                             & \Phi ^i(H_u,H_d)  \\
		bc'                                         & 15. (1,\bar{2},\bar{2},1,1,1,1,1,1,1,1,1,1)                      & \Phi ^i(H_u,H_d)  \\
		bd                                          & 12. (1,\bar{2},1,2,1,1,1,1,1,1,1,1,1)                            & \Xi ^{i}          \\
		bd'                                         & 4. (1,\bar{2},1,\bar{2},1,1,1,1,1,1,1,1,1)                       & \Xi ^{i}          \\
		be                                          & 2. (1,\bar{2},1,1,2,1,1,1,1,1,1,1,1)                             & X_L^{i}           \\
		be'                                         & 4. (1,\bar{2},1,1,\bar{2},1,1,1,1,1,1,1,1)                       & X_L^{i}           \\
		bf_1'                                       & 2. (1,2,1,1,1,2,1,1,1,1,1,1,1)                                   & X_L^{i}           \\
		bf_2'                                       & 2. (1,2,1,1,1,1,2,1,1,1,1,1,1)                                   & X_L^{i}           \\
		ce                                          & 4. (1,1,\bar{2},1,2,1,1,1,1,1,1,1,1)                             & X_R^{i}           \\
		ce'                                         & 2. (1,1,\bar{2},1,\bar{2},1,1,1,1,1,1,1,1)                       & X_R^{i}           \\
		cf_1                                        & 2. (1,1,\bar{2},1,1,2,1,1,1,1,1,1,1)                             & X_R^{i}           \\
		cf_1'                                       & 4. (1,1,\bar{2},1,1,\bar{2},1,1,1,1,1,1,1)                       & X_R^{i}           \\
		cf_2                                        & 2. (1,1,\bar{2},1,1,1,2,1,1,1,1,1,1)                             & X_R^{i}           \\
		cf_2'                                       & 4. (1,1,\bar{2},1,1,1,\bar{2},1,1,1,1,1,1)                       & X_R^{i}           \\
		cg_1                                        & 4. (1,1,\bar{2},1,1,1,1,2,1,1,1,1,1)                             & X_R^{i}           \\
		cg_1'                                       & 4. (1,1,\bar{2},1,1,1,1,\bar{2},1,1,1,1,1)                       & X_R^{i}           \\
		cg_2                                        & 4. (1,1,\bar{2},1,1,1,1,1,2,1,1,1,1)                             & X_R^{i}           \\
		cg_2'                                       & 4. (1,1,\bar{2},1,1,1,1,1,\bar{2},1,1,1,1)                       & X_R^{i}           \\
		ch_4                                        & 1. (1,1,2,1,1,1,1,1,1,1,1,1,\overline{16})                       & X_R^{i}           \\
		ch_4'                                       & 1. (1,1,2,1,1,1,1,1,1,1,1,1,16)                                  & X_R^{i}           \\
		df_1                                        & 1. (1,1,1,\bar{2},1,2,1,1,1,1,1,1,1)                             & X_R^{i}           \\
		df_2'                                       & 1. (1,1,1,\bar{2},1,1,\bar{2},1,1,1,1,1,1)                       & X_R^{i}           \\
		dg_1                                        & 2. (1,1,1,2,1,1,1,\bar{2},1,1,1,1,1)                             & X_R^{i}           \\
		dg_1'                                       & 1. (1,1,1,2,1,1,1,2,1,1,1,1,1)                                   & X_R^{i}           \\
		dg_2                                        & 1. (1,1,1,2,1,1,1,1,\bar{2},1,1,1,1)                             & X_R^{i}           \\
		dg_2'                                       & 2. (1,1,1,2,1,1,1,1,2,1,1,1,1)                                   & X_R^{i}           \\
		\hline
	\end{array}$
	\caption{Particle spectrum of Model~\hyperref[model:r43f1]{r43f1} with gauge symmetry $\SU(4)_{C}\times\SU(2)_{L}\times\SU(2)_{R_1}\times\SU(2)_{R_2}\times\SU(2)^5\times\USp(16)^4$.}
	\label{spec:r43f1}
\end{table}  

The model~\hyperref[model:r43f1]{r43f1} defines a configuration of rank 43 with 1 unit of flux. The construction involves four rigid D6-branes
$\{a,b,c,d\}$ sharing identical fixed points,
$\delta_{ab}^g \neq (4,4,4)$, and realizes the gauge symmetry
$\SU(4)_{C}\times\SU(2)_{L}\times\SU(2)_{R_1}\times\SU(2)_{R_2}\times\SU(2)^5\times\USp(16)^4.$

The matter content is collected in table~\ref{spec:r43f1}, where fields are
organized by their quantum numbers under the gauge symmetry. One finds in the spectrum chiral fermions, scalar fields, and Higgs-like states arising from
brane intersections such as $ab$, $ac$, and $ad'$, yielding fundamental
representations of the corresponding gauge factors. The model~\hyperref[spec:r43f1]{r43f1} features the torus moduli
$\chi_1 = 8 \sqrt{3}$,
$\chi_2 = \frac{8}{\sqrt{3}}$,
$\chi_3 = \frac{8}{\sqrt{3}}$,
and the tree-level gauge coupling relation
$g_a^2=\frac{39}{4}g_b^2=\frac{1911}{292}g_{cd}^2=\frac{3185}{1566}\frac{5 g_Y^2}{3}=2 \sqrt{2} 3^{3/4}\, \pi \,  e^{\phi _4}$.

The resulting chiral spectrum contains left-handed multiplets
$(Q_L,L_L)$ and right-handed multiplets
$(Q_R,L_R)$ arising from distinct D-brane intersections. The left-handed sector encodes the quark and lepton doublets and originate from the $ab'$ (3) sector. The right-handed sector contains the singlet partners and receive contributions from the $ac$ (1), $ac'$ (-6), $ad$ (2) sectors.

The high-scale spectrum contains are 12 scalar fields $\Delta^i$ responsible for breaking the Pati-Salam symmetry
to the Standard Model gauge group. Furthermore, the spectrum contains are 24 Higgs-like fields arising from the $bc$, $bc'$ sectors. The spectrum further contains several chiral exotic states
$X_L^i$ and $X_R^i$ charged under hidden sector gauge groups.

Table~\ref{exotic:r43f1} presents the composite spectrum obtained after confinement in the hidden
sector. Matter charged under hidden non-abelian sectors undergoes confinement dynamics, which induces bound states neutral under the hidden gauge group. These bound states therefore decouple from the low-energy spectrum.

 
\subsection{Model~\texorpdfstring{\hyperref[spec:r7f2]{r7f2}}{r7f2}}

\begin{table}[t]
	\centering
	$\begin{array}{|c|D{.}{\times}{13}|c|}
		\hline
		\text{Model~\hyperref[exotic:r7f2]{r7f2}} & \mc{\text{Quantum Numbers}}                    & \text{Fields}     \\ 
		\hline
		ab                                        & 3. (4,\bar{2},1,1,1)                           & F_L^i(Q_L,L_L)    \\
		ac                                        & 3. (\bar{4},1,2,1,1)                           & F_R^i(Q_R,L_R)    \\
		ad                                        & 2. (\bar{4},1,1,2,1)                           & F_R^{'i}(Q_R,L_R) \\
		ad'                                       & 2. (4,1,1,2,1)                                 & F_R^{'i}(Q_R,L_R) \\
		b_{\overline{\yng(1,1)_{}}}               & 6. (1,\bar{1}_{\overline{\yng(1,1)_{}}},1,1,1) & S_L^i             \\
		c_{\yng(1,1)_{}}                          & 30. (1,1,1_{\yng(1,1)_{}},1,1)                 & S_R^i             \\
		d_{\yng(1,1)_{}}                          & 76. (1,1,1,1_{\yng(1,1)_{}},1)                 & S_R^i             \\
		d_{\yng(2)}                               & 18. (1,1,1,3_{\yng(2)},1)                      & T_R^i             \\
		cd                                        & 6. (1,1,2,\bar{2},1)                           & \Delta ^i         \\
		cd'                                       & 33. (1,1,2,2,1)                                & \Delta ^i         \\
		bc                                        & 4. (1,2,\bar{2},1,1)                           & \Phi ^i(H_u,H_d)  \\
		bc'                                       & 9. (1,\bar{2},\bar{2},1,1)                     & \Phi ^i(H_u,H_d)  \\
		bd                                        & 6. (1,2,1,\bar{2},1)                           & \Xi ^{i}          \\
		be'                                       & 1. (1,2,1,1,2)                                 & X_L^{i}           \\
		ce                                        & 2. (1,1,2,1,\bar{2})                           & X_R^{i}           \\
		ce'                                       & 2. (1,1,2,1,2)                                 & X_R^{i}           \\
		de                                        & 1. (1,1,1,2,\bar{2})                           & X_R^{i}           \\
		de'                                       & 2. (1,1,1,2,2)                                 & X_R^{i}           \\
		\hline
	\end{array}$
	\caption{Particle spectrum of Model~\hyperref[model:r7f2]{r7f2} with gauge symmetry $\SU(4)_{C}\times\SU(2)_{L}\times\SU(2)_{R_1}\times\SU(2)_{R_2}\times\SU(2)$.}
	\label{spec:r7f2}
\end{table}  

The model~\hyperref[model:r7f2]{r7f2} corresponds to a gauge theory of rank 7 with 2 units of flux. The construction involves four rigid D6-branes
$\{a,b,c,d\}$ sharing identical fixed points,
$\delta_{ab}^g \neq (4,4,4)$, and realizes the gauge symmetry
$\SU(4)_{C}\times\SU(2)_{L}\times\SU(2)_{R_1}\times\SU(2)_{R_2}\times\SU(2).$

The matter content is displayed in table~\ref{spec:r7f2}, where fields are
organized by their quantum numbers under the gauge symmetry. The model gives rise to chiral fermions, scalar fields, and Higgs-like states arising from
brane intersections such as $ab$, $ac$, and $ad'$, yielding fundamental
representations of the corresponding gauge factors. The model~\hyperref[spec:r7f2]{r7f2} features the torus moduli
$\chi_1 = 6$,
$\chi_2 = 2$,
$\chi_3 = 5$,
and the tree-level gauge coupling relation
$g_a^2=\frac{2}{3}g_b^2=\frac{884}{385}g_{cd}^2=\frac{4420}{2923}\frac{5 g_Y^2}{3}=\frac{8}{\sqrt{15}}\, \pi \,  e^{\phi _4}$.

One finds in the chiral sector left-handed multiplets
$(Q_L,L_L)$ and right-handed multiplets
$(Q_R,L_R)$ arising from distinct D-brane intersections. The left-handed fields furnish the Standard Model quark and lepton doublets and originate from the $ab$ (3) sector. The right-handed fields account for the corresponding singlet representations and receive contributions from the $ac$ (-3), $ad$ (-2), $ad'$ (2) sectors.

The GUT sector includes are 39 scalar fields $\Delta^i$ responsible for breaking the Pati-Salam symmetry
to the Standard Model gauge group. The electroweak sector includes are 13 Higgs-like fields arising from the $bc$, $bc'$ sectors. The spectrum further contains several chiral exotic states
$X_L^i$ and $X_R^i$ charged under hidden sector gauge groups.

Table~\ref{exotic:r7f2} presents the composite spectrum obtained after confinement in the hidden
sector. Hidden-sector charged states are governed by strong coupling dynamics, which drives bound states neutral under the hidden gauge group. Accordingly, the exotic states do not contribute to the low-energy effective dynamics.

 
\subsection{Model~\texorpdfstring{\hyperref[spec:r10f2]{r10f2}}{r10f2}}

\begin{table}[t]
	\centering
	$\begin{array}{|c|D{.}{\times}{14}|c|}
		\hline
		\text{Model~\hyperref[exotic:r10f2]{r10f2}} & \mc{\text{Quantum Numbers}}      & \text{Fields}     \\ 
		\hline
		ab'                                         & 3. (\bar{4},\bar{2},1,1,1,1)     & F_L^i(Q_L,L_L)    \\
		ac'                                         & 2. (4,1,2,1,1,1)                 & F_R^i(Q_R,L_R)    \\
		ad'                                         & 1. (4,1,1,2,1,1)                 & F_R^{'i}(Q_R,L_R) \\
		b_{\yng(1,1)_{}}                            & 10. (1,1_{\yng(1,1)_{}},1,1,1,1) & S_L^i             \\
		c_{\yng(1,1)_{}}                            & 40. (1,1,1_{\yng(1,1)_{}},1,1,1) & S_R^i             \\
		d_{\yng(1,1)_{}}                            & 40. (1,1,1,1_{\yng(1,1)_{}},1,1) & S_R^i             \\
		cd                                          & 4. (1,1,2,\bar{2},1,1)           & \Delta ^i         \\
		cd'                                         & 24. (1,1,2,2,1,1)                & \Delta ^i         \\
		bc                                          & 4. (1,\bar{2},2,1,1,1)           & \Phi ^i(H_u,H_d)  \\
		bc'                                         & 4. (1,2,2,1,1,1)                 & \Phi ^i(H_u,H_d)  \\
		bd'                                         & 2. (1,2,1,2,1,1)                 & \Xi ^{i}          \\
		be_1                                        & 2. (1,2,1,1,\bar{2},1)           & X_L^{i}           \\
		be_1'                                       & 2. (1,2,1,1,2,1)                 & X_L^{i}           \\
		be_2                                        & 2. (1,2,1,1,1,\bar{4})           & X_L^{i}           \\
		be_2'                                       & 2. (1,2,1,1,1,4)                 & X_L^{i}           \\
		ce_1'                                       & 2. (1,1,2,1,2,1)                 & X_R^{i}           \\
		ce_2'                                       & 2. (1,1,2,1,1,4)                 & X_R^{i}           \\
		de_1                                        & 4. (1,1,1,2,\bar{2},1)           & X_R^{i}           \\
		de_1'                                       & 2. (1,1,1,2,2,1)                 & X_R^{i}           \\
		de_2                                        & 4. (1,1,1,2,1,\bar{4})           & X_R^{i}           \\
		de_2'                                       & 2. (1,1,1,2,1,4)                 & X_R^{i}           \\
		\hline
	\end{array}$
	\caption{Particle spectrum of Model~\hyperref[model:r10f2]{r10f2} with gauge symmetry $\SU(4)_{C}\times\SU(2)_{L}\times\SU(2)_{R_1}\times\SU(2)_{R_2}\times\SU(2)\times\SU(4)$.}
	\label{spec:r10f2}
\end{table}  

The model~\hyperref[model:r10f2]{r10f2} describes a gauge-theoretic construction of rank 10 with 2 units of flux. The construction involves four rigid D6-branes
$\{a,b,c,d\}$ sharing identical fixed points,
$\delta_{ab}^g \neq (4,4,4)$, and realizes the gauge symmetry
$\SU(4)_{C}\times\SU(2)_{L}\times\SU(2)_{R_1}\times\SU(2)_{R_2}\times\SU(2)\times\SU(4).$

The matter content is tabulated in table~\ref{spec:r10f2}, where fields are
organized by their quantum numbers under the gauge symmetry. The spectrum contains chiral fermions, scalar fields, and Higgs-like states arising from
brane intersections such as $ab$, $ac$, and $ad'$, yielding fundamental
representations of the corresponding gauge factors. The model~\hyperref[spec:r10f2]{r10f2} features the torus moduli
$\chi_1 = 2 \sqrt{611}$,
$\chi_2 = \sqrt{\frac{47}{13}}$,
$\chi_3 = \sqrt{\frac{47}{13}}$,
and the tree-level gauge coupling relation
$g_a^2=\frac{6120}{47}g_b^2=\frac{2460}{47}g_{cd}^2=\frac{4100}{1687}\frac{5 g_Y^2}{3}=\frac{8 \sqrt{2} 13^{3/4}}{\sqrt[4]{47}}\, \pi \,  e^{\phi _4}$.

The chiral sector consists of left-handed multiplets
$(Q_L,L_L)$ and right-handed multiplets
$(Q_R,L_R)$ arising from distinct D-brane intersections. The left-handed states accommodate the quark and lepton doublets and originate from the $ab'$ (-3) sector. The right-handed states provide the corresponding singlet partners and receive contributions from the $ac'$ (2), $ad'$ (1) sectors.

At the GUT scale, there are 28 scalar fields $\Delta^i$ responsible for breaking the Pati-Salam symmetry
to the Standard Model gauge group. In addition, there are 8 Higgs-like fields arising from the $bc$, $bc'$ sectors. The spectrum further contains several chiral exotic states
$X_L^i$ and $X_R^i$ charged under hidden sector gauge groups.

Table~\ref{exotic:r10f2} presents the composite spectrum obtained after confinement in the hidden
sector. States charged under non-abelian hidden gauge factors experience strong coupling dynamics, which gives rise to bound states neutral under the hidden gauge group. Hence, the resulting bound states are removed from the low-energy spectrum.

 
\subsection{Model~\texorpdfstring{\hyperref[spec:r35f2]{r35f2}}{r35f2}}

\begin{table}[t]
	\centering
	$\begin{array}{|c|D{.}{\times}{28}|c|}
		\hline
		\text{Model~\hyperref[exotic:r35f2]{r35f2}} & \mc{\text{Quantum Numbers}}                                      & \text{Fields}     \\ 
		\hline
		ab                                          & 3. (4,\bar{2},1,1,1,1,1,1,1,1,1,1,1)                             & F_L^i(Q_L,L_L)    \\
		ab'                                         & 6. (\bar{4},\bar{2},1,1,1,1,1,1,1,1,1,1,1)                       & F_L^i(Q_L,L_L)    \\
		ac                                          & 6. (4,1,\bar{2},1,1,1,1,1,1,1,1,1,1)                             & F_R^i(Q_R,L_R)    \\
		ac'                                         & 1. (\bar{4},1,\bar{2},1,1,1,1,1,1,1,1,1,1)                       & F_R^i(Q_R,L_R)    \\
		ad                                          & 2. (\bar{4},1,1,2,1,1,1,1,1,1,1,1,1)                             & F_R^{'i}(Q_R,L_R) \\
		c_{\overline{\yng(1,1)_{}}}                 & 100. (1,1,\bar{1}_{\overline{\yng(1,1)_{}}},1,1,1,1,1,1,1,1,1,1) & S_R^i             \\
		b_{\yng(2)}                                 & 16. (1,3_{\yng(2)},1,1,1,1,1,1,1,1,1,1,1)                        & T_L^i             \\
		c_{\overline{\yng(2)}}                      & 18. (1,1,\bar{3}_{\overline{\yng(2)}},1,1,1,1,1,1,1,1,1,1)       & T_R^i             \\
		d_{\yng(2)}                                 & 8. (1,1,1,3_{\yng(2)},1,1,1,1,1,1,1,1,1)                         & T_R^i             \\
		cd'                                         & 12. (1,1,\bar{2},\bar{2},1,1,1,1,1,1,1,1,1)                      & \Delta ^i         \\
		bc                                          & 9. (1,\bar{2},2,1,1,1,1,1,1,1,1,1,1)                             & \Phi ^i(H_u,H_d)  \\
		bc'                                         & 15. (1,\bar{2},\bar{2},1,1,1,1,1,1,1,1,1,1)                      & \Phi ^i(H_u,H_d)  \\
		bd                                          & 8. (1,2,1,\bar{2},1,1,1,1,1,1,1,1,1)                             & \Xi ^{i}          \\
		be                                          & 2. (1,2,1,1,\bar{2},1,1,1,1,1,1,1,1)                             & X_L^{i}           \\
		bf_1                                        & 2. (1,\bar{2},1,1,1,2,1,1,1,1,1,1,1)                             & X_L^{i}           \\
		bf_1'                                       & 4. (1,\bar{2},1,1,1,\bar{2},1,1,1,1,1,1,1)                       & X_L^{i}           \\
		bf_2                                        & 4. (1,\bar{2},1,1,1,1,2,1,1,1,1,1,1)                             & X_L^{i}           \\
		bf_2'                                       & 2. (1,\bar{2},1,1,1,1,\bar{2},1,1,1,1,1,1)                       & X_L^{i}           \\
		ce                                          & 4. (1,1,\bar{2},1,2,1,1,1,1,1,1,1,1)                             & X_R^{i}           \\
		ce'                                         & 2. (1,1,\bar{2},1,\bar{2},1,1,1,1,1,1,1,1)                       & X_R^{i}           \\
		cf_1                                        & 4. (1,1,\bar{2},1,1,2,1,1,1,1,1,1,1)                             & X_R^{i}           \\
		cf_1'                                       & 2. (1,1,\bar{2},1,1,\bar{2},1,1,1,1,1,1,1)                       & X_R^{i}           \\
		cf_2                                        & 2. (1,1,\bar{2},1,1,1,2,1,1,1,1,1,1)                             & X_R^{i}           \\
		cf_2'                                       & 4. (1,1,\bar{2},1,1,1,\bar{2},1,1,1,1,1,1)                       & X_R^{i}           \\
		cg_1                                        & 4. (1,1,\bar{2},1,1,1,1,2,1,1,1,1,1)                             & X_R^{i}           \\
		cg_1'                                       & 4. (1,1,\bar{2},1,1,1,1,\bar{2},1,1,1,1,1)                       & X_R^{i}           \\
		cg_2                                        & 4. (1,1,\bar{2},1,1,1,1,1,2,1,1,1,1)                             & X_R^{i}           \\
		cg_2'                                       & 4. (1,1,\bar{2},1,1,1,1,1,\bar{2},1,1,1,1)                       & X_R^{i}           \\
		ch_2                                        & 1. (1,1,2,1,1,1,1,1,1,1,\overline{12},1,1)                       & X_R^{i}           \\
		ch_2'                                       & 1. (1,1,2,1,1,1,1,1,1,1,12,1,1)                                  & X_R^{i}           \\
		df_1                                        & 1. (1,1,1,2,1,\bar{2},1,1,1,1,1,1,1)                             & X_R^{i}           \\
		df_2                                        & 1. (1,1,1,2,1,1,\bar{2},1,1,1,1,1,1)                             & X_R^{i}           \\
		dg_1                                        & 2. (1,1,1,\bar{2},1,1,1,2,1,1,1,1,1)                             & X_R^{i}           \\
		dg_1'                                       & 1. (1,1,1,\bar{2},1,1,1,\bar{2},1,1,1,1,1)                       & X_R^{i}           \\
		dg_2                                        & 1. (1,1,1,\bar{2},1,1,1,1,2,1,1,1,1)                             & X_R^{i}           \\
		dg_2'                                       & 2. (1,1,1,\bar{2},1,1,1,1,\bar{2},1,1,1,1)                       & X_R^{i}           \\
		\hline
	\end{array}$
	\caption{Particle spectrum of Model~\hyperref[model:r35f2]{r35f2} with gauge symmetry $\SU(4)_{C}\times\SU(2)_{L}\times\SU(2)_{R_1}\times\SU(2)_{R_2}\times\SU(2)^5\times\USp(12)^4$.}
	\label{spec:r35f2}
\end{table}  

The model~\hyperref[model:r35f2]{r35f2} constitutes a consistent gauge framework of rank 35 with 2 units of flux. The construction involves four rigid D6-branes
$\{a,b,c,d\}$ sharing identical fixed points,
$\delta_{ab}^g \neq (4,4,4)$, and realizes the gauge symmetry
$\SU(4)_{C}\times\SU(2)_{L}\times\SU(2)_{R_1}\times\SU(2)_{R_2}\times\SU(2)^5\times\USp(12)^4.$

The matter content is organized in table~\ref{spec:r35f2}, where fields are
organized by their quantum numbers under the gauge symmetry. The resulting spectrum exhibits chiral fermions, scalar fields, and Higgs-like states arising from
brane intersections such as $ab$, $ac$, and $ad'$, yielding fundamental
representations of the corresponding gauge factors. The model~\hyperref[spec:r35f2]{r35f2} features the torus moduli
$\chi_1 = 4 \sqrt{30}$,
$\chi_2 = 4 \sqrt{\frac{10}{3}}$,
$\chi_3 = \frac{4 \sqrt{\frac{10}{3}}}{3}$,
and the tree-level gauge coupling relation
$g_a^2=\frac{93}{10}g_b^2=\frac{17329}{17410}g_{cd}^2=\frac{86645}{86888}\frac{5 g_Y^2}{3}=2 \sqrt[4]{\frac{3}{5}} 2^{3/4}\, \pi \,  e^{\phi _4}$.

The chiral matter content comprises left-handed multiplets
$(Q_L,L_L)$ and right-handed multiplets
$(Q_R,L_R)$ arising from distinct D-brane intersections. The left-handed multiplets naturally realize the quark and lepton doublets and originate from the $ab$ (3), $ab'$ (-6) sectors. The right-handed multiplets supply the associated singlet states and receive contributions from the $ac$ (6), $ac'$ (-1), $ad$ (-2) sectors.

At high energies, there are 12 scalar fields $\Delta^i$ responsible for breaking the Pati-Salam symmetry
to the Standard Model gauge group. At lower energies, there are 24 Higgs-like fields arising from the $bc$, $bc'$ sectors. The spectrum further contains several chiral exotic states
$X_L^i$ and $X_R^i$ charged under hidden sector gauge groups.

Table~\ref{exotic:r35f2} presents the composite spectrum obtained after confinement in the hidden
sector. Fields transforming under non-abelian hidden gauge groups are subject to strong coupling effects, which triggers the formation of bound states neutral under the hidden gauge group. Consequently, these states decouple from the low-energy effective spectrum.

 
\subsection{Model~\texorpdfstring{\hyperref[spec:r43af2]{r43af2}}{r43af2}}

\begin{table}[t]
	\centering
	$\begin{array}{|c|D{.}{\times}{28}|c|}
		\hline
		\text{Model~\hyperref[exotic:r43af2]{r43af2}} & \mc{\text{Quantum Numbers}}                                      & \text{Fields}     \\ 
		\hline
		ab                                            & 3. (4,\bar{2},1,1,1,1,1,1,1,1,1,1,1)                             & F_L^i(Q_L,L_L)    \\
		ab'                                           & 6. (\bar{4},\bar{2},1,1,1,1,1,1,1,1,1,1,1)                       & F_L^i(Q_L,L_L)    \\
		ac                                            & 6. (4,1,\bar{2},1,1,1,1,1,1,1,1,1,1)                             & F_R^i(Q_R,L_R)    \\
		ac'                                           & 1. (\bar{4},1,\bar{2},1,1,1,1,1,1,1,1,1,1)                       & F_R^i(Q_R,L_R)    \\
		ad                                            & 2. (\bar{4},1,1,2,1,1,1,1,1,1,1,1,1)                             & F_R^{'i}(Q_R,L_R) \\
		b_{\overline{\yng(1,1)_{}}}                   & 8. (1,\bar{1}_{\overline{\yng(1,1)_{}}},1,1,1,1,1,1,1,1,1,1,1)   & S_L^i             \\
		c_{\overline{\yng(1,1)_{}}}                   & 100. (1,1,\bar{1}_{\overline{\yng(1,1)_{}}},1,1,1,1,1,1,1,1,1,1) & S_R^i             \\
		d_{\overline{\yng(1,1)_{}}}                   & 8. (1,1,1,\bar{1}_{\overline{\yng(1,1)_{}}},1,1,1,1,1,1,1,1,1)   & S_R^i             \\
		b_{\yng(2)}                                   & 12. (1,3_{\yng(2)},1,1,1,1,1,1,1,1,1,1,1)                        & T_L^i             \\
		c_{\overline{\yng(2)}}                        & 18. (1,1,\bar{3}_{\overline{\yng(2)}},1,1,1,1,1,1,1,1,1,1)       & T_R^i             \\
		d_{\yng(2)}                                   & 12. (1,1,1,3_{\yng(2)},1,1,1,1,1,1,1,1,1)                        & T_R^i             \\
		cd                                            & 6. (1,1,2,\bar{2},1,1,1,1,1,1,1,1,1)                             & \Delta ^i         \\
		cd'                                           & 15. (1,1,\bar{2},\bar{2},1,1,1,1,1,1,1,1,1)                      & \Delta ^i         \\
		bc                                            & 8. (1,\bar{2},2,1,1,1,1,1,1,1,1,1,1)                             & \Phi ^i(H_u,H_d)  \\
		bc'                                           & 20. (1,\bar{2},\bar{2},1,1,1,1,1,1,1,1,1,1)                      & \Phi ^i(H_u,H_d)  \\
		bd                                            & 12. (1,2,1,\bar{2},1,1,1,1,1,1,1,1,1)                            & \Xi ^{i}          \\
		bd'                                           & 4. (1,\bar{2},1,\bar{2},1,1,1,1,1,1,1,1,1)                       & \Xi ^{i}          \\
		be                                            & 1. (1,2,1,1,\bar{2},1,1,1,1,1,1,1,1)                             & X_L^{i}           \\
		bf_1                                          & 2. (1,\bar{2},1,1,1,2,1,1,1,1,1,1,1)                             & X_L^{i}           \\
		bf_1'                                         & 4. (1,\bar{2},1,1,1,\bar{2},1,1,1,1,1,1,1)                       & X_L^{i}           \\
		bf_2                                          & 4. (1,\bar{2},1,1,1,1,2,1,1,1,1,1,1)                             & X_L^{i}           \\
		bf_2'                                         & 2. (1,\bar{2},1,1,1,1,\bar{2},1,1,1,1,1,1)                       & X_L^{i}           \\
		ce                                            & 4. (1,1,\bar{2},1,2,1,1,1,1,1,1,1,1)                             & X_R^{i}           \\
		ce'                                           & 2. (1,1,\bar{2},1,\bar{2},1,1,1,1,1,1,1,1)                       & X_R^{i}           \\
		cf_1                                          & 4. (1,1,\bar{2},1,1,2,1,1,1,1,1,1,1)                             & X_R^{i}           \\
		cf_1'                                         & 2. (1,1,\bar{2},1,1,\bar{2},1,1,1,1,1,1,1)                       & X_R^{i}           \\
		cf_2                                          & 2. (1,1,\bar{2},1,1,1,2,1,1,1,1,1,1)                             & X_R^{i}           \\
		cf_2'                                         & 4. (1,1,\bar{2},1,1,1,\bar{2},1,1,1,1,1,1)                       & X_R^{i}           \\
		cg_1                                          & 4. (1,1,\bar{2},1,1,1,1,2,1,1,1,1,1)                             & X_R^{i}           \\
		cg_1'                                         & 4. (1,1,\bar{2},1,1,1,1,\bar{2},1,1,1,1,1)                       & X_R^{i}           \\
		cg_2                                          & 4. (1,1,\bar{2},1,1,1,1,1,2,1,1,1,1)                             & X_R^{i}           \\
		cg_2'                                         & 4. (1,1,\bar{2},1,1,1,1,1,\bar{2},1,1,1,1)                       & X_R^{i}           \\
		ch_2                                          & 1. (1,1,2,1,1,1,1,1,1,1,\overline{16},1,1)                       & X_R^{i}           \\
		ch_2'                                         & 1. (1,1,2,1,1,1,1,1,1,1,16,1,1)                                  & X_R^{i}           \\
		df_1                                          & 1. (1,1,1,2,1,\bar{2},1,1,1,1,1,1,1)                             & X_R^{i}           \\
		df_2                                          & 1. (1,1,1,2,1,1,\bar{2},1,1,1,1,1,1)                             & X_R^{i}           \\
		dg_1                                          & 4. (1,1,1,\bar{2},1,1,1,2,1,1,1,1,1)                             & X_R^{i}           \\
		dg_1'                                         & 2. (1,1,1,\bar{2},1,1,1,\bar{2},1,1,1,1,1)                       & X_R^{i}           \\
		dg_2                                          & 2. (1,1,1,\bar{2},1,1,1,1,2,1,1,1,1)                             & X_R^{i}           \\
		dg_2'                                         & 4. (1,1,1,\bar{2},1,1,1,1,\bar{2},1,1,1,1)                       & X_R^{i}           \\
		\hline
	\end{array}$
	\caption{Particle spectrum of Model~\hyperref[model:r43af2]{r43af2} with gauge symmetry $\SU(4)_{C}\times\SU(2)_{L}\times\SU(2)_{R_1}\times\SU(2)_{R_2}\times\SU(2)^5\times\USp(16)^4$.}
	\label{spec:r43af2}
\end{table}  

The model~\hyperref[model:r43af2]{r43af2} implements a supersymmetric gauge setup of rank 43 with 2 units of flux. The construction involves four rigid D6-branes
$\{a,b,c,d\}$ sharing identical fixed points,
$\delta_{ab}^g \neq (4,4,4)$, and realizes the gauge symmetry
$\SU(4)_{C}\times\SU(2)_{L}\times\SU(2)_{R_1}\times\SU(2)_{R_2}\times\SU(2)^5\times\USp(16)^4.$

The matter content is compiled in table~\ref{spec:r43af2}, where fields are
organized by their quantum numbers under the gauge symmetry. One finds in the spectrum chiral fermions, scalar fields, and Higgs-like states arising from
brane intersections such as $ab$, $ac$, and $ad'$, yielding fundamental
representations of the corresponding gauge factors. The model~\hyperref[spec:r43af2]{r43af2} features the torus moduli
$\chi_1 = 4 \sqrt{195}$,
$\chi_2 = 2 \sqrt{\frac{65}{3}}$,
$\chi_3 = \frac{\sqrt{\frac{65}{3}}}{3}$,
and the tree-level gauge coupling relation
$g_a^2=\frac{1176}{65}g_b^2=\frac{4312}{7475}g_{cd}^2=\frac{21560}{31049}\frac{5 g_Y^2}{3}=4 \sqrt[4]{\frac{3}{65}} \sqrt{2}\, \pi \,  e^{\phi _4}$.

The resulting chiral spectrum contains left-handed multiplets
$(Q_L,L_L)$ and right-handed multiplets
$(Q_R,L_R)$ arising from distinct D-brane intersections. The left-handed sector encodes the quark and lepton doublets and originate from the $ab$ (3), $ab'$ (-6) sectors. The right-handed sector contains the singlet partners and receive contributions from the $ac$ (6), $ac'$ (-1), $ad$ (-2) sectors.

The high-scale spectrum contains are 21 scalar fields $\Delta^i$ responsible for breaking the Pati-Salam symmetry
to the Standard Model gauge group. Furthermore, the spectrum contains are 28 Higgs-like fields arising from the $bc$, $bc'$ sectors. The spectrum further contains several chiral exotic states
$X_L^i$ and $X_R^i$ charged under hidden sector gauge groups.

Table~\ref{exotic:r43af2} presents the composite spectrum obtained after confinement in the hidden
sector. Matter charged under hidden non-abelian sectors undergoes confinement dynamics, which forces the appearance of bound states neutral under the hidden gauge group. As a result, these degrees of freedom are absent from the low-energy effective theory.

 
\subsection{Model~\texorpdfstring{\hyperref[spec:r43bf2]{r43bf2}}{r43bf2}}

\begin{table}[t]
	\centering
	$\begin{array}{|c|D{.}{\times}{28}|c|}
		\hline
		\text{Model~\hyperref[exotic:r43bf2]{r43bf2}} & \mc{\text{Quantum Numbers}}                                    & \text{Fields}     \\ 
		\hline
		ab                                            & 6. (4,\bar{2},1,1,1,1,1,1,1,1,1,1,1)                           & F_L^i(Q_L,L_L)    \\
		ab'                                           & 3. (\bar{4},\bar{2},1,1,1,1,1,1,1,1,1,1,1)                     & F_L^i(Q_L,L_L)    \\
		ac                                            & 6. (\bar{4},1,2,1,1,1,1,1,1,1,1,1,1)                           & F_R^i(Q_R,L_R)    \\
		ac'                                           & 1. (4,1,2,1,1,1,1,1,1,1,1,1,1)                                 & F_R^i(Q_R,L_R)    \\
		ad                                            & 4. (4,1,1,\bar{2},1,1,1,1,1,1,1,1,1)                           & F_R^{'i}(Q_R,L_R) \\
		ad'                                           & 2. (\bar{4},1,1,\bar{2},1,1,1,1,1,1,1,1,1)                     & F_R^{'i}(Q_R,L_R) \\
		b_{\overline{\yng(1,1)_{}}}                   & 8. (1,\bar{1}_{\overline{\yng(1,1)_{}}},1,1,1,1,1,1,1,1,1,1,1) & S_L^i             \\
		c_{\yng(1,1)_{}}                              & 100. (1,1,1_{\yng(1,1)_{}},1,1,1,1,1,1,1,1,1,1)                & S_R^i             \\
		d_{\yng(1,1)_{}}                              & 8. (1,1,1,1_{\yng(1,1)_{}},1,1,1,1,1,1,1,1,1)                  & S_R^i             \\
		b_{\yng(2)}                                   & 12. (1,3_{\yng(2)},1,1,1,1,1,1,1,1,1,1,1)                      & T_L^i             \\
		c_{\yng(2)}                                   & 18. (1,1,3_{\yng(2)},1,1,1,1,1,1,1,1,1,1)                      & T_R^i             \\
		d_{\yng(2)}                                   & 12. (1,1,1,3_{\yng(2)},1,1,1,1,1,1,1,1,1)                      & T_R^i             \\
		cd                                            & 9. (1,1,2,\bar{2},1,1,1,1,1,1,1,1,1)                           & \Delta ^i         \\
		bc                                            & 20. (1,\bar{2},2,1,1,1,1,1,1,1,1,1,1)                          & \Phi ^i(H_u,H_d)  \\
		bc'                                           & 8. (1,\bar{2},\bar{2},1,1,1,1,1,1,1,1,1,1)                     & \Phi ^i(H_u,H_d)  \\
		bd'                                           & 8. (1,2,1,2,1,1,1,1,1,1,1,1,1)                                 & \Xi ^{i}          \\
		be'                                           & 1. (1,2,1,1,2,1,1,1,1,1,1,1,1)                                 & X_L^{i}           \\
		bg_1                                          & 4. (1,\bar{2},1,1,1,1,1,2,1,1,1,1,1)                           & X_L^{i}           \\
		bg_1'                                         & 2. (1,\bar{2},1,1,1,1,1,\bar{2},1,1,1,1,1)                     & X_L^{i}           \\
		bg_2                                          & 2. (1,\bar{2},1,1,1,1,1,1,2,1,1,1,1)                           & X_L^{i}           \\
		bg_2'                                         & 4. (1,\bar{2},1,1,1,1,1,1,\bar{2},1,1,1,1)                     & X_L^{i}           \\
		ce                                            & 4. (1,1,2,1,\bar{2},1,1,1,1,1,1,1,1)                           & X_R^{i}           \\
		ce'                                           & 2. (1,1,2,1,2,1,1,1,1,1,1,1,1)                                 & X_R^{i}           \\
		cf_1                                          & 4. (1,1,2,1,1,\bar{2},1,1,1,1,1,1,1)                           & X_R^{i}           \\
		cf_1'                                         & 4. (1,1,2,1,1,2,1,1,1,1,1,1,1)                                 & X_R^{i}           \\
		cf_2                                          & 4. (1,1,2,1,1,1,\bar{2},1,1,1,1,1,1)                           & X_R^{i}           \\
		cf_2'                                         & 4. (1,1,2,1,1,1,2,1,1,1,1,1,1)                                 & X_R^{i}           \\
		cg_1                                          & 4. (1,1,2,1,1,1,1,\bar{2},1,1,1,1,1)                           & X_R^{i}           \\
		cg_1'                                         & 2. (1,1,2,1,1,1,1,2,1,1,1,1,1)                                 & X_R^{i}           \\
		cg_2                                          & 2. (1,1,2,1,1,1,1,1,\bar{2},1,1,1,1)                           & X_R^{i}           \\
		cg_2'                                         & 4. (1,1,2,1,1,1,1,1,2,1,1,1,1)                                 & X_R^{i}           \\
		ch_3                                          & 1. (1,1,\bar{2},1,1,1,1,1,1,1,1,16,1)                          & X_R^{i}           \\
		ch_3'                                         & 1. (1,1,\bar{2},1,1,1,1,1,1,1,1,\overline{16},1)               & X_R^{i}           \\
		df_1'                                         & 2. (1,1,1,2,1,2,1,1,1,1,1,1,1)                                 & X_R^{i}           \\
		df_2'                                         & 2. (1,1,1,2,1,1,2,1,1,1,1,1,1)                                 & X_R^{i}           \\
		dg_1                                          & 2. (1,1,1,\bar{2},1,1,1,2,1,1,1,1,1)                           & X_R^{i}           \\
		dg_1'                                         & 1. (1,1,1,\bar{2},1,1,1,\bar{2},1,1,1,1,1)                     & X_R^{i}           \\
		dg_2                                          & 2. (1,1,1,\bar{2},1,1,1,1,2,1,1,1,1)                           & X_R^{i}           \\
		dg_2'                                         & 1. (1,1,1,\bar{2},1,1,1,1,\bar{2},1,1,1,1)                     & X_R^{i}           \\
		\hline
	\end{array}$
	\caption{Particle spectrum of Model~\hyperref[model:r43bf2]{r43bf2} with gauge symmetry $\SU(4)_{C}\times\SU(2)_{L}\times\SU(2)_{R_1}\times\SU(2)_{R_2}\times\SU(2)^5\times\USp(16)^4$.}
	\label{spec:r43bf2}
\end{table}  

The model~\hyperref[model:r43bf2]{r43bf2} represents a consistent intersecting-brane model of rank 43 with 2 units of flux. The construction involves four rigid D6-branes
$\{a,b,c,d\}$ sharing identical fixed points,
$\delta_{ab}^g \neq (4,4,4)$, and realizes the gauge symmetry
$\SU(4)_{C}\times\SU(2)_{L}\times\SU(2)_{R_1}\times\SU(2)_{R_2}\times\SU(2)^5\times\USp(16)^4.$

The matter content is detailed in table~\ref{spec:r43bf2}, where fields are
organized by their quantum numbers under the gauge symmetry. The model gives rise to chiral fermions, scalar fields, and Higgs-like states arising from
brane intersections such as $ab$, $ac$, and $ad'$, yielding fundamental
representations of the corresponding gauge factors. The model~\hyperref[spec:r43bf2]{r43bf2} features the torus moduli
$\chi_1 = 3 \sqrt{3}$,
$\chi_2 = 12 \sqrt{3}$,
$\chi_3 = 2 \sqrt{3}$,
and the tree-level gauge coupling relation
$g_a^2=\frac{56}{27}g_b^2=\frac{1736}{3303}g_{cd}^2=\frac{8680}{13381}\frac{5 g_Y^2}{3}=\frac{4 \sqrt{2}}{3 \sqrt[4]{3}}\, \pi \,  e^{\phi _4}$.

One finds in the chiral sector left-handed multiplets
$(Q_L,L_L)$ and right-handed multiplets
$(Q_R,L_R)$ arising from distinct D-brane intersections. The left-handed fields furnish the Standard Model quark and lepton doublets and originate from the $ab$ (6), $ab'$ (-3) sectors. The right-handed fields account for the corresponding singlet representations and receive contributions from the $ac$ (-6), $ac'$ (1), $ad$ (4), $ad'$ (-2) sectors.

The GUT sector includes are 9 scalar fields $\Delta^i$ responsible for breaking the Pati-Salam symmetry
to the Standard Model gauge group. The electroweak sector includes are 28 Higgs-like fields arising from the $bc$, $bc'$ sectors. The spectrum further contains several chiral exotic states
$X_L^i$ and $X_R^i$ charged under hidden sector gauge groups.

Table~\ref{exotic:r43bf2} presents the composite spectrum obtained after confinement in the hidden
sector. Hidden-sector charged states are governed by strong coupling dynamics, which dynamically generates bound states neutral under the hidden gauge group. These bound states therefore decouple from the low-energy spectrum.

 
\subsection{Model~\texorpdfstring{\hyperref[spec:r123f2]{r123f2}}{r123f2}}

\begin{table}[t]
	\centering
	$\begin{array}{|c|D{.}{\times}{28}|c|}
		\hline
		\text{Model~\hyperref[exotic:r123f2]{r123f2}} & \mc{\text{Quantum Numbers}}                                    & \text{Fields}     \\ 
		\hline
		ab                                            & 3. (\bar{4},2,1,1,1,1,1,1,1,1,1,1,1)                           & F_L^i(Q_L,L_L)    \\
		ab'                                           & 6. (4,2,1,1,1,1,1,1,1,1,1,1,1)                                 & F_L^i(Q_L,L_L)    \\
		ac                                            & 4. (\bar{4},1,2,1,1,1,1,1,1,1,1,1,1)                           & F_R^i(Q_R,L_R)    \\
		ad                                            & 1. (4,1,1,\bar{2},1,1,1,1,1,1,1,1,1)                           & F_R^{'i}(Q_R,L_R) \\
		b_{\yng(1,1)_{}}                              & 150. (1,1_{\yng(1,1)_{}},1,1,1,1,1,1,1,1,1,1,1)                & S_L^i             \\
		c_{\yng(1,1)_{}}                              & 12. (1,1,1_{\yng(1,1)_{}},1,1,1,1,1,1,1,1,1,1)                 & S_R^i             \\
		d_{\overline{\yng(1,1)_{}}}                   & 6. (1,1,1,\bar{1}_{\overline{\yng(1,1)_{}}},1,1,1,1,1,1,1,1,1) & S_R^i             \\
		b_{\yng(2)}                                   & 36. (1,3_{\yng(2)},1,1,1,1,1,1,1,1,1,1,1)                      & T_L^i             \\
		c_{\overline{\yng(2)}}                        & 6. (1,1,\bar{3}_{\overline{\yng(2)}},1,1,1,1,1,1,1,1,1,1)      & T_R^i             \\
		d_{\overline{\yng(2)}}                        & 4. (1,1,1,\bar{3}_{\overline{\yng(2)}},1,1,1,1,1,1,1,1,1)      & T_R^i             \\
		cd                                            & 2. (1,1,\bar{2},2,1,1,1,1,1,1,1,1,1)                           & \Delta ^i         \\
		bc                                            & 5. (1,\bar{2},2,1,1,1,1,1,1,1,1,1,1)                           & \Phi ^i(H_u,H_d)  \\
		bc'                                           & 18. (1,2,2,1,1,1,1,1,1,1,1,1,1)                                & \Phi ^i(H_u,H_d)  \\
		bd                                            & 6. (1,2,1,\bar{2},1,1,1,1,1,1,1,1,1)                           & \Xi ^{i}          \\
		bd'                                           & 6. (1,2,1,2,1,1,1,1,1,1,1,1,1)                                 & \Xi ^{i}          \\
		be                                            & 4. (1,2,1,1,\bar{2},1,1,1,1,1,1,1,1)                           & X_L^{i}           \\
		be'                                           & 4. (1,2,1,1,2,1,1,1,1,1,1,1,1)                                 & X_L^{i}           \\
		bf_1                                          & 6. (1,2,1,1,1,\bar{2},1,1,1,1,1,1,1)                           & X_L^{i}           \\
		bf_1'                                         & 4. (1,2,1,1,1,2,1,1,1,1,1,1,1)                                 & X_L^{i}           \\
		bf_2                                          & 6. (1,2,1,1,1,1,\bar{2},1,1,1,1,1,1)                           & X_L^{i}           \\
		bf_2'                                         & 4. (1,2,1,1,1,1,2,1,1,1,1,1,1)                                 & X_L^{i}           \\
		bg_1                                          & 4. (1,2,1,1,1,1,1,\bar{2},1,1,1,1,1)                           & X_L^{i}           \\
		bg_1'                                         & 6. (1,2,1,1,1,1,1,2,1,1,1,1,1)                                 & X_L^{i}           \\
		bg_2                                          & 6. (1,2,1,1,1,1,1,1,\bar{2},1,1,1,1)                           & X_L^{i}           \\
		bg_2'                                         & 4. (1,2,1,1,1,1,1,1,2,1,1,1,1)                                 & X_L^{i}           \\
		bh_4                                          & 1. (1,\bar{2},1,1,1,1,1,1,1,1,1,1,56)                          & X_L^{i}           \\
		bh_4'                                         & 1. (1,\bar{2},1,1,1,1,1,1,1,1,1,1,\overline{56})               & X_L^{i}           \\
		ce                                            & 1. (1,1,\bar{2},1,2,1,1,1,1,1,1,1,1)                           & X_R^{i}           \\
		cf_1                                          & 2. (1,1,2,1,1,\bar{2},1,1,1,1,1,1,1)                           & X_R^{i}           \\
		cf_1'                                         & 2. (1,1,2,1,1,2,1,1,1,1,1,1,1)                                 & X_R^{i}           \\
		cf_2                                          & 3. (1,1,2,1,1,1,\bar{2},1,1,1,1,1,1)                           & X_R^{i}           \\
		cf_2'                                         & 3. (1,1,2,1,1,1,2,1,1,1,1,1,1)                                 & X_R^{i}           \\
		df_1                                          & 1. (1,1,1,\bar{2},1,2,1,1,1,1,1,1,1)                           & X_R^{i}           \\
		df_2                                          & 2. (1,1,1,\bar{2},1,1,2,1,1,1,1,1,1)                           & X_R^{i}           \\
		dg_1                                          & 1. (1,1,1,2,1,1,1,\bar{2},1,1,1,1,1)                           & X_R^{i}           \\
		dg_1'                                         & 1. (1,1,1,2,1,1,1,2,1,1,1,1,1)                                 & X_R^{i}           \\
		\hline
	\end{array}$
	\caption{Particle spectrum of Model~\hyperref[model:r123f2]{r123f2} with gauge symmetry $\SU(4)_{C}\times\SU(2)_{L}\times\SU(2)_{R_1}\times\SU(2)_{R_2}\times\SU(2)^5\times\USp(56)^4$.}
	\label{spec:r123f2}
\end{table}  

The model~\hyperref[model:r123f2]{r123f2} is a gauge theory of rank 123 with 2 units of flux. The construction involves four rigid D6-branes
$\{a,b,c,d\}$ sharing identical fixed points,
$\delta_{ab}^g \neq (4,4,4)$, and realizes the gauge symmetry
$\SU(4)_{C}\times\SU(2)_{L}\times\SU(2)_{R_1}\times\SU(2)_{R_2}\times\SU(2)^5\times\USp(56)^4.$

The matter content is summarized in table~\ref{spec:r123f2}, where fields are
organized by their quantum numbers under the gauge symmetry. The spectrum contains chiral fermions, scalar fields, and Higgs-like states arising from
brane intersections such as $ab$, $ac$, and $ad'$, yielding fundamental
representations of the corresponding gauge factors. The model~\hyperref[spec:r123f2]{r123f2} features the torus moduli
$\chi_1 = 10 \sqrt{14}$,
$\chi_2 = \sqrt{14}$,
$\chi_3 = 3 \sqrt{\frac{7}{2}}$,
and the tree-level gauge coupling relation
$g_a^2=\frac{95}{14}g_b^2=\frac{57}{308}g_{cd}^2=\frac{95}{346}\frac{5 g_Y^2}{3}=\frac{4\ 2^{3/4}}{\sqrt[4]{7} \sqrt{15}}\, \pi \,  e^{\phi _4}$.

The chiral sector consists of left-handed multiplets
$(Q_L,L_L)$ and right-handed multiplets
$(Q_R,L_R)$ arising from distinct D-brane intersections. The left-handed states accommodate the quark and lepton doublets and originate from the $ab$ (-3), $ab'$ (6) sectors. The right-handed states provide the corresponding singlet partners and receive contributions from the $ac$ (-4), $ad$ (1) sectors.

At the GUT scale, there are 2 scalar fields $\Delta^i$ responsible for breaking the Pati-Salam symmetry
to the Standard Model gauge group. In addition, there are 23 Higgs-like fields arising from the $bc$, $bc'$ sectors. The spectrum further contains several chiral exotic states
$X_L^i$ and $X_R^i$ charged under hidden sector gauge groups.

Table~\ref{exotic:r123f2} presents the composite spectrum obtained after confinement in the hidden
sector. States charged under non-abelian hidden gauge factors experience strong coupling dynamics, which leads to the formation of bound states neutral under the hidden gauge group. Accordingly, the exotic states do not contribute to the low-energy effective dynamics.

 
\subsection{Model~\texorpdfstring{\hyperref[spec:r125f2]{r125f2}}{r125f2}}

\begin{table}[t]
	\centering
	$\begin{array}{|c|D{.}{\times}{24}|c|}
		\hline
		\text{Model~\hyperref[exotic:r125f2]{r125f2}} & \mc{\text{Quantum Numbers}}                                & \text{Fields}     \\ 
		\hline
		ab                                            & 6. (\bar{4},2,1,1,1,1,1,1,1,1,1)                           & F_L^i(Q_L,L_L)    \\
		ab'                                           & 3. (4,2,1,1,1,1,1,1,1,1,1)                                 & F_L^i(Q_L,L_L)    \\
		ac                                            & 2. (4,1,\bar{2},1,1,1,1,1,1,1,1)                           & F_R^i(Q_R,L_R)    \\
		ac'                                           & 2. (4,1,2,1,1,1,1,1,1,1,1)                                 & F_R^i(Q_R,L_R)    \\
		ad'                                           & 1. (\bar{4},1,1,\bar{2},1,1,1,1,1,1,1)                     & F_R^{'i}(Q_R,L_R) \\
		b_{\yng(1,1)_{}}                              & 150. (1,1_{\yng(1,1)_{}},1,1,1,1,1,1,1,1,1)                & S_L^i             \\
		c_{\yng(1,1)_{}}                              & 12. (1,1,1_{\yng(1,1)_{}},1,1,1,1,1,1,1,1)                 & S_R^i             \\
		d_{\overline{\yng(1,1)_{}}}                   & 6. (1,1,1,\bar{1}_{\overline{\yng(1,1)_{}}},1,1,1,1,1,1,1) & S_R^i             \\
		b_{\yng(2)}                                   & 36. (1,3_{\yng(2)},1,1,1,1,1,1,1,1,1)                      & T_L^i             \\
		c_{\overline{\yng(2)}}                        & 6. (1,1,\bar{3}_{\overline{\yng(2)}},1,1,1,1,1,1,1,1)      & T_R^i             \\
		d_{\overline{\yng(2)}}                        & 4. (1,1,1,\bar{3}_{\overline{\yng(2)}},1,1,1,1,1,1,1)      & T_R^i             \\
		cd'                                           & 3. (1,1,\bar{2},\bar{2},1,1,1,1,1,1,1)                     & \Delta ^i         \\
		bc                                            & 5. (1,\bar{2},2,1,1,1,1,1,1,1,1)                           & \Phi ^i(H_u,H_d)  \\
		bc'                                           & 18. (1,2,2,1,1,1,1,1,1,1,1)                                & \Phi ^i(H_u,H_d)  \\
		bd                                            & 6. (1,2,1,\bar{2},1,1,1,1,1,1,1)                           & \Xi ^{i}          \\
		bd'                                           & 6. (1,2,1,2,1,1,1,1,1,1,1)                                 & \Xi ^{i}          \\
		be                                            & 4. (1,2,1,1,\bar{2},1,1,1,1,1,1)                           & X_L^{i}           \\
		be'                                           & 4. (1,2,1,1,2,1,1,1,1,1,1)                                 & X_L^{i}           \\
		bf_1                                          & 4. (1,2,1,1,1,\bar{4},1,1,1,1,1)                           & X_L^{i}           \\
		bf_1'                                         & 6. (1,2,1,1,1,4,1,1,1,1,1)                                 & X_L^{i}           \\
		bf_2                                          & 4. (1,2,1,1,1,1,\bar{4},1,1,1,1)                           & X_L^{i}           \\
		bf_2'                                         & 6. (1,2,1,1,1,1,4,1,1,1,1)                                 & X_L^{i}           \\
		bg_2                                          & 1. (1,\bar{2},1,1,1,1,1,1,56,1,1)                          & X_L^{i}           \\
		bg_2'                                         & 1. (1,\bar{2},1,1,1,1,1,1,\overline{56},1,1)               & X_L^{i}           \\
		ce'                                           & 1. (1,1,\bar{2},1,\bar{2},1,1,1,1,1,1)                     & X_R^{i}           \\
		cf_1                                          & 3. (1,1,2,1,1,\bar{4},1,1,1,1,1)                           & X_R^{i}           \\
		cf_1'                                         & 3. (1,1,2,1,1,4,1,1,1,1,1)                                 & X_R^{i}           \\
		cf_2                                          & 2. (1,1,2,1,1,1,\bar{4},1,1,1,1)                           & X_R^{i}           \\
		cf_2'                                         & 2. (1,1,2,1,1,1,4,1,1,1,1)                                 & X_R^{i}           \\
		df_1                                          & 1. (1,1,1,2,1,\bar{4},1,1,1,1,1)                           & X_R^{i}           \\
		df_1'                                         & 1. (1,1,1,2,1,4,1,1,1,1,1)                                 & X_R^{i}           \\
		\hline
	\end{array}$
	\caption{Particle spectrum of Model~\hyperref[model:r125f2]{r125f2} with gauge symmetry $\SU(4)_{C}\times\SU(2)_{L}\times\SU(2)_{R_1}\times\SU(2)_{R_2}\times\SU(2)\times\SU(4)^2\times\USp(56)^4$.}
	\label{spec:r125f2}
\end{table}  

The model~\hyperref[model:r125f2]{r125f2} realizes a gauge sector of rank 125 with 2 units of flux. The construction involves four rigid D6-branes
$\{a,b,c,d\}$ sharing identical fixed points,
$\delta_{ab}^g \neq (4,4,4)$, and realizes the gauge symmetry
$\SU(4)_{C}\times\SU(2)_{L}\times\SU(2)_{R_1}\times\SU(2)_{R_2}\times\SU(2)\times\SU(4)^2\times\USp(56)^4.$

The matter content is presented in table~\ref{spec:r125f2}, where fields are
organized by their quantum numbers under the gauge symmetry. The resulting spectrum exhibits chiral fermions, scalar fields, and Higgs-like states arising from
brane intersections such as $ab$, $ac$, and $ad'$, yielding fundamental
representations of the corresponding gauge factors. The model~\hyperref[spec:r125f2]{r125f2} features the torus moduli
$\chi_1 = \sqrt{29}$,
$\chi_2 = 10 \sqrt{29}$,
$\chi_3 = \frac{2 \sqrt{29}}{3}$,
and the tree-level gauge coupling relation
$g_a^2=\frac{195}{29}g_b^2=\frac{2925}{21808}g_{cd}^2=\frac{4875}{23758}\frac{5 g_Y^2}{3}=\frac{8}{\sqrt{15} \sqrt[4]{29}}\, \pi \,  e^{\phi _4}$.

The chiral matter content comprises left-handed multiplets
$(Q_L,L_L)$ and right-handed multiplets
$(Q_R,L_R)$ arising from distinct D-brane intersections. The left-handed multiplets naturally realize the quark and lepton doublets and originate from the $ab$ (-6), $ab'$ (3) sectors. The right-handed multiplets supply the associated singlet states and receive contributions from the $ac$ (2), $ac'$ (2), $ad'$ (-1) sectors.

At high energies, there are 3 scalar fields $\Delta^i$ responsible for breaking the Pati-Salam symmetry
to the Standard Model gauge group. At lower energies, there are 23 Higgs-like fields arising from the $bc$, $bc'$ sectors. The spectrum further contains several chiral exotic states
$X_L^i$ and $X_R^i$ charged under hidden sector gauge groups.

Table~\ref{exotic:r125f2} presents the composite spectrum obtained after confinement in the hidden
sector. Fields transforming under non-abelian hidden gauge groups are subject to strong coupling effects, which results in the emergence of bound states neutral under the hidden gauge group. Hence, the resulting bound states are removed from the low-energy spectrum.

 
\subsection{Model~\texorpdfstring{\hyperref[spec:r27f3]{r27f3}}{r27f3}}

\begin{table}[t]
	\centering
	$\begin{array}{|c|D{.}{\times}{28}|c|}
		\hline
		\text{Model~\hyperref[exotic:r27f3]{r27f3}} & \mc{\text{Quantum Numbers}}                                      & \text{Fields}     \\ 
		\hline
		ab                                          & 3. (\bar{4},2,1,1,1,1,1,1,1,1,1,1,1)                             & F_L^i(Q_L,L_L)    \\
		ac                                          & 6. (4,1,\bar{2},1,1,1,1,1,1,1,1,1,1)                             & F_R^i(Q_R,L_R)    \\
		ac'                                         & 1. (\bar{4},1,\bar{2},1,1,1,1,1,1,1,1,1,1)                       & F_R^i(Q_R,L_R)    \\
		ad                                          & 2. (4,1,1,\bar{2},1,1,1,1,1,1,1,1,1)                             & F_R^{'i}(Q_R,L_R) \\
		ad'                                         & 4. (\bar{4},1,1,\bar{2},1,1,1,1,1,1,1,1,1)                       & F_R^{'i}(Q_R,L_R) \\
		c_{\overline{\yng(1,1)_{}}}                 & 100. (1,1,\bar{1}_{\overline{\yng(1,1)_{}}},1,1,1,1,1,1,1,1,1,1) & S_R^i             \\
		b_{\yng(2)}                                 & 16. (1,3_{\yng(2)},1,1,1,1,1,1,1,1,1,1,1)                        & T_L^i             \\
		c_{\overline{\yng(2)}}                      & 18. (1,1,\bar{3}_{\overline{\yng(2)}},1,1,1,1,1,1,1,1,1,1)       & T_R^i             \\
		d_{\yng(2)}                                 & 8. (1,1,1,3_{\yng(2)},1,1,1,1,1,1,1,1,1)                         & T_R^i             \\
		cd'                                         & 12. (1,1,\bar{2},\bar{2},1,1,1,1,1,1,1,1,1)                      & \Delta ^i         \\
		bc                                          & 9. (1,\bar{2},2,1,1,1,1,1,1,1,1,1,1)                             & \Phi ^i(H_u,H_d)  \\
		bc'                                         & 15. (1,\bar{2},\bar{2},1,1,1,1,1,1,1,1,1,1)                      & \Phi ^i(H_u,H_d)  \\
		bd                                          & 12. (1,\bar{2},1,2,1,1,1,1,1,1,1,1,1)                            & \Xi ^{i}          \\
		bd'                                         & 4. (1,\bar{2},1,\bar{2},1,1,1,1,1,1,1,1,1)                       & \Xi ^{i}          \\
		be                                          & 2. (1,\bar{2},1,1,2,1,1,1,1,1,1,1,1)                             & X_L^{i}           \\
		be'                                         & 4. (1,\bar{2},1,1,\bar{2},1,1,1,1,1,1,1,1)                       & X_L^{i}           \\
		bf_1                                        & 2. (1,2,1,1,1,\bar{2},1,1,1,1,1,1,1)                             & X_L^{i}           \\
		bf_2'                                       & 2. (1,2,1,1,1,1,2,1,1,1,1,1,1)                                   & X_L^{i}           \\
		ce                                          & 4. (1,1,\bar{2},1,2,1,1,1,1,1,1,1,1)                             & X_R^{i}           \\
		ce'                                         & 2. (1,1,\bar{2},1,\bar{2},1,1,1,1,1,1,1,1)                       & X_R^{i}           \\
		cf_1                                        & 4. (1,1,\bar{2},1,1,2,1,1,1,1,1,1,1)                             & X_R^{i}           \\
		cf_1'                                       & 2. (1,1,\bar{2},1,1,\bar{2},1,1,1,1,1,1,1)                       & X_R^{i}           \\
		cf_2                                        & 2. (1,1,\bar{2},1,1,1,2,1,1,1,1,1,1)                             & X_R^{i}           \\
		cf_2'                                       & 4. (1,1,\bar{2},1,1,1,\bar{2},1,1,1,1,1,1)                       & X_R^{i}           \\
		cg_1                                        & 4. (1,1,\bar{2},1,1,1,1,2,1,1,1,1,1)                             & X_R^{i}           \\
		cg_1'                                       & 4. (1,1,\bar{2},1,1,1,1,\bar{2},1,1,1,1,1)                       & X_R^{i}           \\
		cg_2                                        & 4. (1,1,\bar{2},1,1,1,1,1,2,1,1,1,1)                             & X_R^{i}           \\
		cg_2'                                       & 4. (1,1,\bar{2},1,1,1,1,1,\bar{2},1,1,1,1)                       & X_R^{i}           \\
		ch_2                                        & 1. (1,1,2,1,1,1,1,1,1,1,\bar{8},1,1)                             & X_R^{i}           \\
		ch_2'                                       & 1. (1,1,2,1,1,1,1,1,1,1,8,1,1)                                   & X_R^{i}           \\
		df_1                                        & 1. (1,1,1,\bar{2},1,2,1,1,1,1,1,1,1)                             & X_R^{i}           \\
		df_1'                                       & 2. (1,1,1,\bar{2},1,\bar{2},1,1,1,1,1,1,1)                       & X_R^{i}           \\
		df_2                                        & 1. (1,1,1,\bar{2},1,1,2,1,1,1,1,1,1)                             & X_R^{i}           \\
		df_2'                                       & 2. (1,1,1,\bar{2},1,1,\bar{2},1,1,1,1,1,1)                       & X_R^{i}           \\
		dg_1'                                       & 1. (1,1,1,2,1,1,1,2,1,1,1,1,1)                                   & X_R^{i}           \\
		dg_2                                        & 1. (1,1,1,2,1,1,1,1,\bar{2},1,1,1,1)                             & X_R^{i}           \\
		\hline
	\end{array}$
	\caption{Particle spectrum of Model~\hyperref[model:r27f3]{r27f3} with gauge symmetry $\SU(4)_{C}\times\SU(2)_{L}\times\SU(2)_{R_1}\times\SU(2)_{R_2}\times\SU(2)^5\times\USp(8)^4$.}
	\label{spec:r27f3}
\end{table}  

The model~\hyperref[model:r27f3]{r27f3} defines a configuration of rank 27 with 3 units of flux. The construction involves four rigid D6-branes
$\{a,b,c,d\}$ sharing identical fixed points,
$\delta_{ab}^g \neq (4,4,4)$, and realizes the gauge symmetry
$\SU(4)_{C}\times\SU(2)_{L}\times\SU(2)_{R_1}\times\SU(2)_{R_2}\times\SU(2)^5\times\USp(8)^4.$

The matter content is collected in table~\ref{spec:r27f3}, where fields are
organized by their quantum numbers under the gauge symmetry. One finds in the spectrum chiral fermions, scalar fields, and Higgs-like states arising from
brane intersections such as $ab$, $ac$, and $ad'$, yielding fundamental
representations of the corresponding gauge factors. The model~\hyperref[spec:r27f3]{r27f3} features the torus moduli
$\chi_1 = \frac{8}{3}$,
$\chi_2 = 8$,
$\chi_3 = 24$,
and the tree-level gauge coupling relation
$g_a^2=\frac{5}{4}g_b^2=\frac{1885}{324}g_{cd}^2=\frac{9425}{4742}\frac{5 g_Y^2}{3}=2 \sqrt{2}\, \pi \,  e^{\phi _4}$.

The resulting chiral spectrum contains left-handed multiplets
$(Q_L,L_L)$ and right-handed multiplets
$(Q_R,L_R)$ arising from distinct D-brane intersections. The left-handed sector encodes the quark and lepton doublets and originate from the $ab$ (-3) sector. The right-handed sector contains the singlet partners and receive contributions from the $ac$ (6), $ac'$ (-1), $ad$ (2), $ad'$ (-4) sectors.

The high-scale spectrum contains are 12 scalar fields $\Delta^i$ responsible for breaking the Pati-Salam symmetry
to the Standard Model gauge group. Furthermore, the spectrum contains are 24 Higgs-like fields arising from the $bc$, $bc'$ sectors. The spectrum further contains several chiral exotic states
$X_L^i$ and $X_R^i$ charged under hidden sector gauge groups.

Table~\ref{exotic:r27f3} presents the composite spectrum obtained after confinement in the hidden
sector. Matter charged under hidden non-abelian sectors undergoes confinement dynamics, which induces bound states neutral under the hidden gauge group. Consequently, these states decouple from the low-energy effective spectrum.

\clearpage 
\subsection{Model~\texorpdfstring{\hyperref[spec:r35f3]{r35f3}}{r35f3}}

\begin{table}[t]
	\centering
	$\begin{array}{|c|D{.}{\times}{28}|c|}
		\hline
		\text{Model~\hyperref[exotic:r35f3]{r35f3}} & \mc{\text{Quantum Numbers}}                                    & \text{Fields}     \\ 
		\hline
		ab                                          & 6. (\bar{4},2,1,1,1,1,1,1,1,1,1,1,1)                           & F_L^i(Q_L,L_L)    \\
		ab'                                         & 3. (4,2,1,1,1,1,1,1,1,1,1,1,1)                                 & F_L^i(Q_L,L_L)    \\
		ac                                          & 1. (\bar{4},1,2,1,1,1,1,1,1,1,1,1,1)                           & F_R^i(Q_R,L_R)    \\
		ac'                                         & 6. (4,1,2,1,1,1,1,1,1,1,1,1,1)                                 & F_R^i(Q_R,L_R)    \\
		ad                                          & 4. (\bar{4},1,1,2,1,1,1,1,1,1,1,1,1)                           & F_R^{'i}(Q_R,L_R) \\
		ad'                                         & 2. (4,1,1,2,1,1,1,1,1,1,1,1,1)                                 & F_R^{'i}(Q_R,L_R) \\
		b_{\yng(1,1)_{}}                            & 8. (1,1_{\yng(1,1)_{}},1,1,1,1,1,1,1,1,1,1,1)                  & S_L^i             \\
		c_{\yng(1,1)_{}}                            & 100. (1,1,1_{\yng(1,1)_{}},1,1,1,1,1,1,1,1,1,1)                & S_R^i             \\
		d_{\overline{\yng(1,1)_{}}}                 & 8. (1,1,1,\bar{1}_{\overline{\yng(1,1)_{}}},1,1,1,1,1,1,1,1,1) & S_R^i             \\
		b_{\overline{\yng(2)}}                      & 12. (1,\bar{3}_{\overline{\yng(2)}},1,1,1,1,1,1,1,1,1,1,1)     & T_L^i             \\
		c_{\yng(2)}                                 & 18. (1,1,3_{\yng(2)},1,1,1,1,1,1,1,1,1,1)                      & T_R^i             \\
		d_{\overline{\yng(2)}}                      & 12. (1,1,1,\bar{3}_{\overline{\yng(2)}},1,1,1,1,1,1,1,1,1)     & T_R^i             \\
		cd'                                         & 9. (1,1,2,2,1,1,1,1,1,1,1,1,1)                                 & \Delta ^i         \\
		bc                                          & 8. (1,2,\bar{2},1,1,1,1,1,1,1,1,1,1)                           & \Phi ^i(H_u,H_d)  \\
		bc'                                         & 20. (1,2,2,1,1,1,1,1,1,1,1,1,1)                                & \Phi ^i(H_u,H_d)  \\
		bd'                                         & 8. (1,\bar{2},1,\bar{2},1,1,1,1,1,1,1,1,1)                     & \Xi ^{i}          \\
		be                                          & 1. (1,\bar{2},1,1,2,1,1,1,1,1,1,1,1)                           & X_L^{i}           \\
		bg_1                                        & 4. (1,2,1,1,1,1,1,\bar{2},1,1,1,1,1)                           & X_L^{i}           \\
		bg_1'                                       & 2. (1,2,1,1,1,1,1,2,1,1,1,1,1)                                 & X_L^{i}           \\
		bg_2                                        & 4. (1,2,1,1,1,1,1,1,\bar{2},1,1,1,1)                           & X_L^{i}           \\
		bg_2'                                       & 2. (1,2,1,1,1,1,1,1,2,1,1,1,1)                                 & X_L^{i}           \\
		ce                                          & 4. (1,1,2,1,\bar{2},1,1,1,1,1,1,1,1)                           & X_R^{i}           \\
		ce'                                         & 2. (1,1,2,1,2,1,1,1,1,1,1,1,1)                                 & X_R^{i}           \\
		cf_1                                        & 4. (1,1,2,1,1,\bar{2},1,1,1,1,1,1,1)                           & X_R^{i}           \\
		cf_1'                                       & 4. (1,1,2,1,1,2,1,1,1,1,1,1,1)                                 & X_R^{i}           \\
		cf_2                                        & 4. (1,1,2,1,1,1,\bar{2},1,1,1,1,1,1)                           & X_R^{i}           \\
		cf_2'                                       & 4. (1,1,2,1,1,1,2,1,1,1,1,1,1)                                 & X_R^{i}           \\
		cg_1                                        & 2. (1,1,2,1,1,1,1,\bar{2},1,1,1,1,1)                           & X_R^{i}           \\
		cg_1'                                       & 4. (1,1,2,1,1,1,1,2,1,1,1,1,1)                                 & X_R^{i}           \\
		cg_2                                        & 2. (1,1,2,1,1,1,1,1,\bar{2},1,1,1,1)                           & X_R^{i}           \\
		cg_2'                                       & 4. (1,1,2,1,1,1,1,1,2,1,1,1,1)                                 & X_R^{i}           \\
		ch_4                                        & 1. (1,1,\bar{2},1,1,1,1,1,1,1,1,1,12)                          & X_R^{i}           \\
		ch_4'                                       & 1. (1,1,\bar{2},1,1,1,1,1,1,1,1,1,\overline{12})               & X_R^{i}           \\
		df_1                                        & 2. (1,1,1,\bar{2},1,2,1,1,1,1,1,1,1)                           & X_R^{i}           \\
		df_2                                        & 2. (1,1,1,\bar{2},1,1,2,1,1,1,1,1,1)                           & X_R^{i}           \\
		dg_1                                        & 2. (1,1,1,2,1,1,1,\bar{2},1,1,1,1,1)                           & X_R^{i}           \\
		dg_1'                                       & 1. (1,1,1,2,1,1,1,2,1,1,1,1,1)                                 & X_R^{i}           \\
		dg_2                                        & 1. (1,1,1,2,1,1,1,1,\bar{2},1,1,1,1)                           & X_R^{i}           \\
		dg_2'                                       & 2. (1,1,1,2,1,1,1,1,2,1,1,1,1)                                 & X_R^{i}           \\
		\hline
	\end{array}$
	\caption{Particle spectrum of Model~\hyperref[model:r35f3]{r35f3} with gauge symmetry $\SU(4)_{C}\times\SU(2)_{L}\times\SU(2)_{R_1}\times\SU(2)_{R_2}\times\SU(2)^5\times\USp(12)^4$.}
	\label{spec:r35f3}
\end{table}  

The model~\hyperref[model:r35f3]{r35f3} corresponds to a gauge theory of rank 35 with 3 units of flux. The construction involves four rigid D6-branes
$\{a,b,c,d\}$ sharing identical fixed points,
$\delta_{ab}^g \neq (4,4,4)$, and realizes the gauge symmetry
$\SU(4)_{C}\times\SU(2)_{L}\times\SU(2)_{R_1}\times\SU(2)_{R_2}\times\SU(2)^5\times\USp(12)^4.$

The matter content is displayed in table~\ref{spec:r35f3}, where fields are
organized by their quantum numbers under the gauge symmetry. The model gives rise to chiral fermions, scalar fields, and Higgs-like states arising from
brane intersections such as $ab$, $ac$, and $ad'$, yielding fundamental
representations of the corresponding gauge factors. The model~\hyperref[spec:r35f3]{r35f3} features the torus moduli
$\chi_1 = 12 \sqrt{3}$,
$\chi_2 = 3 \sqrt{3}$,
$\chi_3 = 2 \sqrt{3}$,
and the tree-level gauge coupling relation
$g_a^2=\frac{56}{27}g_b^2=\frac{1736}{3303}g_{cd}^2=\frac{8680}{13381}\frac{5 g_Y^2}{3}=\frac{4 \sqrt{2}}{3 \sqrt[4]{3}}\, \pi \,  e^{\phi _4}$.

One finds in the chiral sector left-handed multiplets
$(Q_L,L_L)$ and right-handed multiplets
$(Q_R,L_R)$ arising from distinct D-brane intersections. The left-handed fields furnish the Standard Model quark and lepton doublets and originate from the $ab$ (-6), $ab'$ (3) sectors. The right-handed fields account for the corresponding singlet representations and receive contributions from the $ac$ (-1), $ac'$ (6), $ad$ (-4), $ad'$ (2) sectors.

The GUT sector includes are 9 scalar fields $\Delta^i$ responsible for breaking the Pati-Salam symmetry
to the Standard Model gauge group. The electroweak sector includes are 28 Higgs-like fields arising from the $bc$, $bc'$ sectors. The spectrum further contains several chiral exotic states
$X_L^i$ and $X_R^i$ charged under hidden sector gauge groups.

Table~\ref{exotic:r35f3} presents the composite spectrum obtained after confinement in the hidden
sector. Hidden-sector charged states are governed by strong coupling dynamics, which drives bound states neutral under the hidden gauge group. As a result, these degrees of freedom are absent from the low-energy effective theory.

 
\subsection{Model~\texorpdfstring{\hyperref[spec:r75f3]{r75f3}}{r75f3}}

\begin{table}[t]
	\centering
	$\begin{array}{|c|D{.}{\times}{28}|c|}
		\hline
		\text{Model~\hyperref[exotic:r75f3]{r75f3}} & \mc{\text{Quantum Numbers}}                                      & \text{Fields}     \\ 
		\hline
		ab                                          & 3. (\bar{4},2,1,1,1,1,1,1,1,1,1,1,1)                             & F_L^i(Q_L,L_L)    \\
		ac                                          & 9. (4,1,\bar{2},1,1,1,1,1,1,1,1,1,1)                             & F_R^i(Q_R,L_R)    \\
		ac'                                         & 2. (\bar{4},1,\bar{2},1,1,1,1,1,1,1,1,1,1)                       & F_R^i(Q_R,L_R)    \\
		ad                                          & 4. (\bar{4},1,1,2,1,1,1,1,1,1,1,1,1)                             & F_R^{'i}(Q_R,L_R) \\
		c_{\overline{\yng(1,1)_{}}}                 & 150. (1,1,\bar{1}_{\overline{\yng(1,1)_{}}},1,1,1,1,1,1,1,1,1,1) & S_R^i             \\
		d_{\yng(1,1)_{}}                            & 6. (1,1,1,1_{\yng(1,1)_{}},1,1,1,1,1,1,1,1,1)                    & S_R^i             \\
		b_{\yng(2)}                                 & 10. (1,3_{\yng(2)},1,1,1,1,1,1,1,1,1,1,1)                        & T_L^i             \\
		c_{\overline{\yng(2)}}                      & 36. (1,1,\bar{3}_{\overline{\yng(2)}},1,1,1,1,1,1,1,1,1,1)       & T_R^i             \\
		d_{\overline{\yng(2)}}                      & 4. (1,1,1,\bar{3}_{\overline{\yng(2)}},1,1,1,1,1,1,1,1,1)        & T_R^i             \\
		cd                                          & 12. (1,1,\bar{2},2,1,1,1,1,1,1,1,1,1)                            & \Delta ^i         \\
		bc                                          & 8. (1,\bar{2},2,1,1,1,1,1,1,1,1,1,1)                             & \Phi ^i(H_u,H_d)  \\
		bc'                                         & 10. (1,\bar{2},\bar{2},1,1,1,1,1,1,1,1,1,1)                      & \Phi ^i(H_u,H_d)  \\
		bd'                                         & 10. (1,\bar{2},1,\bar{2},1,1,1,1,1,1,1,1,1)                      & \Xi ^{i}          \\
		be                                          & 3. (1,\bar{2},1,1,2,1,1,1,1,1,1,1,1)                             & X_L^{i}           \\
		be'                                         & 3. (1,\bar{2},1,1,\bar{2},1,1,1,1,1,1,1,1)                       & X_L^{i}           \\
		bf_1                                        & 3. (1,2,1,1,1,\bar{2},1,1,1,1,1,1,1)                             & X_L^{i}           \\
		bf_2'                                       & 2. (1,2,1,1,1,1,2,1,1,1,1,1,1)                                   & X_L^{i}           \\
		ce                                          & 4. (1,1,\bar{2},1,2,1,1,1,1,1,1,1,1)                             & X_R^{i}           \\
		ce'                                         & 4. (1,1,\bar{2},1,\bar{2},1,1,1,1,1,1,1,1)                       & X_R^{i}           \\
		cf_1                                        & 6. (1,1,\bar{2},1,1,2,1,1,1,1,1,1,1)                             & X_R^{i}           \\
		cf_1'                                       & 4. (1,1,\bar{2},1,1,\bar{2},1,1,1,1,1,1,1)                       & X_R^{i}           \\
		cf_2                                        & 6. (1,1,\bar{2},1,1,1,2,1,1,1,1,1,1)                             & X_R^{i}           \\
		cf_2'                                       & 4. (1,1,\bar{2},1,1,1,\bar{2},1,1,1,1,1,1)                       & X_R^{i}           \\
		cg_1                                        & 6. (1,1,\bar{2},1,1,1,1,2,1,1,1,1,1)                             & X_R^{i}           \\
		cg_1'                                       & 4. (1,1,\bar{2},1,1,1,1,\bar{2},1,1,1,1,1)                       & X_R^{i}           \\
		cg_2                                        & 6. (1,1,\bar{2},1,1,1,1,1,2,1,1,1,1)                             & X_R^{i}           \\
		cg_2'                                       & 4. (1,1,\bar{2},1,1,1,1,1,\bar{2},1,1,1,1)                       & X_R^{i}           \\
		ch_2                                        & 1. (1,1,2,1,1,1,1,1,1,1,\overline{32},1,1)                       & X_R^{i}           \\
		ch_2'                                       & 1. (1,1,2,1,1,1,1,1,1,1,32,1,1)                                  & X_R^{i}           \\
		df_1                                        & 2. (1,1,1,2,1,\bar{2},1,1,1,1,1,1,1)                             & X_R^{i}           \\
		df_1'                                       & 2. (1,1,1,2,1,2,1,1,1,1,1,1,1)                                   & X_R^{i}           \\
		df_2                                        & 1. (1,1,1,2,1,1,\bar{2},1,1,1,1,1,1)                             & X_R^{i}           \\
		df_2'                                       & 1. (1,1,1,2,1,1,2,1,1,1,1,1,1)                                   & X_R^{i}           \\
		dg_2'                                       & 1. (1,1,1,\bar{2},1,1,1,1,\bar{2},1,1,1,1)                       & X_R^{i}           \\
		\hline
	\end{array}$
	\caption{Particle spectrum of Model~\hyperref[model:r75f3]{r75f3} with gauge symmetry $\SU(4)_{C}\times\SU(2)_{L}\times\SU(2)_{R_1}\times\SU(2)_{R_2}\times\SU(2)^5\times\USp(32)^4$.}
	\label{spec:r75f3}
\end{table}  

The model~\hyperref[model:r75f3]{r75f3} describes a gauge-theoretic construction of rank 75 with 3 units of flux. The construction involves four rigid D6-branes
$\{a,b,c,d\}$ sharing identical fixed points,
$\delta_{ab}^g \neq (4,4,4)$, and realizes the gauge symmetry
$\SU(4)_{C}\times\SU(2)_{L}\times\SU(2)_{R_1}\times\SU(2)_{R_2}\times\SU(2)^5\times\USp(32)^4.$

The matter content is tabulated in table~\ref{spec:r75f3}, where fields are
organized by their quantum numbers under the gauge symmetry. The spectrum contains chiral fermions, scalar fields, and Higgs-like states arising from
brane intersections such as $ab$, $ac$, and $ad'$, yielding fundamental
representations of the corresponding gauge factors. The model~\hyperref[spec:r75f3]{r75f3} features the torus moduli
$\chi_1 = 2 \sqrt{\frac{37}{3}}$,
$\chi_2 = \sqrt{\frac{37}{3}}$,
$\chi_3 = 4 \sqrt{111}$,
and the tree-level gauge coupling relation
$g_a^2=\frac{223}{444}g_b^2=\frac{133280}{42587}g_{cd}^2=\frac{666400}{394321}\frac{5 g_Y^2}{3}=\frac{4 \sqrt{2}}{\sqrt[4]{111}}\, \pi \,  e^{\phi _4}$.

The chiral sector consists of left-handed multiplets
$(Q_L,L_L)$ and right-handed multiplets
$(Q_R,L_R)$ arising from distinct D-brane intersections. The left-handed states accommodate the quark and lepton doublets and originate from the $ab$ (-3) sector. The right-handed states provide the corresponding singlet partners and receive contributions from the $ac$ (9), $ac'$ (-2), $ad$ (-4) sectors.

At the GUT scale, there are 12 scalar fields $\Delta^i$ responsible for breaking the Pati-Salam symmetry
to the Standard Model gauge group. In addition, there are 18 Higgs-like fields arising from the $bc$, $bc'$ sectors. The spectrum further contains several chiral exotic states
$X_L^i$ and $X_R^i$ charged under hidden sector gauge groups.

Table~\ref{exotic:r75f3} presents the composite spectrum obtained after confinement in the hidden
sector. States charged under non-abelian hidden gauge factors experience strong coupling dynamics, which gives rise to bound states neutral under the hidden gauge group. These bound states therefore decouple from the low-energy spectrum.

 
\subsection{Model~\texorpdfstring{\hyperref[spec:r76f3]{r76f3}}{r76f3}}

\begin{table}[t]
	\centering
	$\begin{array}{|c|D{.}{\times}{26}|c|}
		\hline
		\text{Model~\hyperref[exotic:r76f3]{r76f3}} & \mc{\text{Quantum Numbers}}                                  & \text{Fields}     \\ 
		\hline
		ab'                                         & 3. (4,2,1,1,1,1,1,1,1,1,1,1)                                 & F_L^i(Q_L,L_L)    \\
		ac                                          & 9. (\bar{4},1,2,1,1,1,1,1,1,1,1,1)                           & F_R^i(Q_R,L_R)    \\
		ac'                                         & 2. (4,1,2,1,1,1,1,1,1,1,1,1)                                 & F_R^i(Q_R,L_R)    \\
		ad                                          & 2. (4,1,1,\bar{2},1,1,1,1,1,1,1,1)                           & F_R^{'i}(Q_R,L_R) \\
		ad'                                         & 2. (4,1,1,2,1,1,1,1,1,1,1,1)                                 & F_R^{'i}(Q_R,L_R) \\
		c_{\yng(1,1)_{}}                            & 150. (1,1,1_{\yng(1,1)_{}},1,1,1,1,1,1,1,1,1)                & S_R^i             \\
		d_{\overline{\yng(1,1)_{}}}                 & 6. (1,1,1,\bar{1}_{\overline{\yng(1,1)_{}}},1,1,1,1,1,1,1,1) & S_R^i             \\
		b_{\yng(2)}                                 & 10. (1,3_{\yng(2)},1,1,1,1,1,1,1,1,1,1)                      & T_L^i             \\
		c_{\yng(2)}                                 & 36. (1,1,3_{\yng(2)},1,1,1,1,1,1,1,1,1)                      & T_R^i             \\
		d_{\overline{\yng(2)}}                      & 4. (1,1,1,\bar{3}_{\overline{\yng(2)}},1,1,1,1,1,1,1,1)      & T_R^i             \\
		cd                                          & 6. (1,1,2,\bar{2},1,1,1,1,1,1,1,1)                           & \Delta ^i         \\
		cd'                                         & 8. (1,1,2,2,1,1,1,1,1,1,1,1)                                 & \Delta ^i         \\
		bc                                          & 10. (1,\bar{2},2,1,1,1,1,1,1,1,1,1)                          & \Phi ^i(H_u,H_d)  \\
		bc'                                         & 8. (1,\bar{2},\bar{2},1,1,1,1,1,1,1,1,1)                     & \Phi ^i(H_u,H_d)  \\
		bd'                                         & 6. (1,2,1,2,1,1,1,1,1,1,1,1)                                 & \Xi ^{i}          \\
		bf_1                                        & 2. (1,\bar{2},1,1,1,1,2,1,1,1,1,1)                           & X_L^{i}           \\
		bf_1'                                       & 2. (1,\bar{2},1,1,1,1,\bar{2},1,1,1,1,1)                     & X_L^{i}           \\
		bf_2                                        & 3. (1,\bar{2},1,1,1,1,1,4,1,1,1,1)                           & X_L^{i}           \\
		bf_2'                                       & 3. (1,\bar{2},1,1,1,1,1,\bar{4},1,1,1,1)                     & X_L^{i}           \\
		ce_1                                        & 4. (1,1,2,1,\bar{2},1,1,1,1,1,1,1)                           & X_R^{i}           \\
		ce_1'                                       & 6. (1,1,2,1,2,1,1,1,1,1,1,1)                                 & X_R^{i}           \\
		ce_2                                        & 6. (1,1,2,1,1,\bar{2},1,1,1,1,1,1)                           & X_R^{i}           \\
		ce_2'                                       & 4. (1,1,2,1,1,2,1,1,1,1,1,1)                                 & X_R^{i}           \\
		cf_1                                        & 4. (1,1,2,1,1,1,\bar{2},1,1,1,1,1)                           & X_R^{i}           \\
		cf_1'                                       & 4. (1,1,2,1,1,1,2,1,1,1,1,1)                                 & X_R^{i}           \\
		cf_2                                        & 4. (1,1,2,1,1,1,1,\bar{4},1,1,1,1)                           & X_R^{i}           \\
		cf_2'                                       & 4. (1,1,2,1,1,1,1,4,1,1,1,1)                                 & X_R^{i}           \\
		cg_3                                        & 1. (1,1,\bar{2},1,1,1,1,1,1,1,32,1)                          & X_R^{i}           \\
		cg_3'                                       & 1. (1,1,\bar{2},1,1,1,1,1,1,1,\overline{32},1)               & X_R^{i}           \\
		de_2                                        & 1. (1,1,1,2,1,\bar{2},1,1,1,1,1,1)                           & X_R^{i}           \\
		de_2'                                       & 1. (1,1,1,2,1,2,1,1,1,1,1,1)                                 & X_R^{i}           \\
		df_1                                        & 1. (1,1,1,\bar{2},1,1,2,1,1,1,1,1)                           & X_R^{i}           \\
		df_2                                        & 2. (1,1,1,\bar{2},1,1,1,4,1,1,1,1)                           & X_R^{i}           \\
		\hline
	\end{array}$
	\caption{Particle spectrum of Model~\hyperref[model:r76f3]{r76f3} with gauge symmetry $\SU(4)_{C}\times\SU(2)_{L}\times\SU(2)_{R_1}\times\SU(2)_{R_2}\times\SU(2)^3\times\SU(4)\times\USp(32)^4$.}
	\label{spec:r76f3}
\end{table}  

The model~\hyperref[model:r76f3]{r76f3} constitutes a consistent gauge framework of rank 76 with 3 units of flux. The construction involves four rigid D6-branes
$\{a,b,c,d\}$ sharing identical fixed points,
$\delta_{ab}^g \neq (4,4,4)$, and realizes the gauge symmetry
$\SU(4)_{C}\times\SU(2)_{L}\times\SU(2)_{R_1}\times\SU(2)_{R_2}\times\SU(2)^3\times\SU(4)\times\USp(32)^4.$

The matter content is organized in table~\ref{spec:r76f3}, where fields are
organized by their quantum numbers under the gauge symmetry. The resulting spectrum exhibits chiral fermions, scalar fields, and Higgs-like states arising from
brane intersections such as $ab$, $ac$, and $ad'$, yielding fundamental
representations of the corresponding gauge factors. The model~\hyperref[spec:r76f3]{r76f3} features the torus moduli
$\chi_1 = \sqrt{\frac{129}{2}}$,
$\chi_2 = 2 \sqrt{\frac{86}{3}}$,
$\chi_3 = \sqrt{\frac{86}{3}}$,
and the tree-level gauge coupling relation
$g_a^2=\frac{419}{129}g_b^2=\frac{200606}{108403}g_{cd}^2=\frac{1003030}{726421}\frac{5 g_Y^2}{3}=\frac{8\ 2^{3/4}}{\sqrt[4]{129}}\, \pi \,  e^{\phi _4}$.

The chiral matter content comprises left-handed multiplets
$(Q_L,L_L)$ and right-handed multiplets
$(Q_R,L_R)$ arising from distinct D-brane intersections. The left-handed multiplets naturally realize the quark and lepton doublets and originate from the $ab'$ (3) sector. The right-handed multiplets supply the associated singlet states and receive contributions from the $ac$ (-9), $ac'$ (2), $ad$ (2), $ad'$ (2) sectors.

At high energies, there are 14 scalar fields $\Delta^i$ responsible for breaking the Pati-Salam symmetry
to the Standard Model gauge group. At lower energies, there are 18 Higgs-like fields arising from the $bc$, $bc'$ sectors. The spectrum further contains several chiral exotic states
$X_L^i$ and $X_R^i$ charged under hidden sector gauge groups.

Table~\ref{exotic:r76f3} presents the composite spectrum obtained after confinement in the hidden
sector. Fields transforming under non-abelian hidden gauge groups are subject to strong coupling effects, which triggers the formation of bound states neutral under the hidden gauge group. Accordingly, the exotic states do not contribute to the low-energy effective dynamics.

 
\subsection{Model~\texorpdfstring{\hyperref[spec:r20f4]{r20f4}}{r20f4}}

\begin{table}[t]
	\centering
	$\begin{array}{|c|D{.}{\times}{24}|c|}
		\hline
		\text{Model~\hyperref[exotic:r20f4]{r20f4}} & \mc{\text{Quantum Numbers}}            & \text{Fields}     \\ 
		\hline
		ab'                                         & 3. (4,2,1,1,1,1,1,1,1)                 & F_L^i(Q_L,L_L)    \\
		ac                                          & 2. (4,1,\bar{2},1,1,1,1,1,1)           & F_R^i(Q_R,L_R)    \\
		ad                                          & 8. (\bar{4},1,1,2,1,1,1,1,1)           & F_R^{'i}(Q_R,L_R) \\
		ad'                                         & 3. (4,1,1,2,1,1,1,1,1)                 & F_R^{'i}(Q_R,L_R) \\
		b_{\yng(1,1)_{}}                            & 8. (1,1_{\yng(1,1)_{}},1,1,1,1,1,1,1)  & S_L^i             \\
		c_{\yng(1,1)_{}}                            & 2. (1,1,1_{\yng(1,1)_{}},1,1,1,1,1,1)  & S_R^i             \\
		d_{\yng(1,1)_{}}                            & 96. (1,1,1,1_{\yng(1,1)_{}},1,1,1,1,1) & S_R^i             \\
		d_{\yng(2)}                                 & 12. (1,1,1,3_{\yng(2)},1,1,1,1,1)      & T_R^i             \\
		cd                                          & 8. (1,1,2,\bar{2},1,1,1,1,1)           & \Delta ^i         \\
		cd'                                         & 4. (1,1,\bar{2},\bar{2},1,1,1,1,1)     & \Delta ^i         \\
		bc                                          & 2. (1,\bar{2},2,1,1,1,1,1,1)           & \Phi ^i(H_u,H_d)  \\
		bf_1                                        & 3. (1,2,1,1,1,1,\bar{4},1,1)           & X_L^{i}           \\
		bf_2'                                       & 2. (1,2,1,1,1,1,1,4,1)                 & X_L^{i}           \\
		ce_1'                                       & 2. (1,1,2,1,4,1,1,1,1)                 & X_R^{i}           \\
		ce_2'                                       & 2. (1,1,2,1,1,4,1,1,1)                 & X_R^{i}           \\
		cf_1'                                       & 2. (1,1,\bar{2},1,1,1,\bar{4},1,1)     & X_R^{i}           \\
		cf_2'                                       & 1. (1,1,\bar{2},1,1,1,1,\bar{4},1)     & X_R^{i}           \\
		de_1                                        & 4. (1,1,1,2,\bar{4},1,1,1,1)           & X_R^{i}           \\
		de_1'                                       & 3. (1,1,1,2,4,1,1,1,1)                 & X_R^{i}           \\
		de_2                                        & 4. (1,1,1,2,1,\bar{4},1,1,1)           & X_R^{i}           \\
		de_2'                                       & 3. (1,1,1,2,1,4,1,1,1)                 & X_R^{i}           \\
		df_1                                        & 3. (1,1,1,2,1,1,\bar{4},1,1)           & X_R^{i}           \\
		df_1'                                       & 8. (1,1,1,2,1,1,4,1,1)                 & X_R^{i}           \\
		df_2                                        & 6. (1,1,1,2,1,1,1,\bar{4},1)           & X_R^{i}           \\
		df_2'                                       & 4. (1,1,1,2,1,1,1,4,1)                 & X_R^{i}           \\
		\hline
	\end{array}$
	\caption{Particle spectrum of Model~\hyperref[model:r20f4]{r20f4} with gauge symmetry $\SU(4)_{C}\times\SU(2)_{L}\times\SU(2)_{R_1}\times\SU(2)_{R_2}\times\SU(4)^3\times\SU(4)\times\USp(4)$.}
	\label{spec:r20f4}
\end{table}  

The model~\hyperref[model:r20f4]{r20f4} implements a supersymmetric gauge setup of rank 20 with 4 units of flux. The construction involves four rigid D6-branes
$\{a,b,c,d\}$ sharing identical fixed points,
$\delta_{ab}^g \neq (4,4,4)$, and realizes the gauge symmetry
$\SU(4)_{C}\times\SU(2)_{L}\times\SU(2)_{R_1}\times\SU(2)_{R_2}\times\SU(4)^3\times\SU(4)\times\USp(4).$

The matter content is compiled in table~\ref{spec:r20f4}, where fields are
organized by their quantum numbers under the gauge symmetry. One finds in the spectrum chiral fermions, scalar fields, and Higgs-like states arising from
brane intersections such as $ab$, $ac$, and $ad'$, yielding fundamental
representations of the corresponding gauge factors. The model~\hyperref[spec:r20f4]{r20f4} features the torus moduli
$\chi_1 = \frac{3 \sqrt{\frac{41}{10}}}{2}$,
$\chi_2 = \sqrt{410}$,
$\chi_3 = \sqrt{\frac{82}{5}}$,
and the tree-level gauge coupling relation
$g_a^2=\frac{58}{41}g_b^2=\frac{153}{410}g_{cd}^2=\frac{255}{512}\frac{5 g_Y^2}{3}=\frac{8\ 2^{3/4}}{\sqrt{3} \sqrt[4]{205}}\, \pi \,  e^{\phi _4}$.

The resulting chiral spectrum contains left-handed multiplets
$(Q_L,L_L)$ and right-handed multiplets
$(Q_R,L_R)$ arising from distinct D-brane intersections. The left-handed sector encodes the quark and lepton doublets and originate from the $ab'$ (3) sector. The right-handed sector contains the singlet partners and receive contributions from the $ac$ (2), $ad$ (-8), $ad'$ (3) sectors.

The high-scale spectrum contains are 12 scalar fields $\Delta^i$ responsible for breaking the Pati-Salam symmetry
to the Standard Model gauge group. Furthermore, the spectrum contains are 2 Higgs-like fields arising from the $bc$ sector. The spectrum further contains several chiral exotic states
$X_L^i$ and $X_R^i$ charged under hidden sector gauge groups.

Table~\ref{exotic:r20f4} presents the composite spectrum obtained after confinement in the hidden
sector. Matter charged under hidden non-abelian sectors undergoes confinement dynamics, which forces the appearance of bound states neutral under the hidden gauge group. Hence, the resulting bound states are removed from the low-energy spectrum.

 
\subsection{Model~\texorpdfstring{\hyperref[spec:r26f4]{r26f4}}{r26f4}}

\begin{table}[t]
	\centering
	$\begin{array}{|c|D{.}{\times}{30}|c|}
		\hline
		\text{Model~\hyperref[exotic:r26f4]{r26f4}} & \mc{\text{Quantum Numbers}}                                      & \text{Fields}     \\ 
		\hline
		ab                                          & 3. (\bar{4},2,1,1,1,1,1,1,1,1,1,1,1,1)                           & F_L^i(Q_L,L_L)    \\
		ac                                          & 4. (\bar{4},1,2,1,1,1,1,1,1,1,1,1,1,1)                           & F_R^i(Q_R,L_R)    \\
		ac'                                         & 9. (4,1,2,1,1,1,1,1,1,1,1,1,1,1)                                 & F_R^i(Q_R,L_R)    \\
		ad'                                         & 2. (\bar{4},1,1,\bar{2},1,1,1,1,1,1,1,1,1,1)                     & F_R^{'i}(Q_R,L_R) \\
		b_{\yng(1,1)_{}}                            & 4. (1,1_{\yng(1,1)_{}},1,1,1,1,1,1,1,1,1,1,1,1)                  & S_L^i             \\
		c_{\yng(1,1)_{}}                            & 108. (1,1,1_{\yng(1,1)_{}},1,1,1,1,1,1,1,1,1,1,1)                & S_R^i             \\
		d_{\overline{\yng(1,1)_{}}}                 & 2. (1,1,1,\bar{1}_{\overline{\yng(1,1)_{}}},1,1,1,1,1,1,1,1,1,1) & S_R^i             \\
		c_{\yng(2)}                                 & 16. (1,1,3_{\yng(2)},1,1,1,1,1,1,1,1,1,1,1)                      & T_R^i             \\
		cd                                          & 5. (1,1,\bar{2},2,1,1,1,1,1,1,1,1,1,1)                           & \Delta ^i         \\
		bc                                          & 3. (1,2,\bar{2},1,1,1,1,1,1,1,1,1,1,1)                           & \Phi ^i(H_u,H_d)  \\
		bc'                                         & 6. (1,\bar{2},\bar{2},1,1,1,1,1,1,1,1,1,1,1)                     & \Phi ^i(H_u,H_d)  \\
		bd                                          & 2. (1,\bar{2},1,2,1,1,1,1,1,1,1,1,1,1)                           & \Xi ^{i}          \\
		bf_1'                                       & 2. (1,\bar{2},1,1,1,1,\bar{2},1,1,1,1,1,1,1)                     & X_L^{i}           \\
		bf_2'                                       & 1. (1,\bar{2},1,1,1,1,1,\bar{2},1,1,1,1,1,1)                     & X_L^{i}           \\
		bg_1'                                       & 3. (1,2,1,1,1,1,1,1,4,1,1,1,1,1)                                 & X_L^{i}           \\
		bg_2'                                       & 2. (1,2,1,1,1,1,1,1,1,4,1,1,1,1)                                 & X_L^{i}           \\
		ce_1                                        & 2. (1,1,2,1,\bar{2},1,1,1,1,1,1,1,1,1)                           & X_R^{i}           \\
		ce_1'                                       & 3. (1,1,2,1,2,1,1,1,1,1,1,1,1,1)                                 & X_R^{i}           \\
		ce_2                                        & 3. (1,1,2,1,1,\bar{2},1,1,1,1,1,1,1,1)                           & X_R^{i}           \\
		ce_2'                                       & 2. (1,1,2,1,1,2,1,1,1,1,1,1,1,1)                                 & X_R^{i}           \\
		cf_1                                        & 2. (1,1,2,1,1,1,\bar{2},1,1,1,1,1,1,1)                           & X_R^{i}           \\
		cf_1'                                       & 3. (1,1,2,1,1,1,2,1,1,1,1,1,1,1)                                 & X_R^{i}           \\
		cf_2                                        & 2. (1,1,2,1,1,1,1,\bar{2},1,1,1,1,1,1)                           & X_R^{i}           \\
		cf_2'                                       & 3. (1,1,2,1,1,1,1,2,1,1,1,1,1,1)                                 & X_R^{i}           \\
		cg_1                                        & 9. (1,1,2,1,1,1,1,1,\bar{4},1,1,1,1,1)                           & X_R^{i}           \\
		cg_1'                                       & 4. (1,1,2,1,1,1,1,1,4,1,1,1,1,1)                                 & X_R^{i}           \\
		cg_2                                        & 6. (1,1,2,1,1,1,1,1,1,\bar{4},1,1,1,1)                           & X_R^{i}           \\
		cg_2'                                       & 6. (1,1,2,1,1,1,1,1,1,4,1,1,1,1)                                 & X_R^{i}           \\
		ch_4                                        & 1. (1,1,\bar{2},1,1,1,1,1,1,1,1,1,1,4)                           & X_R^{i}           \\
		ch_4'                                       & 1. (1,1,\bar{2},1,1,1,1,1,1,1,1,1,1,\bar{4})                     & X_R^{i}           \\
		de_1                                        & 1. (1,1,1,2,\bar{2},1,1,1,1,1,1,1,1,1)                           & X_R^{i}           \\
		de_2'                                       & 1. (1,1,1,2,1,2,1,1,1,1,1,1,1,1)                                 & X_R^{i}           \\
		dg_1                                        & 2. (1,1,1,\bar{2},1,1,1,1,4,1,1,1,1,1)                           & X_R^{i}           \\
		dg_2'                                       & 1. (1,1,1,\bar{2},1,1,1,1,1,\bar{4},1,1,1,1)                     & X_R^{i}           \\
		\hline
	\end{array}$
	\caption{Particle spectrum of Model~\hyperref[model:r26f4]{r26f4} with gauge symmetry $\SU(4)_{C}\times\SU(2)_{L}\times\SU(2)_{R_1}\times\SU(2)_{R_2}\times\SU(2)^4\times\SU(4)^2\times\USp(8)\times\USp(4)^3$.}
	\label{spec:r26f4}
\end{table}  

The model~\hyperref[model:r26f4]{r26f4} represents a consistent intersecting-brane model of rank 26 with 4 units of flux. The construction involves four rigid D6-branes
$\{a,b,c,d\}$ sharing identical fixed points,
$\delta_{ab}^g \neq (4,4,4)$, and realizes the gauge symmetry
$\SU(4)_{C}\times\SU(2)_{L}\times\SU(2)_{R_1}\times\SU(2)_{R_2}\times\SU(2)^4\times\SU(4)^2\times\USp(8)\times\USp(4)^3.$

The matter content is detailed in table~\ref{spec:r26f4}, where fields are
organized by their quantum numbers under the gauge symmetry. The model gives rise to chiral fermions, scalar fields, and Higgs-like states arising from
brane intersections such as $ab$, $ac$, and $ad'$, yielding fundamental
representations of the corresponding gauge factors. The model~\hyperref[spec:r26f4]{r26f4} features the torus moduli
$\chi_1 = 2 \sqrt{\frac{17}{3}}$,
$\chi_2 = \sqrt{51}$,
$\chi_3 = \frac{10 \sqrt{\frac{17}{3}}}{3}$,
and the tree-level gauge coupling relation
$g_a^2=\frac{95}{102}g_b^2=\frac{285}{442}g_{cd}^2=\frac{475}{632}\frac{5 g_Y^2}{3}=\frac{8 \sqrt[4]{\frac{3}{17}}}{\sqrt{5}}\, \pi \,  e^{\phi _4}$.

One finds in the chiral sector left-handed multiplets
$(Q_L,L_L)$ and right-handed multiplets
$(Q_R,L_R)$ arising from distinct D-brane intersections. The left-handed fields furnish the Standard Model quark and lepton doublets and originate from the $ab$ (-3) sector. The right-handed fields account for the corresponding singlet representations and receive contributions from the $ac$ (-4), $ac'$ (9), $ad'$ (-2) sectors.

The GUT sector includes are 5 scalar fields $\Delta^i$ responsible for breaking the Pati-Salam symmetry
to the Standard Model gauge group. The electroweak sector includes are 9 Higgs-like fields arising from the $bc$, $bc'$ sectors. The spectrum further contains several chiral exotic states
$X_L^i$ and $X_R^i$ charged under hidden sector gauge groups.

Table~\ref{exotic:r26f4} presents the composite spectrum obtained after confinement in the hidden
sector. Hidden-sector charged states are governed by strong coupling dynamics, which dynamically generates bound states neutral under the hidden gauge group. Consequently, these states decouple from the low-energy effective spectrum.

 
\subsection{Model~\texorpdfstring{\hyperref[spec:r33f4]{r33f4}}{r33f4}}

\begin{table}[t]
	\centering
	$\begin{array}{|c|D{.}{\times}{31}|c|}
		\hline
		\text{Model~\hyperref[exotic:r33f4]{r33f4}} & \mc{\text{Quantum Numbers}}                         & \text{Fields}     \\ 
		\hline
		ab                                          & 3. (\bar{4},2,1,1,1,1,1,1,1,1,1,1,1,1,1)            & F_L^i(Q_L,L_L)    \\
		ac                                          & 1. (\bar{4},1,2,1,1,1,1,1,1,1,1,1,1,1,1)            & F_R^i(Q_R,L_R)    \\
		ac'                                         & 6. (4,1,2,1,1,1,1,1,1,1,1,1,1,1,1)                  & F_R^i(Q_R,L_R)    \\
		ad'                                         & 2. (\bar{4},1,1,\bar{2},1,1,1,1,1,1,1,1,1,1,1)      & F_R^{'i}(Q_R,L_R) \\
		b_{\yng(1,1)_{}}                            & 4. (1,1_{\yng(1,1)_{}},1,1,1,1,1,1,1,1,1,1,1,1,1)   & S_L^i             \\
		c_{\yng(1,1)_{}}                            & 100. (1,1,1_{\yng(1,1)_{}},1,1,1,1,1,1,1,1,1,1,1,1) & S_R^i             \\
		d_{\yng(1,1)_{}}                            & 4. (1,1,1,1_{\yng(1,1)_{}},1,1,1,1,1,1,1,1,1,1,1)   & S_R^i             \\
		c_{\yng(2)}                                 & 18. (1,1,3_{\yng(2)},1,1,1,1,1,1,1,1,1,1,1,1)       & T_R^i             \\
		cd                                          & 3. (1,1,\bar{2},2,1,1,1,1,1,1,1,1,1,1,1)            & \Delta ^i         \\
		bc                                          & 4. (1,2,\bar{2},1,1,1,1,1,1,1,1,1,1,1,1)            & \Phi ^i(H_u,H_d)  \\
		bd                                          & 2. (1,\bar{2},1,2,1,1,1,1,1,1,1,1,1,1,1)            & \Xi ^{i}          \\
		be_1'                                       & 1. (1,\bar{2},1,1,\bar{2},1,1,1,1,1,1,1,1,1,1)      & X_L^{i}           \\
		be_2                                        & 1. (1,\bar{2},1,1,1,2,1,1,1,1,1,1,1,1,1)            & X_L^{i}           \\
		be_3'                                       & 1. (1,\bar{2},1,1,1,1,\bar{2},1,1,1,1,1,1,1,1)      & X_L^{i}           \\
		bf_1'                                       & 2. (1,2,1,1,1,1,1,2,1,1,1,1,1,1,1)                  & X_L^{i}           \\
		bf_2                                        & 2. (1,2,1,1,1,1,1,1,\bar{2},1,1,1,1,1,1)            & X_L^{i}           \\
		ce_1                                        & 2. (1,1,2,1,\bar{2},1,1,1,1,1,1,1,1,1,1)            & X_R^{i}           \\
		ce_1'                                       & 4. (1,1,2,1,2,1,1,1,1,1,1,1,1,1,1)                  & X_R^{i}           \\
		ce_2                                        & 4. (1,1,2,1,1,\bar{2},1,1,1,1,1,1,1,1,1)            & X_R^{i}           \\
		ce_2'                                       & 2. (1,1,2,1,1,2,1,1,1,1,1,1,1,1,1)                  & X_R^{i}           \\
		ce_3                                        & 2. (1,1,2,1,1,1,\bar{2},1,1,1,1,1,1,1,1)            & X_R^{i}           \\
		ce_3'                                       & 4. (1,1,2,1,1,1,2,1,1,1,1,1,1,1,1)                  & X_R^{i}           \\
		cf_1                                        & 4. (1,1,2,1,1,1,1,\bar{2},1,1,1,1,1,1,1)            & X_R^{i}           \\
		cf_1'                                       & 2. (1,1,2,1,1,1,1,2,1,1,1,1,1,1,1)                  & X_R^{i}           \\
		cf_2                                        & 2. (1,1,2,1,1,1,1,1,\bar{2},1,1,1,1,1,1)            & X_R^{i}           \\
		cf_2'                                       & 4. (1,1,2,1,1,1,1,1,2,1,1,1,1,1,1)                  & X_R^{i}           \\
		cg_1                                        & 4. (1,1,2,1,1,1,1,1,1,\bar{4},1,1,1,1,1)            & X_R^{i}           \\
		cg_1'                                       & 4. (1,1,2,1,1,1,1,1,1,4,1,1,1,1,1)                  & X_R^{i}           \\
		cg_2                                        & 4. (1,1,2,1,1,1,1,1,1,1,\bar{4},1,1,1,1)            & X_R^{i}           \\
		cg_2'                                       & 4. (1,1,2,1,1,1,1,1,1,1,4,1,1,1,1)                  & X_R^{i}           \\
		ch_3                                        & 1. (1,1,\bar{2},1,1,1,1,1,1,1,1,1,1,8,1)            & X_R^{i}           \\
		ch_3'                                       & 1. (1,1,\bar{2},1,1,1,1,1,1,1,1,1,1,\bar{8},1)      & X_R^{i}           \\
		df_1                                        & 1. (1,1,1,\bar{2},1,1,1,2,1,1,1,1,1,1,1)            & X_R^{i}           \\
		df_2                                        & 1. (1,1,1,\bar{2},1,1,1,1,2,1,1,1,1,1,1)            & X_R^{i}           \\
		dg_1'                                       & 2. (1,1,1,2,1,1,1,1,1,4,1,1,1,1,1)                  & X_R^{i}           \\
		dg_2'                                       & 2. (1,1,1,2,1,1,1,1,1,1,4,1,1,1,1)                  & X_R^{i}           \\
		\hline
	\end{array}$
	\caption{Particle spectrum of Model~\hyperref[model:r33f4]{r33f4} with gauge symmetry $\SU(4)_{C}\times\SU(2)_{L}\times\SU(2)_{R_1}\times\SU(2)_{R_2}\times\SU(2)^5\times\SU(4)^2\times\USp(8)^4$.}
	\label{spec:r33f4}
\end{table}  

The model~\hyperref[model:r33f4]{r33f4} is a gauge theory of rank 33 with 4 units of flux. The construction involves four rigid D6-branes
$\{a,b,c,d\}$ sharing identical fixed points,
$\delta_{ab}^g \neq (4,4,4)$, and realizes the gauge symmetry
$\SU(4)_{C}\times\SU(2)_{L}\times\SU(2)_{R_1}\times\SU(2)_{R_2}\times\SU(2)^5\times\SU(4)^2\times\USp(8)^4.$

The matter content is summarized in table~\ref{spec:r33f4}, where fields are
organized by their quantum numbers under the gauge symmetry. The spectrum contains chiral fermions, scalar fields, and Higgs-like states arising from
brane intersections such as $ab$, $ac$, and $ad'$, yielding fundamental
representations of the corresponding gauge factors. The model~\hyperref[spec:r33f4]{r33f4} features the torus moduli
$\chi_1 = \sqrt{11}$,
$\chi_2 = 4 \sqrt{11}$,
$\chi_3 = 2 \sqrt{11}$,
and the tree-level gauge coupling relation
$g_a^2=\frac{24}{11}g_b^2=\frac{120}{187}g_{cd}^2=\frac{200}{267}\frac{5 g_Y^2}{3}=\frac{4 \sqrt{2}}{\sqrt[4]{11}}\, \pi \,  e^{\phi _4}$.

The chiral sector consists of left-handed multiplets
$(Q_L,L_L)$ and right-handed multiplets
$(Q_R,L_R)$ arising from distinct D-brane intersections. The left-handed states accommodate the quark and lepton doublets and originate from the $ab$ (-3) sector. The right-handed states provide the corresponding singlet partners and receive contributions from the $ac$ (-1), $ac'$ (6), $ad'$ (-2) sectors.

At the GUT scale, there are 3 scalar fields $\Delta^i$ responsible for breaking the Pati-Salam symmetry
to the Standard Model gauge group. In addition, there are 4 Higgs-like fields arising from the $bc$ sector. The spectrum further contains several chiral exotic states
$X_L^i$ and $X_R^i$ charged under hidden sector gauge groups.

Table~\ref{exotic:r33f4} presents the composite spectrum obtained after confinement in the hidden
sector. States charged under non-abelian hidden gauge factors experience strong coupling dynamics, which leads to the formation of bound states neutral under the hidden gauge group. As a result, these degrees of freedom are absent from the low-energy effective theory.



\section{Asymptotic freedom}\label{sec:asymptoticfreedom}

Constructing the Standard Model from rigid cycles, in addition to eliminating adjoint chiral multiplets, is also advantageous for realizing a gauge theory that is asymptotically free, characterized by a negative one-loop beta function. This setup facilitates the convergence of the gauge couplings in the MSSM and also enables gaugino condensation through the non-perturbative superpotential of the form
\begin{align}
	W_{\text{eff}} & \sim \frac{M_\text{S}\,\beta_1^g}{32 \pi^2} \exp\left( \frac{8\pi^2}{g_\text{YM}^2\, \beta_1^g} \right), 
\end{align}
where the gauge couplings depend on the complex structure (or Kähler) moduli in type IIA (or type IIB) string theory. This effective superpotential may, in principle, stabilize some of the closed-string moduli, potentially in combination with other mechanisms such as background fluxes. 

In general, the beta functions are sensitive to the entire massless spectrum, including additional light non-chiral states that are not captured by intersection numbers or topological invariants. Therefore, it is crucial to have complete control over the full spectrum of the theory. To this end, let us examine the light spectrum arising from fractional branes, which are either constructed by splitting the bulk branes or are otherwise generic rigid branes.

\subsection{Fractional branes from splitting bulk branes}

Consider a bulk D-brane $a$ supporting a $\U(N)$ gauge group. This brane contains three adjoint chiral multiplets, in addition to other matter from intersections with other D-branes. Neglecting the latter, the one-loop beta function is
\begin{align}
	b_1^{\U(N)} = -3N + 3 \times N = 0, 
\end{align}
indicating that bulk D-branes have vanishing or positive beta functions.

One might attempt to improve this by splitting the bulk brane into four rigid fractional constituents ${b_1, b_2, b_3, b_4}$ transforming in the regular representation of $\mathbb{Z}_2 \times \mathbb{Z}_2$, forming, for example, the D-brane stack $b$. This decomposition yields a gauge group $\U(N)^4$ with no massless adjoint fields. However, additional nonchiral matter may arise between pairs of these fractional D-branes, and in fact, this is generally the case. This spectrum can be computed from the boundary state overlaps:
\begin{align} \label{eq:overlap}
	\tilde{A}_{b_i,b_j} & = \int_0^\infty dl\, \langle b_i| e^{-2\pi l H_{\text{cl}}} |b_j \rangle + \int_0^\infty dl\, \langle b_j| e^{-2\pi l H_{\text{cl}}} |b_i \rangle, \quad i \neq j,                                                                                                                            
\end{align}
whose loop channel representation is
\begin{align}
	A_{b_i,b_j} = \int_0^\infty \frac{dt}{t} \, \text{Tr}_{ij+ji} \left( \frac{1 + \theta + \omega + \theta \omega}{4} \, e^{-2\pi t H_0} \right). 
\end{align}
The twisted sector projections result in a single massless hypermultiplet, composed of scalar states associated with the oscillators $\psi^i_{-\frac{1}{2}}|0\rangle$, $\overline{\psi}^i_{-\frac{1}{2}}|0\rangle$, for $I \in \{1,2,3\}$. For $i = j$, the non-compact oscillators $\psi^\mu_{-\frac{1}{2}}|0\rangle$, $\overline{\psi}^\mu_{-\frac{1}{2}}|0\rangle$ survive the projection, leading to an $\mathcal{N}=1$ vector multiplet.

Each $\U(N)$ factor thereby receives $6N$ chiral supermultiplets in the fundamental representation, resulting in a one-loop beta function:
\begin{align}
	b_1^{\U(N)} = -3N + 6N \times \frac{1}{2} = 0. 
\end{align}
Hence, splitting a $\U(N)$ bulk brane into fractional branes does not improve the asymptotic behavior of the beta function. Consequently, rigid D-branes obtained through this method cannot yield asymptotically free gauge groups.

\subsection{Beta functions for generic rigid fractional branes}

Consider a rigid fractional D-brane $a$ supporting a $\U(4)$ gauge group. Since the brane is rigid, no adjoint chiral multiplets arise in the $aa$ sector. However, the $aa'$ sector contributes one hypermultiplet in the antisymmetric representation $\mathbf{6}$ of $\SU(4)$, corresponding to two chiral multiplets \cite{Blumenhagen:2005tn}.

Additional massless states charged under $\SU(4)$ may arise from bifundamental vector-like pairs in the $ab$ sectors, where $b$ denotes another fractional D-brane. Their multiplicities are determined by first computing the spectrum in the covering theory and subsequently imposing the orbifold projection, implemented via twisted-sector contributions in the loop channel.

The one-loop $\mathcal{N}=1$ beta function for a gauge group $G$ is given by
\begin{equation}
	\beta^G = -3\,C_2(G) + \sum_{\text{chiral}} T(R) \,,
\end{equation}
where $C_2(G)$ is the quadratic Casimir of the adjoint representation and $T(R)$ is the Dynkin index of a chiral multiplet in representation $R$.  

For the rigid $\SU(4)$ brane stack considered here, we have
\begin{equation*}
	C_2(\SU(4)) = 4, \qquad T(\mathbf{4}) = \tfrac{1}{2}, \qquad T(\mathbf{6}) = 1.
\end{equation*}
The antisymmetric hypermultiplet arising from the $aa'$ sector contributes $2 \times T(\mathbf{6}) = 2 \times 1$, while chiral multiplets at intersections with other branes contribute $N_a^{\text{chiral}} \times T(\mathbf{4}) = \frac{1}{2} N_a^{\text{chiral}}$. Combining these contributions yields
\begin{equation}\label{eq:betaSU4}
	\beta^{\SU(4)} = -3 \times 4 + \frac{1}{2}\, N_a^{\text{chiral}} + 2 \times 1 \,.
\end{equation} 
Here $N_a^{\text{chiral}}$ denotes twice the total number of chiral multiplets arising at intersections of the $\SU(4)$ stack with other visible-sector branes,
\begin{equation}
	N_a^{\text{chiral}} = 2 \Bigl( |I_{ab}| + |I_{ab'}| + |I_{ac}| + |I_{ac'}| + |I_{ad}| + |I_{ad'}| \Bigr).
\end{equation}
 
Applying \eqref{eq:betaSU4} to our set of rigid fractional brane models,
we find that only a small subset yields negative beta functions.
In particular,
Models~\hyperref[spec:r15f1]{r15f1},
\hyperref[spec:r17f1]{r17f1},
and \hyperref[spec:r10f2]{r10f2}
satisfy $\beta^{\SU(4)}<0$ and are therefore asymptotically free.

All remaining models have $\beta^{\SU(4)} \ge 0$ and do not exhibit
asymptotic freedom.
The individual beta function coefficients for each model are listed in
Appendix~\ref{appC}.

\section{Conclusion}\label{sec:conclusion}  

In this work, we have constructed a new class of three-family $\mathcal{N}=1$ supersymmetric Pati–Salam flux vacua in the Type IIB framework on the $\mathbb{T}^6/(\mathbb{Z}_2 \times \mathbb{Z}_2)$ orientifold, which is T-dual to the Type IIA $\mathbb{T}^6/(\mathbb{Z}_2 \times \mathbb{Z}_2^\prime)$ orientifold with discrete torsion. The models are realized using magnetized D-branes in the presence of exotic $\mathrm{O}3^{++}$-planes and quantized background $G_3$ flux. 

A central feature of these constructions is the simultaneous stabilizations of both open- and closed-string moduli. On the open-string side, the worldvolume magnetic fluxes generate a superpotential that fixes D-brane position and Wilson line moduli, thereby eliminating all adjoint chiral multiplets, in direct analogy with the rigid D6-brane mechanism employed in our previous work~\cite{Mansha:2025yxm}. On the closed-string side, supersymmetric $(2,1)$ imaginary self-dual $G_3$ flux generates a Gukov–Vafa–Witten superpotential that stabilizes the complex structure moduli and the axio-dilaton at tree level. The Kähler moduli remain unfixed at this stage due to the no-scale structure, but can be stabilized non-perturbatively via mechanisms such as gaugino condensation or Euclidean D3-brane instantons in the hidden sector. This stabilization of closed-string moduli is a novel feature of the present flux constructions and was not achieved in our previous rigid D6-brane models.

All models satisfy the full set of string consistency conditions, including $\mathcal{N}=1$ supersymmetry, RR tadpole cancellation in the presence of flux, and K-theory constraints. The resulting chiral spectra contain exactly three generations of the SM matter after Pati–Salam symmetry breaking, which can be implemented via a supersymmetry-preserving Higgs mechanism. We found that only a subset of models exhibit asymptotic freedom in the strong sector, making it a significant constraint for constructing phenomenologically viable models with fractional D-branes.

We have presented the complete perturbative particle spectra and shown that all the exotic vector-like states can dynamically decouple through strong gauge dynamics in the hidden sector, leading to phenomenologically viable low-energy theories. These results demonstrate that the rigid intersecting D-brane constructions with flux provide a robust framework for simultaneously addressing open- and closed-string moduli stabilization while realizing realistic chiral particle physics models. At best, these models remain semi-realistic, as questions related to scale separation, swampland constraints arising from coupling to gravity, and the sign of the cosmological constant remain crucial but are beyond the scope of the present work.

This framework opens several promising directions for future research. A detailed phenomenological analysis, including the computation of Yukawa couplings, supersymmetry breaking soft terms, and the study of flavor structures, remains an important next step. Unlike the non-rigid case \cite{Sabir:2024cgt, Sabir:2024jsx}, the interpretation of twisted sector contributions to Yukawas in the rigid setup is still an open problem. Moreover, the systematic investigations of flux configurations and the landscape of consistent vacua in this magnetized D-brane setup may reveal additional three-family models with realistic low-energy physics. Extending the construction to non-factorizable or tilted tori may further enrich the class of phenomenologically viable models. Such extensions can also provide deeper insights into moduli stabilization in Type IIB compactifications. Exploring the full landscape of possible configurations, following the approach of \cite{He:2021gug}, is expected to be significantly more challenging but may uncover additional structures relevant for realistic model building.

\acknowledgments{AM is supported by the Guangdong Basic and Applied Basic Research Foundation (Grant No. 2021B1515130007), Shenzhen Natural Science Fund (the Stable Support Plan Program 20220810130956001). TL is supported in part by the National Key Research and Development Program of China Grant No. 2020YFC2201504, by the Projects No. 11875062, No. 11947302, No. 12047503, and No. 12275333 supported by the National Natural Science Foundation of China, by the Key Research Program of the Chinese Academy of Sciences, Grant No. XDPB15, by the Scientific Instrument Developing Project of the Chinese Academy of Sciences, Grant No. YJKYYQ20190049, and by the International Partnership Program of Chinese Academy of Sciences for Grand Challenges, Grant No. 112311KYSB20210012. MS is supported in part by the National Natural Science Foundation of China (Grant No. 12475105).}

\appendix

\section{Three-family Pati-Salam flux models on rigid cycles}\label{appA} 

In this appendix, we list the three independent three-family $\mathcal{N}=1$ supersymmetric flux Pati–Salam models constructed from rigid intersecting D-branes on the type IIA $\mathbb{T}^6/(\mathbb{Z}_2 \times \mathbb{Z}_2')$ orientifold with discrete torsion. The models are named according to the rank of their gauge groups together with quantized flux. For each model, we provide the brane stacks with their multiplicities, wrapping numbers $(n^i, m^i)$ along the three two-tori, and the type of branes, including fractional and bulk D-branes with worldvolume fluxes. The complete gauge group for each model, the corresponding two-torus complex structure moduli and the tree-level string-scale gauge coupling relations are specified in the captions of the tables.

\begin{table}[th]
	\centering
	$\begin{array}{|c|c|c|l|}
		\hline
		\text{Model~\hyperref[spec:r15f1]{r15f1}} & N & (n^1, m^1) \times (n^2, m^2) \times (n^3, m^3) & \text{Type of brane}        \\ 
		\hline 
		a                                         & 4 & (0, -1)\times (0, 1)\times (1, 0)              & \text{frac. D7$_3$}         \\
		b                                         & 2 & (3, -2)\times (-3, 1)\times (2, -1)            & \text{frac. D9 w. flux}     \\
		c                                         & 2 & (-2, 1)\times (1, 0)\times (-2, -1)            & \text{frac. D7$_2$ w. flux} \\
		d                                         & 2 & (-1, 0)\times (1, 1)\times (-2, 1)             & \text{frac. D7$_1$ w. flux} \\
		e                                         & 2 & (0, -1)\times (1, 0)\times (0, 1)              & \text{bulk D7$_2$}          \\
		f_1                                       & 2 & (0, -1)\times (0, 1)\times (1, 0)              & \text{bulk D7$_3$}          \\
		f_2                                       & 2 & (0, -1)\times (0, -1)\times (-1, 0)            & \text{bulk D7$_3$}          \\
		g_1                                       & 4 & (1, 0)\times (0, -1)\times (0, 1)              & \text{bulk D7$_1$}          \\
		g_2                                       & 4 & (-1, 0)\times (0, 1)\times (0, 1)              & \text{bulk D7$_1$}          \\
		\hline
	\end{array}$
	\caption{Model~\hyperref[spec:r15f1]{r15f1} with the gauge group $\SU(4)_{C}\times\SU(2)_{L}\times\SU(2)_{R_1}\times\SU(2)_{R_2}\times\SU(2)^3\times\SU(4)^2$, the torus moduli $\chi_1=2 \sqrt{6}$, $\chi_2=\sqrt{6}$, $\chi_3= 2 \sqrt{6}$ and the gauge coupling relation $g_a^2=\frac{35}{2}g_b^2=\frac{7}{9}g_{cd}^2=\frac{35}{41}\frac{5 g_Y^2}{3}=\frac{4\ 2^{3/4}}{\sqrt[4]{3}}\, \pi \,  e^{\phi _4}$.}
	\label{model:r15f1}
\end{table}

\begin{table}[th]
	\centering
	$\begin{array}{|c|c|c|l|}
		\hline
		\text{Model~\hyperref[spec:r17f1]{r17f1}} & N & (n^1, m^1) \times (n^2, m^2) \times (n^3, m^3) & \text{Type of brane}    \\ 
		\hline 
		a                                         & 4 & (1, 0)\times (0, -1)\times (0, 1)              & \text{frac. D7$_1$}     \\
		b                                         & 2 & (-2, -1)\times (1, -1)\times (-4, 1)           & \text{frac. D9 w. flux} \\
		c                                         & 2 & (4, 1)\times (-3, -1)\times (1, 1)             & \text{frac. D9 w. flux} \\
		d                                         & 2 & (-4, 1)\times (1, -1)\times (3, -1)            & \text{frac. D9 w. flux} \\
		e_1                                       & 4 & (0, -1)\times (1, 0)\times (0, 1)              & \text{bulk D7$_2$}      \\
		e_2                                       & 2 & (0, 1)\times (-1, 0)\times (0, 1)              & \text{bulk D7$_2$}      \\
		f_1                                       & 2 & (1, 0)\times (1, 0)\times (1, 0)               & \text{bulk D3}          \\
		f_2                                       & 2 & (-1, 0)\times (-1, 0)\times (1, 0)             & \text{bulk D3}          \\
		f_3                                       & 2 & (1, 0)\times (-1, 0)\times (-1, 0)             & \text{bulk D3}          \\
		f_4                                       & 2 & (-1, 0)\times (1, 0)\times (-1, 0)             & \text{bulk D3}          \\
		\hline
	\end{array}$
	\caption{Model~\hyperref[spec:r17f1]{r17f1} with the gauge group $\SU(4)_{C}\times\SU(2)_{L}\times\SU(2)_{R_1}\times\SU(2)_{R_2}\times\SU(4)\times\SU(2)\times\USp(4)^4$, the torus moduli $\chi_1=2 \sqrt{611}$, $\chi_2=\sqrt{\frac{47}{13}}$, $\chi_3= \sqrt{\frac{47}{13}}$ and the gauge coupling relation $g_a^2=\frac{6120}{47}g_b^2=\frac{2460}{47}g_{cd}^2=\frac{4100}{1687}\frac{5 g_Y^2}{3}=\frac{8 \sqrt{2} 13^{3/4}}{\sqrt[4]{47}}\, \pi \,  e^{\phi _4}$.}
	\label{model:r17f1}
\end{table}

\begin{table}[th]
	\centering
	$\begin{array}{|c|c|c|l|}
		\hline
		\text{Model~\hyperref[spec:r43f1]{r43f1}} & N & (n^1, m^1) \times (n^2, m^2) \times (n^3, m^3) & \text{Type of brane}        \\ 
		\hline 
		a                                         & 4 & (1, 0)\times (0, -1)\times (0, 1)              & \text{frac. D7$_1$}         \\
		b                                         & 2 & (-4, -1)\times (1, 0)\times (-4, 3)            & \text{frac. D7$_2$ w. flux} \\
		c                                         & 2 & (4, 1)\times (-3, -1)\times (4, 1)             & \text{frac. D9 w. flux}     \\
		d                                         & 2 & (-2, 1)\times (2, 3)\times (-1, 0)             & \text{frac. D7$_3$ w. flux} \\
		e                                         & 2 & (0, -1)\times (0, 1)\times (1, 0)              & \text{bulk D7$_3$}          \\
		f_1                                       & 2 & (1, 0)\times (0, -1)\times (0, 1)              & \text{bulk D7$_1$}          \\
		f_2                                       & 2 & (-1, 0)\times (0, 1)\times (0, 1)              & \text{bulk D7$_1$}          \\
		g_1                                       & 2 & (0, -1)\times (1, 0)\times (0, 1)              & \text{bulk D7$_2$}          \\
		g_2                                       & 2 & (0, 1)\times (-1, 0)\times (0, 1)              & \text{bulk D7$_2$}          \\
		h_1                                       & 8 & (1, 0)\times (1, 0)\times (1, 0)               & \text{bulk D3}              \\
		h_2                                       & 8 & (-1, 0)\times (-1, 0)\times (1, 0)             & \text{bulk D3}              \\
		h_3                                       & 8 & (1, 0)\times (-1, 0)\times (-1, 0)             & \text{bulk D3}              \\
		h_4                                       & 8 & (-1, 0)\times (1, 0)\times (-1, 0)             & \text{bulk D3}              \\
		\hline
	\end{array}$
	\caption{Model~\hyperref[spec:r43f1]{r43f1} with the gauge group $\SU(4)_{C}\times\SU(2)_{L}\times\SU(2)_{R_1}\times\SU(2)_{R_2}\times\SU(2)^5\times\USp(16)^4$, the torus moduli $\chi_1=8 \sqrt{3}$, $\chi_2=\frac{8}{\sqrt{3}}$, $\chi_3= \frac{8}{\sqrt{3}}$ and the gauge coupling relation $g_a^2=\frac{39}{4}g_b^2=\frac{1911}{292}g_{cd}^2=\frac{3185}{1566}\frac{5 g_Y^2}{3}=2 \sqrt{2} 3^{3/4}\, \pi \,  e^{\phi _4}$.}
	\label{model:r43f1}
\end{table}

\begin{table}[th]
	\centering
	$\begin{array}{|c|c|c|l|}
		\hline
		\text{Model~\hyperref[spec:r7f2]{r7f2}} & N & (n^1, m^1) \times (n^2, m^2) \times (n^3, m^3) & \text{Type of brane}        \\ 
		\hline 
		a                                       & 4 & (0, -1)\times (1, 0)\times (0, 1)              & \text{frac. D7$_2$}         \\
		b                                       & 2 & (1, 0)\times (-2, 1)\times (-5, -1)            & \text{frac. D7$_1$ w. flux} \\
		c                                       & 2 & (-4, 1)\times (2, -1)\times (1, -1)            & \text{frac. D9 w. flux}     \\
		d                                       & 2 & (3, -2)\times (-2, 1)\times (3, -1)            & \text{frac. D9 w. flux}     \\
		e                                       & 2 & (0, -1)\times (0, 1)\times (1, 0)              & \text{bulk D7$_3$}          \\
		\hline
	\end{array}$
	\caption{Model~\hyperref[spec:r7f2]{r7f2} with the gauge group $\SU(4)_{C}\times\SU(2)_{L}\times\SU(2)_{R_1}\times\SU(2)_{R_2}\times\SU(2)$, the torus moduli $\chi_1=6$, $\chi_2=2$, $\chi_3= 5$ and the gauge coupling relation $g_a^2=\frac{2}{3}g_b^2=\frac{884}{385}g_{cd}^2=\frac{4420}{2923}\frac{5 g_Y^2}{3}=\frac{8}{\sqrt{15}}\, \pi \,  e^{\phi _4}$.}
	\label{model:r7f2}
\end{table}

\begin{table}[th]
	\centering
	$\begin{array}{|c|c|c|l|}
		\hline
		\text{Model~\hyperref[spec:r10f2]{r10f2}} & N & (n^1, m^1) \times (n^2, m^2) \times (n^3, m^3) & \text{Type of brane}    \\ 
		\hline 
		a                                         & 4 & (1, 0)\times (0, -1)\times (0, 1)              & \text{frac. D7$_1$}     \\
		b                                         & 2 & (-2, 1)\times (-4, -1)\times (1, 1)            & \text{frac. D9 w. flux} \\
		c                                         & 2 & (4, -1)\times (1, -1)\times (-3, 1)            & \text{frac. D9 w. flux} \\
		d                                         & 2 & (-4, 1)\times (3, -1)\times (1, -1)            & \text{frac. D9 w. flux} \\
		e_1                                       & 2 & (0, -1)\times (0, -1)\times (-1, 0)            & \text{bulk D7$_3$}      \\
		e_2                                       & 4 & (0, -1)\times (0, 1)\times (1, 0)              & \text{bulk D7$_3$}      \\
		\hline
	\end{array}$
	\caption{Model~\hyperref[spec:r10f2]{r10f2} with the gauge group $\SU(4)_{C}\times\SU(2)_{L}\times\SU(2)_{R_1}\times\SU(2)_{R_2}\times\SU(4)\times\SU(2)$, the torus moduli $\chi_1=2 \sqrt{611}$, $\chi_2=\sqrt{\frac{47}{13}}$, $\chi_3= \sqrt{\frac{47}{13}}$ and the gauge coupling relation $g_a^2=\frac{6120}{47}g_b^2=\frac{2460}{47}g_{cd}^2=\frac{4100}{1687}\frac{5 g_Y^2}{3}=\frac{8 \sqrt{2} 13^{3/4}}{\sqrt[4]{47}}\, \pi \,  e^{\phi _4}$.}
	\label{model:r10f2}
\end{table}

\begin{table}[th]
	\centering
	$\begin{array}{|c|c|c|l|}
		\hline
		\text{Model~\hyperref[spec:r35f2]{r35f2}} & N & (n^1, m^1) \times (n^2, m^2) \times (n^3, m^3) & \text{Type of brane}        \\ 
		\hline 
		a                                         & 4 & (0, -1)\times (1, 0)\times (0, 1)              & \text{frac. D7$_2$}         \\
		b                                         & 2 & (-4, -1)\times (-4, 3)\times (1, 0)            & \text{frac. D7$_3$ w. flux} \\
		c                                         & 2 & (4, 1)\times (4, 1)\times (-3, -1)             & \text{frac. D9 w. flux}     \\
		d                                         & 2 & (-1, 0)\times (-2, -1)\times (2, -3)           & \text{frac. D7$_1$ w. flux} \\
		e                                         & 2 & (1, 0)\times (0, -1)\times (0, 1)              & \text{bulk D7$_1$}          \\
		f_1                                       & 2 & (0, -1)\times (1, 0)\times (0, 1)              & \text{bulk D7$_2$}          \\
		f_2                                       & 2 & (0, 1)\times (-1, 0)\times (0, 1)              & \text{bulk D7$_2$}          \\
		g_1                                       & 2 & (0, -1)\times (0, 1)\times (1, 0)              & \text{bulk D7$_3$}          \\
		g_2                                       & 2 & (0, -1)\times (0, -1)\times (-1, 0)            & \text{bulk D7$_3$}          \\
		h_1                                       & 6 & (1, 0)\times (1, 0)\times (1, 0)               & \text{bulk D3}              \\
		h_2                                       & 6 & (-1, 0)\times (-1, 0)\times (1, 0)             & \text{bulk D3}              \\
		h_3                                       & 6 & (1, 0)\times (-1, 0)\times (-1, 0)             & \text{bulk D3}              \\
		h_4                                       & 6 & (-1, 0)\times (1, 0)\times (-1, 0)             & \text{bulk D3}              \\
		\hline
	\end{array}$
	\caption{Model~\hyperref[spec:r35f2]{r35f2} with the gauge group $\SU(4)_{C}\times\SU(2)_{L}\times\SU(2)_{R_1}\times\SU(2)_{R_2}\times\SU(2)^5\times\USp(12)^4$, the torus moduli $\chi_1=4 \sqrt{30}$, $\chi_2=4 \sqrt{\frac{10}{3}}$, $\chi_3= \frac{4 \sqrt{\frac{10}{3}}}{3}$ and the gauge coupling relation $g_a^2=\frac{93}{10}g_b^2=\frac{17329}{17410}g_{cd}^2=\frac{86645}{86888}\frac{5 g_Y^2}{3}=2 \sqrt[4]{\frac{3}{5}} 2^{3/4}\, \pi \,  e^{\phi _4}$.}
	\label{model:r35f2}
\end{table}

\begin{table}[th]
	\centering
	$\begin{array}{|c|c|c|l|}
		\hline
		\text{Model~\hyperref[spec:r43af2]{r43af2}} & N & (n^1, m^1) \times (n^2, m^2) \times (n^3, m^3) & \text{Type of brane}        \\ 
		\hline 
		a                                           & 4 & (0, -1)\times (1, 0)\times (0, 1)              & \text{frac. D7$_2$}         \\
		b                                           & 2 & (-4, -1)\times (-2, 3)\times (1, 0)            & \text{frac. D7$_3$ w. flux} \\
		c                                           & 2 & (4, 1)\times (4, 1)\times (-3, -1)             & \text{frac. D9 w. flux}     \\
		d                                           & 2 & (-1, 0)\times (-4, -1)\times (2, -3)           & \text{frac. D7$_1$ w. flux} \\
		e                                           & 2 & (1, 0)\times (0, -1)\times (0, 1)              & \text{bulk D7$_1$}          \\
		f_1                                         & 2 & (0, -1)\times (1, 0)\times (0, 1)              & \text{bulk D7$_2$}          \\
		f_2                                         & 2 & (0, 1)\times (-1, 0)\times (0, 1)              & \text{bulk D7$_2$}          \\
		g_1                                         & 2 & (0, -1)\times (0, 1)\times (1, 0)              & \text{bulk D7$_3$}          \\
		g_2                                         & 2 & (0, -1)\times (0, -1)\times (-1, 0)            & \text{bulk D7$_3$}          \\
		h_1                                         & 8 & (1, 0)\times (1, 0)\times (1, 0)               & \text{bulk D3}              \\
		h_2                                         & 8 & (-1, 0)\times (-1, 0)\times (1, 0)             & \text{bulk D3}              \\
		h_3                                         & 8 & (1, 0)\times (-1, 0)\times (-1, 0)             & \text{bulk D3}              \\
		h_4                                         & 8 & (-1, 0)\times (1, 0)\times (-1, 0)             & \text{bulk D3}              \\
		\hline
	\end{array}$
	\caption{Model~\hyperref[spec:r43af2]{r43af2} with the gauge group $\SU(4)_{C}\times\SU(2)_{L}\times\SU(2)_{R_1}\times\SU(2)_{R_2}\times\SU(2)^5\times\USp(16)^4$, the torus moduli $\chi_1=4 \sqrt{195}$, $\chi_2=2 \sqrt{\frac{65}{3}}$, $\chi_3= \frac{\sqrt{\frac{65}{3}}}{3}$ and the gauge coupling relation $g_a^2=\frac{1176}{65}g_b^2=\frac{4312}{7475}g_{cd}^2=\frac{21560}{31049}\frac{5 g_Y^2}{3}=4 \sqrt[4]{\frac{3}{65}} \sqrt{2}\, \pi \,  e^{\phi _4}$.}
	\label{model:r43af2}
\end{table}

\begin{table}[th]
	\centering
	$\begin{array}{|c|c|c|l|}
		\hline
		\text{Model~\hyperref[spec:r43bf2]{r43bf2}} & N & (n^1, m^1) \times (n^2, m^2) \times (n^3, m^3) & \text{Type of brane}        \\ 
		\hline 
		a                                           & 4 & (0, -1)\times (0, 1)\times (1, 0)              & \text{frac. D7$_3$}         \\
		b                                           & 2 & (1, 0)\times (-4, -1)\times (-2, 3)            & \text{frac. D7$_1$ w. flux} \\
		c                                           & 2 & (-3, 1)\times (4, -1)\times (4, -1)            & \text{frac. D9 w. flux}     \\
		d                                           & 2 & (2, 1)\times (-1, 0)\times (-4, 3)             & \text{frac. D7$_2$ w. flux} \\
		e                                           & 2 & (0, -1)\times (1, 0)\times (0, 1)              & \text{bulk D7$_2$}          \\
		f_1                                         & 2 & (1, 0)\times (0, -1)\times (0, 1)              & \text{bulk D7$_1$}          \\
		f_2                                         & 2 & (-1, 0)\times (0, 1)\times (0, 1)              & \text{bulk D7$_1$}          \\
		g_1                                         & 2 & (0, -1)\times (0, 1)\times (1, 0)              & \text{bulk D7$_3$}          \\
		g_2                                         & 2 & (0, -1)\times (0, -1)\times (-1, 0)            & \text{bulk D7$_3$}          \\
		h_1                                         & 8 & (1, 0)\times (1, 0)\times (1, 0)               & \text{bulk D3}              \\
		h_2                                         & 8 & (-1, 0)\times (-1, 0)\times (1, 0)             & \text{bulk D3}              \\
		h_3                                         & 8 & (1, 0)\times (-1, 0)\times (-1, 0)             & \text{bulk D3}              \\
		h_4                                         & 8 & (-1, 0)\times (1, 0)\times (-1, 0)             & \text{bulk D3}              \\
		\hline
	\end{array}$
	\caption{Model~\hyperref[spec:r43bf2]{r43bf2} with the gauge group $\SU(4)_{C}\times\SU(2)_{L}\times\SU(2)_{R_1}\times\SU(2)_{R_2}\times\SU(2)^5\times\USp(16)^4$, the torus moduli $\chi_1=3 \sqrt{3}$, $\chi_2=12 \sqrt{3}$, $\chi_3= 2 \sqrt{3}$ and the gauge coupling relation $g_a^2=\frac{56}{27}g_b^2=\frac{1736}{3303}g_{cd}^2=\frac{8680}{13381}\frac{5 g_Y^2}{3}=\frac{4 \sqrt{2}}{3 \sqrt[4]{3}}\, \pi \,  e^{\phi _4}$.}
	\label{model:r43bf2}
\end{table}

\begin{table}[th]
	\centering
	$\begin{array}{|c|c|c|l|}
		\hline
		\text{Model~\hyperref[spec:r123f2]{r123f2}} & N  & (n^1, m^1) \times (n^2, m^2) \times (n^3, m^3) & \text{Type of brane}        \\ 
		\hline 
		a                                           & 4  & (0, -1)\times (1, 0)\times (0, 1)              & \text{frac. D7$_2$}         \\
		b                                           & 2  & (5, -1)\times (-4, 1)\times (4, -1)            & \text{frac. D9 w. flux}     \\
		c                                           & 2  & (-5, 1)\times (1, 2)\times (-1, 0)             & \text{frac. D7$_3$ w. flux} \\
		d                                           & 2  & (-1, 0)\times (1, -1)\times (-3, -2)           & \text{frac. D7$_1$ w. flux} \\
		e                                           & 2  & (1, 0)\times (0, -1)\times (0, 1)              & \text{bulk D7$_1$}          \\
		f_1                                         & 2  & (0, -1)\times (1, 0)\times (0, 1)              & \text{bulk D7$_2$}          \\
		f_2                                         & 2  & (0, 1)\times (-1, 0)\times (0, 1)              & \text{bulk D7$_2$}          \\
		g_1                                         & 2  & (0, -1)\times (0, 1)\times (1, 0)              & \text{bulk D7$_3$}          \\
		g_2                                         & 2  & (0, -1)\times (0, -1)\times (-1, 0)            & \text{bulk D7$_3$}          \\
		h_1                                         & 28 & (1, 0)\times (1, 0)\times (1, 0)               & \text{bulk D3}              \\
		h_2                                         & 28 & (-1, 0)\times (-1, 0)\times (1, 0)             & \text{bulk D3}              \\
		h_3                                         & 28 & (1, 0)\times (-1, 0)\times (-1, 0)             & \text{bulk D3}              \\
		h_4                                         & 28 & (-1, 0)\times (1, 0)\times (-1, 0)             & \text{bulk D3}              \\
		\hline
	\end{array}$
	\caption{Model~\hyperref[spec:r123f2]{r123f2} with the gauge group $\SU(4)_{C}\times\SU(2)_{L}\times\SU(2)_{R_1}\times\SU(2)_{R_2}\times\SU(2)^5\times\USp(56)^4$, the torus moduli $\chi_1=10 \sqrt{14}$, $\chi_2=\sqrt{14}$, $\chi_3= 3 \sqrt{\frac{7}{2}}$ and the gauge coupling relation $g_a^2=\frac{95}{14}g_b^2=\frac{57}{308}g_{cd}^2=\frac{95}{346}\frac{5 g_Y^2}{3}=\frac{4\ 2^{3/4}}{\sqrt[4]{7} \sqrt{15}}\, \pi \,  e^{\phi _4}$.}
	\label{model:r123f2}
\end{table}

\begin{table}[th]
	\centering
	$\begin{array}{|c|c|c|l|}
		\hline
		\text{Model~\hyperref[spec:r125f2]{r125f2}} & N  & (n^1, m^1) \times (n^2, m^2) \times (n^3, m^3) & \text{Type of brane}        \\ 
		\hline 
		a                                           & 4  & (0, -1)\times (0, 1)\times (1, 0)              & \text{frac. D7$_3$}         \\
		b                                           & 2  & (4, -1)\times (5, -1)\times (-4, 1)            & \text{frac. D9 w. flux}     \\
		c                                           & 2  & (-1, -2)\times (-5, 1)\times (1, 0)            & \text{frac. D7$_3$ w. flux} \\
		d                                           & 2  & (-3, -2)\times (-1, 0)\times (1, -1)           & \text{frac. D7$_2$ w. flux} \\
		e                                           & 2  & (0, -1)\times (1, 0)\times (0, 1)              & \text{bulk D7$_2$}          \\
		f_1                                         & 4  & (1, 0)\times (0, -1)\times (0, 1)              & \text{bulk D7$_1$}          \\
		f_2                                         & 4  & (-1, 0)\times (0, 1)\times (0, 1)              & \text{bulk D7$_1$}          \\
		g_1                                         & 28 & (1, 0)\times (1, 0)\times (1, 0)               & \text{bulk D3}              \\
		g_2                                         & 28 & (-1, 0)\times (-1, 0)\times (1, 0)             & \text{bulk D3}              \\
		g_3                                         & 28 & (1, 0)\times (-1, 0)\times (-1, 0)             & \text{bulk D3}              \\
		g_4                                         & 28 & (-1, 0)\times (1, 0)\times (-1, 0)             & \text{bulk D3}              \\
		\hline
	\end{array}$
	\caption{Model~\hyperref[spec:r125f2]{r125f2} with the gauge group $\SU(4)_{C}\times\SU(2)_{L}\times\SU(2)_{R_1}\times\SU(2)_{R_2}\times\SU(2)\times\SU(4)^2\times\USp(56)^4$, the torus moduli $\chi_1=\sqrt{29}$, $\chi_2=10 \sqrt{29}$, $\chi_3= \frac{2 \sqrt{29}}{3}$ and the gauge coupling relation $g_a^2=\frac{195}{29}g_b^2=\frac{2925}{21808}g_{cd}^2=\frac{4875}{23758}\frac{5 g_Y^2}{3}=\frac{8}{\sqrt{15} \sqrt[4]{29}}\, \pi \,  e^{\phi _4}$.}
	\label{model:r125f2}
\end{table}

\begin{table}[th]
	\centering
	$\begin{array}{|c|c|c|l|}
		\hline
		\text{Model~\hyperref[spec:r27f3]{r27f3}} & N & (n^1, m^1) \times (n^2, m^2) \times (n^3, m^3) & \text{Type of brane}        \\ 
		\hline 
		a                                         & 4 & (0, -1)\times (1, 0)\times (0, 1)              & \text{frac. D7$_2$}         \\
		b                                         & 2 & (-4, 3)\times (-4, -1)\times (1, 0)            & \text{frac. D7$_3$ w. flux} \\
		c                                         & 2 & (4, 1)\times (4, 1)\times (-3, -1)             & \text{frac. D9 w. flux}     \\
		d                                         & 2 & (-1, 0)\times (-2, 3)\times (2, 1)             & \text{frac. D7$_1$ w. flux} \\
		e                                         & 2 & (1, 0)\times (0, -1)\times (0, 1)              & \text{bulk D7$_1$}          \\
		f_1                                       & 2 & (0, -1)\times (1, 0)\times (0, 1)              & \text{bulk D7$_2$}          \\
		f_2                                       & 2 & (0, 1)\times (-1, 0)\times (0, 1)              & \text{bulk D7$_2$}          \\
		g_1                                       & 2 & (0, -1)\times (0, 1)\times (1, 0)              & \text{bulk D7$_3$}          \\
		g_2                                       & 2 & (0, -1)\times (0, -1)\times (-1, 0)            & \text{bulk D7$_3$}          \\
		h_1                                       & 4 & (1, 0)\times (1, 0)\times (1, 0)               & \text{bulk D3}              \\
		h_2                                       & 4 & (-1, 0)\times (-1, 0)\times (1, 0)             & \text{bulk D3}              \\
		h_3                                       & 4 & (1, 0)\times (-1, 0)\times (-1, 0)             & \text{bulk D3}              \\
		h_4                                       & 4 & (-1, 0)\times (1, 0)\times (-1, 0)             & \text{bulk D3}              \\
		\hline
	\end{array}$
	\caption{Model~\hyperref[spec:r27f3]{r27f3} with the gauge group $\SU(4)_{C}\times\SU(2)_{L}\times\SU(2)_{R_1}\times\SU(2)_{R_2}\times\SU(2)^5\times\USp(8)^4$, the torus moduli $\chi_1=\frac{8}{3}$, $\chi_2=8$, $\chi_3= 24$ and the gauge coupling relation $g_a^2=\frac{5}{4}g_b^2=\frac{1885}{324}g_{cd}^2=\frac{9425}{4742}\frac{5 g_Y^2}{3}=2 \sqrt{2}\, \pi \,  e^{\phi _4}$.}
	\label{model:r27f3}
\end{table}

\begin{table}[th]
	\centering
	$\begin{array}{|c|c|c|l|}
		\hline
		\text{Model~\hyperref[spec:r35f3]{r35f3}} & N & (n^1, m^1) \times (n^2, m^2) \times (n^3, m^3) & \text{Type of brane}        \\ 
		\hline 
		a                                         & 4 & (0, -1)\times (0, 1)\times (1, 0)              & \text{frac. D7$_3$}         \\
		b                                         & 2 & (-4, 1)\times (1, 0)\times (-2, -3)            & \text{frac. D7$_2$ w. flux} \\
		c                                         & 2 & (4, -1)\times (-3, 1)\times (4, -1)            & \text{frac. D9 w. flux}     \\
		d                                         & 2 & (-1, 0)\times (2, -1)\times (-4, -3)           & \text{frac. D7$_1$ w. flux} \\
		e                                         & 2 & (1, 0)\times (0, -1)\times (0, 1)              & \text{bulk D7$_1$}          \\
		f_1                                       & 2 & (0, -1)\times (1, 0)\times (0, 1)              & \text{bulk D7$_2$}          \\
		f_2                                       & 2 & (0, 1)\times (-1, 0)\times (0, 1)              & \text{bulk D7$_2$}          \\
		g_1                                       & 2 & (0, -1)\times (0, 1)\times (1, 0)              & \text{bulk D7$_3$}          \\
		g_2                                       & 2 & (0, -1)\times (0, -1)\times (-1, 0)            & \text{bulk D7$_3$}          \\
		h_1                                       & 6 & (1, 0)\times (1, 0)\times (1, 0)               & \text{bulk D3}              \\
		h_2                                       & 6 & (-1, 0)\times (-1, 0)\times (1, 0)             & \text{bulk D3}              \\
		h_3                                       & 6 & (1, 0)\times (-1, 0)\times (-1, 0)             & \text{bulk D3}              \\
		h_4                                       & 6 & (-1, 0)\times (1, 0)\times (-1, 0)             & \text{bulk D3}              \\
		\hline
	\end{array}$
	\caption{Model~\hyperref[spec:r35f3]{r35f3} with the gauge group $\SU(4)_{C}\times\SU(2)_{L}\times\SU(2)_{R_1}\times\SU(2)_{R_2}\times\SU(2)^5\times\USp(12)^4$, the torus moduli $\chi_1=12 \sqrt{3}$, $\chi_2=3 \sqrt{3}$, $\chi_3= 2 \sqrt{3}$ and the gauge coupling relation $g_a^2=\frac{56}{27}g_b^2=\frac{1736}{3303}g_{cd}^2=\frac{8680}{13381}\frac{5 g_Y^2}{3}=\frac{4 \sqrt{2}}{3 \sqrt[4]{3}}\, \pi \,  e^{\phi _4}$.}
	\label{model:r35f3}
\end{table}

\begin{table}[th]
	\centering
	$\begin{array}{|c|c|c|l|}
		\hline
		\text{Model~\hyperref[spec:r75f3]{r75f3}} & N  & (n^1, m^1) \times (n^2, m^2) \times (n^3, m^3) & \text{Type of brane}        \\ 
		\hline 
		a                                         & 4  & (1, 0)\times (0, -1)\times (0, 1)              & \text{frac. D7$_1$}         \\
		b                                         & 2  & (-5, -1)\times (-5, 2)\times (1, 0)            & \text{frac. D7$_3$ w. flux} \\
		c                                         & 2  & (4, 1)\times (5, 1)\times (-4, -1)             & \text{frac. D9 w. flux}     \\
		d                                         & 2  & (-1, -2)\times (-1, 0)\times (3, -1)           & \text{frac. D7$_2$ w. flux} \\
		e                                         & 2  & (0, -1)\times (1, 0)\times (0, 1)              & \text{bulk D7$_2$}          \\
		f_1                                       & 2  & (1, 0)\times (0, -1)\times (0, 1)              & \text{bulk D7$_1$}          \\
		f_2                                       & 2  & (-1, 0)\times (0, 1)\times (0, 1)              & \text{bulk D7$_1$}          \\
		g_1                                       & 2  & (0, -1)\times (0, 1)\times (1, 0)              & \text{bulk D7$_3$}          \\
		g_2                                       & 2  & (0, -1)\times (0, -1)\times (-1, 0)            & \text{bulk D7$_3$}          \\
		h_1                                       & 16 & (1, 0)\times (1, 0)\times (1, 0)               & \text{bulk D3}              \\
		h_2                                       & 16 & (-1, 0)\times (-1, 0)\times (1, 0)             & \text{bulk D3}              \\
		h_3                                       & 16 & (1, 0)\times (-1, 0)\times (-1, 0)             & \text{bulk D3}              \\
		h_4                                       & 16 & (-1, 0)\times (1, 0)\times (-1, 0)             & \text{bulk D3}              \\
		\hline
	\end{array}$
	\caption{Model~\hyperref[spec:r75f3]{r75f3} with the gauge group $\SU(4)_{C}\times\SU(2)_{L}\times\SU(2)_{R_1}\times\SU(2)_{R_2}\times\SU(2)^5\times\USp(32)^4$, the torus moduli $\chi_1=2 \sqrt{\frac{37}{3}}$, $\chi_2=\sqrt{\frac{37}{3}}$, $\chi_3= 4 \sqrt{111}$ and the gauge coupling relation $g_a^2=\frac{223}{444}g_b^2=\frac{133280}{42587}g_{cd}^2=\frac{666400}{394321}\frac{5 g_Y^2}{3}=\frac{4 \sqrt{2}}{\sqrt[4]{111}}\, \pi \,  e^{\phi _4}$.}
	\label{model:r75f3}
\end{table}

\begin{table}[th]
	\centering
	$\begin{array}{|c|c|c|l|}
		\hline
		\text{Model~\hyperref[spec:r76f3]{r76f3}} & N  & (n^1, m^1) \times (n^2, m^2) \times (n^3, m^3) & \text{Type of brane}        \\ 
		\hline 
		a                                         & 4  & (0, -1)\times (1, 0)\times (0, 1)              & \text{frac. D7$_2$}         \\
		b                                         & 2  & (1, 0)\times (-5, -1)\times (-5, 2)            & \text{frac. D7$_1$ w. flux} \\
		c                                         & 2  & (-4, 1)\times (4, -1)\times (5, -1)            & \text{frac. D9 w. flux}     \\
		d                                         & 2  & (3, 2)\times (-1, 0)\times (-1, 1)             & \text{frac. D7$_2$ w. flux} \\
		e_1                                       & 2  & (1, 0)\times (0, -1)\times (0, 1)              & \text{bulk D7$_1$}          \\
		e_2                                       & 2  & (-1, 0)\times (0, 1)\times (0, 1)              & \text{bulk D7$_1$}          \\
		f_1                                       & 2  & (0, -1)\times (0, -1)\times (-1, 0)            & \text{bulk D7$_3$}          \\
		f_2                                       & 4  & (0, -1)\times (0, 1)\times (1, 0)              & \text{bulk D7$_3$}          \\
		g_1                                       & 16 & (1, 0)\times (1, 0)\times (1, 0)               & \text{bulk D3}              \\
		g_2                                       & 16 & (-1, 0)\times (-1, 0)\times (1, 0)             & \text{bulk D3}              \\
		g_3                                       & 16 & (1, 0)\times (-1, 0)\times (-1, 0)             & \text{bulk D3}              \\
		g_4                                       & 16 & (-1, 0)\times (1, 0)\times (-1, 0)             & \text{bulk D3}              \\
		\hline
	\end{array}$
	\caption{Model~\hyperref[spec:r76f3]{r76f3} with the gauge group $\SU(4)_{C}\times\SU(2)_{L}\times\SU(2)_{R_1}\times\SU(2)_{R_2}\times\SU(2)^3\times\SU(4)\times\USp(32)^4$, the torus moduli $\chi_1=\sqrt{\frac{129}{2}}$, $\chi_2=2 \sqrt{\frac{86}{3}}$, $\chi_3= \sqrt{\frac{86}{3}}$ and the gauge coupling relation $g_a^2=\frac{419}{129}g_b^2=\frac{200606}{108403}g_{cd}^2=\frac{1003030}{726421}\frac{5 g_Y^2}{3}=\frac{8\ 2^{3/4}}{\sqrt[4]{129}}\, \pi \,  e^{\phi _4}$.}
	\label{model:r76f3}
\end{table}

\begin{table}[th]
	\centering
	$\begin{array}{|c|c|c|l|}
		\hline
		\text{Model~\hyperref[spec:r20f4]{r20f4}} & N & (n^1, m^1) \times (n^2, m^2) \times (n^3, m^3) & \text{Type of brane}        \\ 
		\hline 
		a                                         & 4 & (0, 1)\times (0, -1)\times (1, 0)              & \text{frac. D7$_3$}         \\
		b                                         & 2 & (-1, 0)\times (5, -1)\times (-1, -1)           & \text{frac. D7$_1$ w. flux} \\
		c                                         & 2 & (-3, -1)\times (1, 0)\times (-4, 1)            & \text{frac. D7$_2$ w. flux} \\
		d                                         & 2 & (3, -1)\times (-7, 1)\times (2, -1)            & \text{frac. D9 w. flux}     \\
		e_1                                       & 4 & (1, 0)\times (0, -1)\times (0, 1)              & \text{bulk D7$_1$}          \\
		e_2                                       & 4 & (-1, 0)\times (0, 1)\times (0, 1)              & \text{bulk D7$_1$}          \\
		f_1                                       & 4 & (0, -1)\times (0, 1)\times (1, 0)              & \text{bulk D7$_3$}          \\
		f_2                                       & 4 & (0, -1)\times (0, -1)\times (-1, 0)            & \text{bulk D7$_3$}          \\
		g                                         & 2 & (1, 0)\times (1, 0)\times (1, 0)               & \text{bulk D3}              \\
		\hline
	\end{array}$
	\caption{Model~\hyperref[spec:r20f4]{r20f4} with the gauge group $\SU(4)_{C}\times\SU(2)_{L}\times\SU(2)_{R_1}\times\SU(2)_{R_2}\times\SU(4)^3\times\SU(4)\times\USp(4)$, the torus moduli $\chi_1=\frac{3 \sqrt{\frac{41}{10}}}{2}$, $\chi_2=\sqrt{410}$, $\chi_3= \sqrt{\frac{82}{5}}$ and the gauge coupling relation $g_a^2=\frac{58}{41}g_b^2=\frac{153}{410}g_{cd}^2=\frac{255}{512}\frac{5 g_Y^2}{3}=\frac{8\ 2^{3/4}}{\sqrt{3} \sqrt[4]{205}}\, \pi \,  e^{\phi _4}$.}
	\label{model:r20f4}
\end{table}

\begin{table}[th]
	\centering
	$\begin{array}{|c|c|c|l|}
		\hline
		\text{Model~\hyperref[spec:r26f4]{r26f4}} & N & (n^1, m^1) \times (n^2, m^2) \times (n^3, m^3) & \text{Type of brane}        \\ 
		\hline 
		a                                         & 4 & (1, 0)\times (0, 1)\times (0, -1)              & \text{frac. D7$_1$}         \\
		b                                         & 2 & (-3, -1)\times (1, 0)\times (-5, 1)            & \text{frac. D7$_2$ w. flux} \\
		c                                         & 2 & (2, -1)\times (-5, 1)\times (5, -1)            & \text{frac. D9 w. flux}     \\
		d                                         & 2 & (-2, 1)\times (3, 1)\times (-1, 0)             & \text{frac. D7$_3$ w. flux} \\
		e_1                                       & 2 & (0, -1)\times (1, 0)\times (0, 1)              & \text{bulk D7$_2$}          \\
		e_2                                       & 2 & (0, 1)\times (-1, 0)\times (0, 1)              & \text{bulk D7$_2$}          \\
		f_1                                       & 2 & (0, -1)\times (0, 1)\times (1, 0)              & \text{bulk D7$_3$}          \\
		f_2                                       & 2 & (0, -1)\times (0, -1)\times (-1, 0)            & \text{bulk D7$_3$}          \\
		g_1                                       & 4 & (1, 0)\times (0, -1)\times (0, 1)              & \text{bulk D7$_1$}          \\
		g_2                                       & 4 & (-1, 0)\times (0, 1)\times (0, 1)              & \text{bulk D7$_1$}          \\
		h_1                                       & 4 & (1, 0)\times (1, 0)\times (1, 0)               & \text{bulk D3}              \\
		h_2                                       & 2 & (-1, 0)\times (-1, 0)\times (1, 0)             & \text{bulk D3}              \\
		h_3                                       & 2 & (1, 0)\times (-1, 0)\times (-1, 0)             & \text{bulk D3}              \\
		h_4                                       & 2 & (-1, 0)\times (1, 0)\times (-1, 0)             & \text{bulk D3}              \\
		\hline
	\end{array}$
	\caption{Model~\hyperref[spec:r26f4]{r26f4} with the gauge group $\SU(4)_{C}\times\SU(2)_{L}\times\SU(2)_{R_1}\times\SU(2)_{R_2}\times\SU(2)^4\times\SU(4)^2\times\USp(8)\times\USp(4)^3$, the torus moduli $\chi_1=2 \sqrt{\frac{17}{3}}$, $\chi_2=\sqrt{51}$, $\chi_3= \frac{10 \sqrt{\frac{17}{3}}}{3}$ and the gauge coupling relation $g_a^2=\frac{95}{102}g_b^2=\frac{285}{442}g_{cd}^2=\frac{475}{632}\frac{5 g_Y^2}{3}=\frac{8 \sqrt[4]{\frac{3}{17}}}{\sqrt{5}}\, \pi \,  e^{\phi _4}$.}
	\label{model:r26f4}
\end{table}

\begin{table}[th]
	\centering
	$\begin{array}{|c|c|c|l|}
		\hline
		\text{Model~\hyperref[spec:r33f4]{r33f4}} & N & (n^1, m^1) \times (n^2, m^2) \times (n^3, m^3) & \text{Type of brane}        \\ 
		\hline 
		a                                         & 4 & (0, 1)\times (0, -1)\times (1, 0)              & \text{frac. D7$_3$}         \\
		b                                         & 2 & (1, 0)\times (-4, 1)\times (-2, -1)            & \text{frac. D7$_1$ w. flux} \\
		c                                         & 2 & (-3, 1)\times (4, -1)\times (4, -1)            & \text{frac. D9 w. flux}     \\
		d                                         & 2 & (2, 1)\times (-1, 0)\times (-4, 1)             & \text{frac. D7$_2$ w. flux} \\
		e_1                                       & 2 & (0, 1)\times (1, 0)\times (0, -1)              & \text{bulk D7$_2$}          \\
		e_2                                       & 2 & (0, -1)\times (1, 0)\times (0, 1)              & \text{bulk D7$_2$}          \\
		e_3                                       & 2 & (0, -1)\times (-1, 0)\times (0, -1)            & \text{bulk D7$_2$}          \\
		f_1                                       & 2 & (0, -1)\times (0, 1)\times (1, 0)              & \text{bulk D7$_3$}          \\
		f_2                                       & 2 & (0, -1)\times (0, -1)\times (-1, 0)            & \text{bulk D7$_3$}          \\
		g_1                                       & 4 & (1, 0)\times (0, -1)\times (0, 1)              & \text{bulk D7$_1$}          \\
		g_2                                       & 4 & (-1, 0)\times (0, 1)\times (0, 1)              & \text{bulk D7$_1$}          \\
		h_1                                       & 4 & (1, 0)\times (1, 0)\times (1, 0)               & \text{bulk D3}              \\
		h_2                                       & 4 & (-1, 0)\times (-1, 0)\times (1, 0)             & \text{bulk D3}              \\
		h_3                                       & 4 & (1, 0)\times (-1, 0)\times (-1, 0)             & \text{bulk D3}              \\
		h_4                                       & 4 & (-1, 0)\times (1, 0)\times (-1, 0)             & \text{bulk D3}              \\
		\hline
	\end{array}$
	\caption{Model~\hyperref[spec:r33f4]{r33f4} with the gauge group $\SU(4)_{C}\times\SU(2)_{L}\times\SU(2)_{R_1}\times\SU(2)_{R_2}\times\SU(2)^5\times\SU(4)^2\times\USp(8)^4$, the torus moduli $\chi_1=\sqrt{11}$, $\chi_2=4 \sqrt{11}$, $\chi_3= 2 \sqrt{11}$ and the gauge coupling relation $g_a^2=\frac{24}{11}g_b^2=\frac{120}{187}g_{cd}^2=\frac{200}{267}\frac{5 g_Y^2}{3}=\frac{4 \sqrt{2}}{\sqrt[4]{11}}\, \pi \,  e^{\phi _4}$.}
	\label{model:r33f4}
\end{table}

\section{Decoupling of exotic particles}\label{appB} 
In this appendix we describe the decoupling of chiral exotic states through strong coupling dynamics in the hidden sector, cf. \cite{Cvetic:2004ui}. The exotic matter fields are charged under asymptotically free hidden gauge groups which become strongly coupled at an intermediate scale. As a result, these states confine and reorganize into gauge invariant composite operators, removing the chiral exotics from the low energy spectrum.

The exotic fields arise at intersections involving the confining stacks and transform in fundamental or bifundamental representations. Upon confinement, gauge invariant composites are formed with quantum numbers fixed by the strong dynamics. In particular, pairs of hidden sector doublets combine as $2\otimes 2 = 3 \oplus 1$, yielding composite states in real or singlet representations of the confining groups. These composites are therefore non chiral and decouple from the infrared theory. Importantly, this mechanism does not introduce new gauge or mixed anomalies. Anomaly matching is automatically satisfied, while the visible sector gauge symmetry remains unbroken. Consequently, the resulting low energy effective theory is anomaly consistent and free of chiral exotics without the need for additional Higgsing or explicit mass terms.

\begin{table*}[th]
	\centering \footnotesize

	\caption{The composite particle spectrum of Model~\hyperref[spec:r15f1]{r15f1} formed due to the strong forces in hidden sector.} 
	\label{exotic:r15f1}
\end{table*}

\begin{table*}[th]
	\centering \footnotesize
	%
	\caption{The composite particle spectrum of Model~\hyperref[spec:r17f1]{r17f1} formed due to the strong forces in hidden sector.} 
	\label{exotic:r17f1}
\end{table*}

\begin{table*}[th]
	\centering \footnotesize
	%
	\caption{The composite particle spectrum of Model~\hyperref[spec:r43f1]{r43f1} formed due to the strong forces in hidden sector.} 
	\label{exotic:r43f1}
\end{table*}

\begin{table*}[th]
	\centering \footnotesize
	\begin{tabular}{c}$\begin{array}{|c|c|D{.}{\times}{0}|c|}
		\hline \multicolumn{1}{|c|}{\text{Model~\hyperref[model:r7f2]{r7f2}}} & \multicolumn{3}{c|}{\SU(4)_{C}\times\SU(2)_{L}\times\SU(2)_{R_1}\times\SU(2)_{R_2}\times\SU(2)} \\
		\hline 
		\text{Confining Force} & \text{Intersection} & \mc{\text{Exotic Particle}}          & \text{Confined Particle Spectrum} \\
		\hline\SU(2)_{e}       & ce                  & 2\times \left(1,1,2,1,\bar{2}\right) & \left(1,1,2^2,1,1\right)          \\
		                       & de                  & 1\times \left(1,1,1,2,\bar{2}\right) & \left(1,1,1,2^2,1\right)          \\
		                       & be'                 & 1\times (1,2,1,1,2)                  & \left(1,2^2,1,1,1\right)          \\
		                       & ce'                 & 2\times (1,1,2,1,2)                  & \left(1,1,2^2,1,1\right)          \\
		                       & de'                 & 2\times (1,1,1,2,2)                  & \left(1,1,1,2^2,1\right)          \\\hline
		\end{array}$
	\end{tabular}
	\caption{The composite particle spectrum of Model~\hyperref[spec:r7f2]{r7f2} formed due to the strong forces in hidden sector.} 
	\label{exotic:r7f2}
\end{table*}

\begin{table*}[th]
	\centering \footnotesize
	\begin{tabular}{c}$\begin{array}{|c|c|D{.}{\times}{0}|c|}
		\hline \multicolumn{1}{|c|}{\text{Model~\hyperref[model:r10f2]{r10f2}}} & \multicolumn{3}{c|}{\SU(4)_{C}\times\SU(2)_{L}\times\SU(2)_{R_1}\times\SU(2)_{R_2}\times\SU(2)\times\SU(4)} \\
		\hline 
		\text{Confining Force} & \text{Intersection} & \mc{\text{Exotic Particle}}            & \text{Confined Particle Spectrum} \\
		\hline\SU(2)_{e_1}     & be_1                & 2\times \left(1,2,1,1,\bar{2},1\right) & \left(1,2^2,1,1,1,1\right)        \\
		                       & de_1                & 4\times \left(1,1,1,2,\bar{2},1\right) & \left(1,1,1,2^2,1,1\right)        \\
		                       & be_1'               & 2\times (1,2,1,1,2,1)                  & \left(1,2^2,1,1,1,1\right)        \\
		                       & ce_1'               & 2\times (1,1,2,1,2,1)                  & \left(1,1,2^2,1,1,1\right)        \\
		                       & de_1'               & 2\times (1,1,1,2,2,1)                  & \left(1,1,1,2^2,1,1\right)        \\
		\hline
		\SU(4)_{e_2}           & be_2                & 2\times \left(1,2,1,1,1,\bar{4}\right) & \left(1,2^2,1,1,1,1\right)        \\
		                       & de_2                & 4\times \left(1,1,1,2,1,\bar{4}\right) & \left(1,1,1,2^2,1,1\right)        \\
		                       & be_2'               & 2\times (1,2,1,1,1,4)                  & \left(1,2^2,1,1,1,1\right)        \\
		                       & ce_2'               & 2\times (1,1,2,1,1,4)                  & \left(1,1,2^2,1,1,1\right)        \\
		                       & de_2'               & 2\times (1,1,1,2,1,4)                  & \left(1,1,1,2^2,1,1\right)        \\\hline
		\end{array}$
	\end{tabular}%
	\caption{The composite particle spectrum of Model~\hyperref[spec:r43af2]{r43af2} formed due to the strong forces in hidden sector.} 
	\label{exotic:r43af2}
\end{table*}

\begin{table*}[th]
	\centering \footnotesize
	%
	\caption{The composite particle spectrum of Model~\hyperref[spec:r43bf2]{r43bf2} formed due to the strong forces in hidden sector.} 
	\label{exotic:r43bf2}
\end{table*}

\begin{table*}[th]
	\centering \footnotesize
	%
	\caption{The composite particle spectrum of Model~\hyperref[spec:r33f4]{r33f4} formed due to the strong forces in hidden sector.} 
	\label{exotic:r33f4}
\end{table*}

\section{Beta function calculation for the SU(4) fractional brane}\label{appC} 

\begin{align}
	\text{Model~\hyperref[spec:r15f1]{r15f1}}: \quad \beta^{\SU(4)}   & = -3 \times 4 + 16 \times \frac{1}{2} + 2 \times 1 = -2 \,, \nonumber \\
	\text{Model~\hyperref[spec:r17f1]{r17f1}}: \quad \beta^{\SU(4)}   & = -3 \times 4 + 12 \times \frac{1}{2} + 2 \times 1 = -4 \,, \nonumber \\
	\text{Model~\hyperref[spec:r43f1]{r43f1}}: \quad \beta^{\SU(4)}   & = -3 \times 4 + 24 \times \frac{1}{2} + 2 \times 1 = 2 \,, \nonumber  \\
	\text{Model~\hyperref[spec:r7f2]{r07f2}}: \quad \beta^{\SU(4)}    & = -3 \times 4 + 20 \times \frac{1}{2} + 2 \times 1 = 0 \,, \nonumber  \\
	\text{Model~\hyperref[spec:r10f2]{r10f2}}: \quad \beta^{\SU(4)}   & = -3 \times 4 + 12 \times \frac{1}{2} + 2 \times 1 = -4 \,, \nonumber \\ 
	\text{Model~\hyperref[spec:r35f2]{r35f2}}: \quad \beta^{\SU(4)}   & = -3 \times 4 + 36 \times \frac{1}{2} + 2 \times 1 = 8 \,, \nonumber  \\
	\text{Model~\hyperref[spec:r43af2]{r43af2}}: \quad \beta^{\SU(4)} & = -3 \times 4 + 36 \times \frac{1}{2} + 2 \times 1 = 8 \,, \nonumber  \\
	\text{Model~\hyperref[spec:r43bf2]{r43bf2}}: \quad \beta^{\SU(4)} & = -3 \times 4 + 44 \times \frac{1}{2} + 2 \times 1 = 12 \,, \nonumber \\
	\text{Model~\hyperref[spec:r123f2]{r123f2}}: \quad \beta^{\SU(4)} & = -3 \times 4 + 28 \times \frac{1}{2} + 2 \times 1 = 4 \,, \nonumber  \\
	\text{Model~\hyperref[spec:r125f2]{r125f2}}: \quad \beta^{\SU(4)} & = -3 \times 4 + 28 \times \frac{1}{2} + 2 \times 1 = 4 \,, \nonumber  \\
	\text{Model~\hyperref[spec:r27f3]{r27f3}}: \quad \beta^{\SU(4)}   & = -3 \times 4 + 32 \times \frac{1}{2} + 2 \times 1 = 6 \,, \nonumber  \\ 
	\text{Model~\hyperref[spec:r35f3]{r35f3}}: \quad \beta^{\SU(4)}   & = -3 \times 4 + 44 \times \frac{1}{2} + 2 \times 1 = 12 \,, \nonumber \\ 
	\text{Model~\hyperref[spec:r75f3]{r75f3}}: \quad \beta^{\SU(4)}   & = -3 \times 4 + 36 \times \frac{1}{2} + 2 \times 1 = 8 \,, \nonumber  \\ 
	\text{Model~\hyperref[spec:r76f3]{r76f3}}: \quad \beta^{\SU(4)}   & = -3 \times 4 + 36 \times \frac{1}{2} + 2 \times 1 = 8 \,, \nonumber  \\ 
	\text{Model~\hyperref[spec:r20f4]{r20f4}}: \quad \beta^{\SU(4)}   & = -3 \times 4 + 32 \times \frac{1}{2} + 2 \times 1 = 6 \,, \nonumber  \\
	\text{Model~\hyperref[spec:r26f4]{r26f4}}: \quad \beta^{\SU(4)}   & = -3 \times 4 + 36 \times \frac{1}{2} + 2 \times 1 = 8 \,, \nonumber  \\ 
	\text{Model~\hyperref[spec:r33f4]{r33f4}}: \quad \beta^{\SU(4)}   & = -3 \times 4 + 24 \times \frac{1}{2} + 2 \times 1 = 2 \,.            
\end{align} 
  
\FloatBarrier
 
\bibliographystyle{JHEP}

\providecommand{\href}[2]{#2}\begingroup\raggedright\endgroup
  
\end{document}